\documentclass[fleqn,usenatbib]{mnras}

\usepackage{newtxtext,newtxmath}

\usepackage[T1]{fontenc}

\DeclareRobustCommand{\VAN}[3]{#2}
\let\VANthebibliography\thebibliography
\def\thebibliography{\DeclareRobustCommand{\VAN}[3]{##3}\VANthebibliography}

\usepackage{graphicx}	
\usepackage{amsmath}	
\usepackage{xcolor}
\usepackage{orcidlink}

\newcommand{\cf}[1]{{\textcolor{black}{#1}}}

\newcommand{\rf}[1]{{\textcolor{black}{#1}}}

\title[Atmospheric Parameters from BP/RP using UNNs]{Stellar Atmospheric Parameters From Gaia BP/RP Spectra using Uncertain Neural Networks}

\author[C. P. Fallows and J. L. Sanders]{
Connor P. Fallows\thanks{connor.fallows.20@ucl.ac.uk}
and Jason L. Sanders \orcidlink{0000-0003-4593-6788}
\\
University College London, Gower St., London WC1E 6BT, UK
}

\date{Accepted XXX. Received YYY; in original form ZZZ}

\pubyear{2023}

\begin{document}
\label{firstpage}
\pagerange{\pageref{firstpage}--\pageref{lastpage}}
\maketitle

\begin{abstract}
With the plentiful information available in the Gaia BP/RP spectra, there is significant scope for applying discriminative models to extract stellar atmospheric parameters and abundances. We describe an approach to leverage an `Uncertain Neural Network' model trained on APOGEE data to provide high-quality predictions with robust estimates for per-prediction uncertainty. We report median formal uncertainties of \rf{0.068 dex, 69.1K, 0.14 dex, 0.031 dex, 0.040 dex, and 0.029} dex for [Fe/H], $T_\mathrm{eff}$, $\log g$, [C/Fe], [N/Fe], and [$\alpha$/M] respectively. We validate these predictions against our APOGEE training data, LAMOST, and Gaia GSP-Phot stellar parameters, and see a strong correlation between our predicted parameters and those derived from these surveys. 
We investigate the information content of the spectra by considering the `attention' our model pays to different spectral features compared to expectations from synthetic spectra calculations.
Our model's predictions are applied to the Gaia dataset, and we produce a publicly available catalogue of our model's predictions.
\end{abstract}




\section{Introduction}

Our understanding of the structure and evolution of the Milky Way is driven by the analysis of its constituent stellar populations. Stars can be separated into distinct populations using their measured properties, such as effective temperature, surface gravity, mass, and age. The greater the number of properties identified, and the more accurately these properties are estimated, the more precisely a star's population --  and hence its evolutionary history --  can be known.

Further information on stellar populations is provided by elemental abundances.
\rf{Most basically, the stellar metallicity, $\mathrm{[M/H]}$, tracks the build-up of metals returned to the interstellar medium through a range of stellar evolutionary channels. Abundance ratios, e.g. $[\alpha/\mathrm{M}]$, give specific information on the relative importance of different evolutionary processes. Although a given star can stray far from its birth site \citep[through processes such as radial migration, ][]{Sellwood2002, Shonrich2009, Dantas2023}, the elemental fingerprint is largely preserved and acts like a time capsule that can be used to trace the star back to its birth environment \citep[e.g.][]{Frankel2020}. Stars with a shared origin can then be successfully identified through their location in abundance space \citep[e.g.][]{Zasowski2016, Bovy2016, Naidu2020}. 
The time capsule idea fails for lighter elements like carbon and nitrogen, which are processed within stars throughout their lifetimes.} However, this turns out to be an advantage, as carbon and nitrogen abundances then reveal mass and age information of red giant stars \citep{Masseron2015, Martig2016, Casali2019}, allowing a deeper probe into the formation history of different Galactic populations. Therefore, accurate stellar parameters and abundances provide an essential insight into how Milky Way populations have developed and allow us to trace the formation of our Galaxy's stellar populations, structure, and components throughout its history. In this way, we can place the Milky Way's evolution into the general context of galactic evolution across the Universe \citep{Bland-Hawth2016, Barbuy2018}

Accurate stellar parameters and abundances are perhaps best extracted from detailed spectra, where the shapes and depths of individual elemental and molecular lines are clearly measurable \citep{Hotlzman2015, Buder2019}. Recent years have seen the industrialization of this practice with the rise of large-scale spectroscopic surveys such as APOGEE \citep{Abdurrouf2022}, GALAH \citep{Buder2019} and LAMOST \citep{Cui2012} providing spectra and accurate stellar parameters and abundances for around $10$ million stars.
However, such observations are necessarily targeted and time-intensive thus giving a restricted view of the entire Galaxy. Recently, there have been efforts to utilise the power of data-driven approaches to transfer knowledge gained from smaller samples of well-studied stars to much bigger samples of typically lower-quality data that crucially cover larger swathes of the Milky Way \citep{Xiang2019, LiZ2022, Andrae2023b, LiX2023}.

Perhaps the largest stellar datasets come from photometric surveys. Careful choice of stellar colour combinations from these surveys can reveal information on stellar parameters and abundances. For example, the SkyMapper \citep{SkyMapper, Chiti2021} and PRISTINE \citep{PRISTINE_old, Martin2023} surveys target the near-UV Ca H \& K region with narrow-band filters to provide \rf{precise} metallicity estimates with uncertainties $\lesssim0.2\,\mathrm{dex}$. This wavelength range is strongly affected by extinction, so for studying high-extinction regions, some authors have demonstrated the power of using near- and mid-infrared photometric surveys with data-driven techniques to estimate metallicities \citep{Schlaufman2014, Koposov2015, Li2016, Casey2018, Grady2021, Fallows2022} with uncertainties of $0.2-0.4\,\mathrm{dex}$.

As part of the third data release of the Gaia survey \citep{GaiaDR3}, we now have access to low-resolution spectroscopic measurements for more than 220 million objects, referred to herein as the Gaia `BP/RP' spectra. These spectra have a wavelength-dependent resolution of between $30$ and $100$ across the optical wavelength range of $330\,\mathrm{nm}-1050\,\mathrm{nm}$ \citep{Carrasco2021,DeAngeli2022,Montegriffo2022}. Despite their low resolution, there is sufficient information in the spectra for accurate estimates of the `basic' stellar parameters of effective temperature, surface gravity and metallicity \citep{Liu2012, Witten2022, Xylakis2022, Andrae2023a, Martin2023} and the possibility that more detailed abundance measurements of e.g. carbon and alpha elements can also be made \citep{Gavel2021, Witten2022, Lucey2023, Sanders2023, Li2023_AspGap}. Furthermore, the BP/RP spectra are provided for all Gaia sources down to $G=17.65\,\mathrm{mag}$ across the entire sky. In this way, the Gaia BP/RP dataset represents a compromise between the accurate, focused view offered by higher-resolution spectroscopic studies and the more panoramic view offered by photometric surveys. 

The quantity and complexity of the Gaia BP/RP data mean they are perhaps best exploited through machine-learning/data-driven techniques. For example, 
\citet{Andrae2023b} and \citet{Belokurov2022} have demonstrated the use of discriminative ensemble learning techniques such as XGBoost \citep{XGBoost} 
using APOGEE and LAMOST stellar training labels to derive accurate 
metallicity, surface gravity and effective temperature from the Gaia BP/RP spectra. \rf{Discriminate neural networks (NN) further extend these capabilities, allowing the use of spectroscopic, photometric, and astrometric datasets simultaneously to make predictions \citep{Guiglion2020, Sprague2022}.}

More sophisticated generative approaches using neural networks \citep[]{Zhang2023, Li2023_AspGap, Guiglon2023} and transformer models \citep{NNTransformer, Leung2023} provide an alternative route to stellar parameter estimation and allow the models to be used in different modes for different tasks (e.g. mock population generation).

One advantage the generative approaches have over the discriminative approaches is the conceptual simplicity with which different sources of uncertainty can be included in the modelling.
Using ensemble learning, it is difficult to propagate input (or aleatoric) uncertainty 
\citep{Wang2018}, thus limiting how reliably these models can be trained when measurement uncertainties are dominant.
A further source of uncertainty is the range of possible models that are consistent with the training set (so-called epistemic uncertainty).
Bayesian NNs \citep{Denker1990,Wilson2020} \cf{approach this by marginalising over the weights and biases of the model, allowing the NN to produce a probability distribution for each prediction.
}\cf{While a full Bayesian methodology allows all uncertainties to be propagated within the network, these models are often complex to design and expensive to train. Therefore, many prior works have implemented methods to approximate Bayesian marginalisation, significantly reducing the complexity of the model, e,g. NN drop-out algorithms \citep{Gal2015, Leung2019a}, network ensembles \citep{Lakshmi2016}, or leveraging the backpropagation of the NN itself \citep{Blundell2015} to approximate the Bayesian marginalisation across the model.}

\cf{The prior work on Gaia BP/RP spectra has left an opportunity to build a lightweight discriminative machine-learning method that can apply Bayesian methodology to predict high-quality stellar parameters and \rf{precise} uncertainties. This is the aim of our work. We construct a feed-forward neural network to predict APOGEE stellar parameters and abundances given the BP/RP spectra complemented with photometry from Gaia, 2MASS and WISE. We account for the uncertainties in the inputs and outputs as well as additional model uncertainty. Given the success of prior works in measuring or forecasting more detailed abundance information from the BP/RP spectra \citep{Gavel2021, Witten2022, Li2023_AspGap}, we investigate the degree to which our model can measure some detailed elemental abundances and validate these measurements through comparisons to other survey results and synthetic spectra.} Section~\ref{sec:data} outlines our data selection. Section~\ref{sec:nn} describes our Neural Network setup, our approach to incorporate uncertainties into our predictions, and input data processing steps. Section~\ref{sec:validation} outlines our model's validation comparisons to external survey data, and Section~\ref{sec:spectral_info} details our approach to extract our model's attention to elemental features in the BP/RP spectra. Our conclusions are described in Section~\ref{sec:conclusions}.

\section{Data Selection}\label{sec:data}

\cf{Our approach requires two forms of data to train the model: input features that are fed into the model, and reference stellar parameters we use to train the model's predictions. For our input features, we select BP/RP spectra from Gaia DR3 \citep{GaiaDR3}, as well as photometry from Gaia, 2MASS \citep{Skrutskie2006}, and unWISE \citep{Schlafly2019}. We select parameters from SDSS/APOGEE \citep{Abdurrouf2022} for our model to learn to estimate.}

\subsection{Input Data}

The primary source of our input data is Gaia Data Release 3 \citep[DR3,][]{GaiaDR3}. From this, we select the three photometric magnitudes ($G$, $G_\mathrm{BP}$, $G_\mathrm{RP}$) and the low resolution BP/RP spectra. The spectra are represented in a continuous form as an array of basis coefficients from which the (internally- or flux-calibrated) spectra can be constructed through multiplication with a set of basis functions \citep{Carrasco2021}. $55$ basis coefficients are provided for each of the $G_\mathrm{BP}$ and $G_\mathrm{RP}$ wavelength ranges. We choose to leave the BP/RP data in its continuous form rather than convert it to sampled spectra. 
Note that we do not apply the basis coefficient truncation described in \citet{Carrasco2021}, and thus use the full $110$ basis coefficients for every source. As our method incorporates the data uncertainties into the model's training, the model will ignore any insignificant coefficients.

\cf{We remove sources with systematic photometric issues (blending) in $G_\mathrm{BP}$ and $G_\mathrm{RP}$ photometry by limiting to objects with \texttt{bp\_rp\_excess\_factor} < 3.0. Further, we filter for sources with good astrometric solutions by selecting objects with \texttt{ruwe} $\leq$ 1.4. As previously stated, we incorporate uncertainties into our model's training process, so we do not filter out high-uncertainty observations in any parameter.}

We complement the Gaia data with photometry from the 2MASS survey \citep{Skrutskie2006}, with its three primary band-passes, $J$, $H$, and $K_s$ (peak sensitivity of 1235nm, 1662nm, and 2159nm respectively), and the WISE survey \citep{Wright2010} with its four broad-band filters, $W1$, $W2$, $W3$, and $W4$ (peak sensitivity of 3.4$\mu$m, 4.6$\mu$m, 12$\mu$m, and 22$\mu$m respectively). \cf{We use the unWISE catalogue as the source of WISE photometry \citep{Schlafly2019} due to its greater depth in high-density regions of the Galaxy compared to previous WISE reductions. }\cf{From the set of 2MASS and WISE bands, we use only the $J$, $H$, $W1$, and $W2$ bands due to their known correlations with stellar parameters. \citet{Majewski2003} note that the 2MASS bands can be used to separate main-sequence and giant stars, while \citet{Grady2021} describe a strong correlation between the $W1-W2$ colour and stellar metallicity.}

\subsection{BP/RP coefficient normalisation}
\label{sec:coeff_norm}

\cf{We want our model's predictions to be insensitive to a star's apparent brightness, as otherwise this would introduce distance information into the input data. Trends in brightness and distance are likely to arise from selection effects in the chosen survey, rather than physical changes in the properties of the stars. As we want the model to learn physically important trends from the BP/RP spectra, we therefore need to normalise the coefficients to remove this apparent brightness information.}
\cf{To achieve this,} we normalise the input features relative to the first term in the RP spectra. For a set of 
unnormalized BP/RP coefficients $\tilde{x}_\mathrm{BP/RP}$. We choose the primary (0th) coefficient of the RP spectra as a normalisation factor, $\tilde{x}_{\mathrm{BP/RP},0}$, such that for the $i^{\mathrm{th}}$ coefficient, $\tilde{x}_{\mathrm{BP/RP}, i}$, the normalised coefficients, $x_{\mathrm{BP/RP}, i}$ are defined as
\begin{equation}
    x_{\mathrm{BP/RP}, i} \equiv \frac{\tilde{x}_{\mathrm{BP/RP}, i}}{\tilde{x}_{0}}.
\end{equation}
This normalises the BP/RP coefficients relative to the primary RP coefficient, with the $0$th RP coefficient now being equal to one for all stars. In the rare case where the $0$th RP coefficient is equal to zero, we instead select $x_{\mathrm{RP},0}$ to be the uncertainty in the 0th RP coefficient instead to maintain a similar order of magnitude in the normalisation. 
The covariance matrix of the normalized coefficients, ${\Sigma}^x_{i, j}$, is computed from the coefficient covariance matrix reported in Gaia DR3 $\tilde{{\Sigma}}^x_{i, j}$ as 
\begin{equation}
    {\Sigma}^x_{i,j} =  \frac{1}{{{\tilde x}}^2_{\mathrm{RP},0}}
    \left( {\tilde\Sigma}^x_{i,j}
    -{{x}}_{\mathrm{RP}, i} {\tilde\Sigma}^x_{0,j}
    -{{x}}_{\mathrm{RP}, j} {\tilde\Sigma}^x_{0,i}
    +{{x}}_{\mathrm{RP}, i} {{x}}_{\mathrm{RP}, j} {\tilde\Sigma}^x_{0,0}\right),
\end{equation}
for the RP coefficients and simply as ${{\Sigma}}^x_{i,j}=\tilde{\Sigma}^x_{i,j}/\tilde x_{\mathrm{RP},0}^2$ for the BP coefficients.

\cf{Similar to \cite{Andrae2023b}, we do not apply any extinction correction to the BP/RP coefficients. We thus rely on the model internally learning how to extinction correct the data to provide reliable predictions of effective temperature, for example. Furthermore, as the extinction is learned dynamically by the model from the training data, its predictions are not reliant on external extinction models (with their own uncertainties and biases to account for). This allows the model's predictions to be more robust, as well as removing unnecessary complexity from applying extinction laws.}

\subsection{Training Sample}

\cf{Our primary sample for the model's training (i.e. the output values the model will predict) must be high-quality stellar parameters. We therefore select our training sample from medium-resolution spectroscopic data:} the Apache Point Observatory Galactic Evolution Experiment 2 (APOGEE-2) survey, which is one of the sub-programs available as part of the 17th data release of the Sloan Digital Sky Survey (SDSS) \citep{Abdurrouf2022}. With medium-resolution ($R\sim$22,500), high signal-to-noise ($>100$) infrared spectra \citep{Majewski2017}, the APOGEE catalogue contains more than 650,000 stars within the Milky Way and its satellite galaxies. We use the calibrated parameter fit values from the APSCAP pipeline as our model's target parameters: \texttt{alpha\_m, teff, logg, c\_fe, n\_fe}, and \texttt{fe\_h}. The APOGEE data is filtered for `good' observations using the following flag bitmasks: \texttt{aspcapflag} bit 23; \texttt{starflag} bits 3, 4, 9, 12, 13, \& 16; and \texttt{extratarg} bit 4. 
This specific bitmask filtering removes objects with poorer observations in our chosen APOGEE parameters, but retains those with other measurement issues (such as spurious observations in parameters beyond our focus) that will not impact our method. From this filtering, we ensure our chosen parameter measurements are good, and increase the potential sample size available to train the model.
Combining these quality cuts with the previously described Gaia DR3 quality cuts results in a training sample of $3,845,370$ objects with Gaia, 2MASS, UNWISE, and APOGEE data. \rf{Of this training sample, $19,288$ objects have [Fe/H]$ < -1$ and $118$ objects have [Fe/H]$> 0.5$.}

\section{Methodology}\label{sec:nn}

We apply machine learning techniques to predict six stellar parameters, $\boldsymbol{y}$, from input normalized BP/RP spectra and broadband photometric colours, $\boldsymbol{x}$. The stellar parameters,  $\boldsymbol{y}$, are effective temperature ($T_\mathrm{{eff}}$), surface gravity ($\log g$), iron [Fe/H], carbon [C/Fe], nitrogen [N/Fe], and alpha [$\alpha$/M]  abundance. The specific input features, $\boldsymbol{x}$, are: 55 normalized RP coefficients from the Gaia BP/RP spectra, $\boldsymbol{x}_\mathrm{RP}$, 55 normalized BP coefficients from the BP/RP spectra, $\boldsymbol{x}_\mathrm{BP}$, 4 photometric colours (Gaia $G_\mathrm{BP}-G_\mathrm{RP}$, WISE $W1-W2$, 2MASS $J-H$, and $H-W2$ from 2MASS and WISE respectively), and a feature for each object's pseudo-absolute magnitude following
\begin{equation}
    M_{\mathrm{pseudo}} = \varpi 10^{0.2 G_{\mathrm{RP}}}.
	\label{eq:psuedo_mag}
\end{equation}
where $\varpi$ is an Gaia parallax, and $G_\mathrm{RP}$ is Gaia RP magnitude. This provides the model a proxy measure of an object's absolute magnitude, which improves predictions for parameters like $\log g$ (see Appendix~\ref{ap:pseudo_mag} for further details). This totals $115$ input features per object. 

\subsection{An uncertain neural network}
\label{sec:unc_metrics}
We employ a regression Neural Network (NN) machine learning algorithm for our model, $\mathcal{M}\equiv\boldsymbol{y}_\mathrm{pred}(\boldsymbol{x})$ with modifications to incorporate different sources of uncertainty, as we will discuss. These models are extremely effective for identifying trends in multi-dimensional datasets. Further, NNs are flexible in their design and allow us to easily modify and tune the specific structure of the model to implement Bayesian uncertainty propagation. Our NN is constructed using the \texttt{PyTorch} \citep{PyTorch} module for Python, and is constructed from 4 layers of interconnected nodes. Our initial input layer matches the size of our input feature matrix ($115$ nodes). 
Our NN contains two hidden linear layers of $128$ interconnected nodes, each with dropout applied at a probability of 20\% (each node has a 1/5 chance of being `dropped' each pass as discussed further in Sec.~\ref{sec:dropout_unc}). Finally, our output layer contains 12 nodes: 6 predictions for each feature of $\boldsymbol{y}$ and 6 nodes for the model's excess uncertainty in each feature of $\boldsymbol{y}$ (Sec.~\ref{sec:excess}). The NN is optimised using the \texttt{Adam} optimiser function \citep{ADAM}, with a learning rate of $5\times10^{-4}$, and a weight decay of $1\times10^{-5}$. We describe in detail our chosen loss function below.

Due to the size of the training dataset derived from APOGEE objects, we train our model in batches: 10,000 objects per batch, trained for 20,000 iterations per batch. \rf{We note that these batches are selected in a fixed order for training, to limit the introduction of additional scatter between separate training runs.} We use 2500 of these iterations to seed the NN's gradient for use in calculating input uncertainties (Sec.~\ref{sec:unc_propagation}), with the remaining iterations used to train the model with all uncertainties accounted for. \rf{We do not apply any form of early stop to the training process as dropout should prevent model overfitting \citep{Hinton2012}.}

Machine learning uncertainties can be divided into two components: aleatoric uncertainty and epistemic uncertainty. Aleatoric uncertainty is otherwise known as statistical uncertainty and is the uncertainty from random scatter within the model's data. Epistemic uncertainty, however, is the systematic uncertainty of the model, and describes the scatter produced from the architecture of the model itself. \cf{Such a distinction is important due to the aleatoric uncertainty changing per object in our sample, while the epistemic remains constant so long as the model is unchanged. Separating these two uncertainty metrics therefore allows us to probe the source of uncertainty in the model's predictions.}
First, we discuss the incorporation of epistemic uncertainty in our modelling. The aleatoric/measurement uncertainties will be discussed when we introduce the loss function. 

\subsubsection{Model Uncertainty}
\label{sec:dropout_unc}

Epistemic uncertainty can be introduced to the NN's predictions due to the specific training of the model being used. As the model is trained, the weights and biases within the network are tuned to produce the `best' predictions. However, for similar models trained on the same sample of data, there is no guarantee their internal architecture will be identical. \rf{For example, the random initialisation of the network's weights and biases will differ between models, altering the initial conditions of the NNs. This may cause variation in how these networks optimise during training, causing predictions to diverge between similarly trained NNs. We consider this divergence as a scatter in our model's prediction.}

\cf{We estimate this by emulating the outputs of an ensemble of similar models, allowing the average scatter in their predictions to be measured. This emulated ensemble is produced by introducing `drop-out' into our model's design. Drop-out \citep{Hinton2012} is the process of `zero-ing' a random fraction of layer nodes during forward passes through the network. These `zeroed' nodes are effectively removed from the model's architecture, forcing the NN's training to optimise its predictions without these removed nodes. Drop-out is usually implemented to limit over-fitting, forcing the model's `learning' to be more evenly distributed across the network as individual nodes cannot be relied upon for accurate predictions.}

\cf{As noted by \citet{Gal2015}, the inclusion of drop-out in NNs can be used to make a single model act like an ensemble of similarly accurate models. To build our model uncertainty, we repeat each prediction 50 times with drop-out enabled, producing 50 predictions which are randomly scattered by the model's architecture. From this array of predictions, we can calculate a median prediction (which we output as the overall prediction) and an uncertainty (which we report with the prediction).}

\cf{Note that we choose a drop-out fraction of 0.2 applied to each layer of the network (such that 20\% of nodes in each layer are zeroed for each forward pass), which we found to be large enough to reduce over-fitting and return a large difference between subsequent model predictions, while retaining enough nodes to allow the model to train accurately on reasonable timescales.}

\subsubsection{Excess Uncertainty}
\label{sec:excess}

\cf{We introduce an additional source of model uncertainty which we dub the `excess uncertainty'. Even with perfect data, there are unknown `hidden variables' that may impact the final prediction. These `variables' cause predictions from the model to vary, producing a range of outputs even when using identical model inputs. As an example, for a star in our training data, there may be unaccounted-for oddities in the object itself, or by observations from APOGEE and/or Gaia. Thus, while the observations may be recorded with low uncertainty, this `hidden' scatter may still alter the model's predictions. While this `excess' uncertainty is expected to be small for most objects from our chosen surveys, we must account for its presence to fully describe the uncertainties in the model.}

We follow the approach of \citet{Leung2019a}, who include a `predictive' uncertainty to account for this excess uncertainty component. We therefore include six additional nodes in our model's output layer, for each of our six predicted stellar parameters. These extra nodes are used by the NN to represent our excess uncertainty. The model learns this metric during the training process, as the output from these nodes is included as an uncertainty within the NN loss function. We output this value alongside our model's predictions, to return our excess uncertainty, $\sigma^2_{\mathrm{excess}}$.

\subsubsection{Loss function}
\label{sec:unc_propagation}
To train our model, we express our loss function as a Gaussian likelihood that propagates the sources of aleatoric and epistemic uncertainties through the network. We model the likelihood, $p(\boldsymbol{y} | \boldsymbol{x}, \mathcal{M})$, of the output parameters $\boldsymbol{y}$ with covariances $\boldsymbol{\Sigma}^y$ given a set of inputs $\boldsymbol{x}$ with covariances $\boldsymbol{\Sigma}^x$ and a (NN) model $\mathcal{M}$. As a reminder, $\boldsymbol{x}$ is the input feature array including normalized BP/RP coefficients and broadband colours and $\boldsymbol{y}$ is the APOGEE stellar parameters. Note that $x$ and $y$ are normalised before model training using either z-score standardization \citep{zScore}, or the BP/RP normalisation method described in Sec.~\ref{sec:coeff_norm}. The likelihood term can be written as
\begin{multline}
\label{eq:prob_gauss}
p(\boldsymbol{y}|\boldsymbol{x}, \boldsymbol{\Sigma}^y, \boldsymbol{\Sigma}^x, \mathcal{M})
\propto p(\boldsymbol{x}, \boldsymbol{y}|\boldsymbol{\Sigma}^y, \boldsymbol{\Sigma}^x, \mathcal{M}), \\ 
\quad = \prod_i \int \mathrm{d} ^n\hat{\boldsymbol{x}} \,p(\boldsymbol{y}_i|\hat{\boldsymbol{x}},\boldsymbol{\Sigma}^y_i,\mathcal{M}) p(\boldsymbol{x}_i|\hat{\boldsymbol{x}}, \boldsymbol{\Sigma}^x_i) p(\hat{\boldsymbol{x}}), \\
\quad = \prod_i\int \mathrm{d}^n
\hat{\boldsymbol{x}}\, 
\mathcal{N}(\boldsymbol{y}_i | \boldsymbol{y}_{\mathrm{pred}}(\hat {\boldsymbol{x}}), \boldsymbol{\Sigma}^y_i+\boldsymbol{\Sigma}_{\mathrm{exc}, i}(\hat {\boldsymbol{x}}))
\mathcal{N}(\boldsymbol{x}_i| \hat{\boldsymbol{x}}, \boldsymbol{\Sigma}^x_i)
p(\hat{\boldsymbol{x}}).
\end{multline}
In the first line, we have dropped the model-independent term $p(\boldsymbol{x})$ whilst in the second line we have expanded the likelihood as a product of independent per-star likelihoods. For each of these terms, we have marginalized over the `true' values of the input features, $\hat{\boldsymbol{x}}$. In the third line, we have explicitly written out the uncertainty terms as Gaussian/normal distributions $\mathcal{N}(x|\mu,\Sigma)$.
The model $\mathcal{M}$ is given by $\boldsymbol{y}_\mathrm{pred}(\boldsymbol{x})$ and has an excess covariance at each input given by the covariance matrix $\boldsymbol{\Sigma}_\mathrm{exc}(\boldsymbol{x})$ (as described in the previous section).

\cf{The two Gaussian components of this function can be tied to the two sources of aleatoric uncertainty in the data. The former describes the 
uncertainties in the stellar parameters $\boldsymbol{y}$ whilst the latter instead describes the distribution of the noisy input features, $\boldsymbol{x}$.}
\cf{In the region of $\boldsymbol{x}_i$, the integral in equation~\eqref{eq:prob_gauss} must be computed numerically. We make simplifying assumptions to make this integral computationally tractable. 
First, the prior distribution of the `true' input features, $p(\hat{\boldsymbol{x}})$, is assumed to be uniform across the uncertainty interval although we note that a Gaussian prior could be simply incorporated into our methodology. 
Secondly, in the limit where the uncertainties are small, $\boldsymbol{y}_\mathrm{pred}(\hat{\boldsymbol{x}})$ can be expanded as a Taylor series about $\boldsymbol{y}_\mathrm{pred}(\boldsymbol{x})$ as}
\begin{equation}
    \label{eq:Taylor_exp}
    \boldsymbol{y}_\mathrm{pred}(\hat{\boldsymbol{x}})\approx\boldsymbol{y}_\mathrm{pred}(\boldsymbol{x}_i)+\frac{\partial \boldsymbol{y}_\mathrm{pred}}{\partial\boldsymbol{x}}\Big|_{\boldsymbol{x}_i}(\hat{\boldsymbol{x}}-\boldsymbol{x}_i)+\cdots.
\end{equation}
Neglecting the higher-order terms produces a linear relation between $\boldsymbol{y}_\mathrm{pred}(\hat{\boldsymbol{x}})$ and $\hat{\boldsymbol{x}}$ in the vicinity of $\boldsymbol{x}_i$. These assumptions allow the integral in equation~\eqref{eq:prob_gauss} to be analytically computed as
\begin{equation}
    p(\boldsymbol{y}|\boldsymbol{x}, \boldsymbol{\Sigma}^y, \boldsymbol{\Sigma}^x, \mathcal{M}) \approx \prod_i\mathcal{N}\Big(
    \boldsymbol{y}_i | \boldsymbol{y}_{\mathrm{pred}}( {\boldsymbol{x}_i}), \boldsymbol{\Sigma}_{\mathrm{input},i}+\boldsymbol{\Sigma}^y_i+\boldsymbol{\Sigma}_{\mathrm{exc}}(\boldsymbol{x}_i)\Big),
\end{equation}
where
\begin{equation}
    \boldsymbol{\Sigma}_{\mathrm{input}} = \left( \frac{\partial \boldsymbol{y}_{\mathrm{pred}}}{\partial \boldsymbol{x}} \right)\boldsymbol{\Sigma}^x \left(\frac{\partial \boldsymbol{y}_{\mathrm{pred}}}{\partial \boldsymbol{x}} \right)^\mathrm{T}.
	\label{eq:input_uncs}
\end{equation}
Derivatives are computed using the NN model's \texttt{autograd} function. 
For simplicity, we ignore the off-diagonal terms in all output covariance matrices so define
\begin{align}
    \label{eq:sigma_diag_1}
    & \sigma^{2}_{y} \equiv \mathrm{diag}(\boldsymbol{\Sigma}^y) ;\quad
    \sigma^{2}_{\mathrm{excess}} \equiv \mathrm{diag}(\boldsymbol{\Sigma}_{\mathrm{exc}});\quad
    \sigma^{2}_{\mathrm{input}} \equiv \mathrm{diag}(\boldsymbol{\Sigma}_{\mathrm{input}}).
\end{align}
This discussion has neglected the model uncertainty arising from different NN configurations explored through dropout and described in Section~\ref{sec:dropout_unc}. This can be formally considered as an additional marginalization over possible models $\mathcal{M}$ in equation~\eqref{eq:prob_gauss} (potentially folded with some prior $p(\mathcal{M})$) but in practice involves summing in quadrature the additional scatter in the predictions for each model $\mathcal{M}$.

In conclusion, our loss function, $J$, is constructed as the logarithm of the likelihood in equation~\eqref{eq:prob_gauss} averaged over all objects in our sample:
\begin{equation}
    \begin{split}
    J
    & \equiv \frac{1}{n} \log(p(\boldsymbol{y}|\boldsymbol{x}, \boldsymbol{\Sigma}^y, \boldsymbol{\Sigma}^x, \mathcal{M})) \\ & = \frac{1}{n} \sum_{i=1}^{n} \frac{1}{2} |\boldsymbol{y}_{\mathrm{pred}}(\boldsymbol{x}_i) - \boldsymbol{y}_{i}|^2 e^{-s_{i}} + \frac{1}{2} s_{i}, 
    \end{split}
	\label{eq:one}
\end{equation}
where
\begin{equation}
    \label{eq:error_combo}
    s_{i} = \ln(\sigma^{2}_{y, i} + \sigma^{2}_{\mathrm{model}, i} + \sigma^{2}_{\mathrm{input}, i} +
    \sigma^{2}_{\mathrm{excess}, i}).
\end{equation}
\cf{For objects with larger uncertainties, the value of $s_i$ increases allowing for more flexibility in the model's predictions without affecting the loss function significantly.
The second term acts to normalise the loss function relative to the model's uncertainty estimations.}
\cf{We finally note that when making predictions for unseen data we report the median $\boldsymbol{y}_\mathrm{pred}(\boldsymbol{x})$ averaged over 50 dropout realisations with corresponding total uncertainty, $\sigma^{2}_{\mathrm{total}, i}$ as the sum in quadrature of our model, input, and excess uncertainty metrics, i.e.}
\begin{equation}
    \label{eq:err_total}
    \sigma^{2}_{\mathrm{total}, i} = \sigma^{2}_{\mathrm{model}, i} + \sigma^{2}_{\mathrm{input}, i} +
    \sigma^{2}_{\mathrm{excess}, i}.
\end{equation}

\section{Model Accuracy}\label{sec:validation}

With our model trained and capable of reporting predictions and their associated uncertainties, we validate its performance against atmospheric parameters from other methods. We compare the NN's predictions against the APOGEE dataset \citep{Abdurrouf2022}, the LAMOST spectroscopic data \citep{Cui2012}, and the Gaia GSP-Phot parameters \citep{Liu2012}.

 \begin{table}
    \centering
    \caption{\rf{Statistics for our NN vs APOGEE comparisons. We report the model's median prediction per parameter ($\mu_{\mathrm{NN}}$) and reported uncertainty ($\sigma_{\mathrm{total}}$), alongside the median residuals, and the root mean squared error (RMSE), $\sqrt{ \frac{1}{n} \Sigma^n_{i=1}  (y_{\mathrm{pred}, i}-y_i)^2}$, of the comparison.}}
    \begin{tabular}{lllll}
                        & $\mu_{\mathrm{NN}}$ &  $\sigma_{\mathrm{total}}$ &  Residuals & RMSE \\ \hline
        [Fe/H] (dex)          & $-0.22$ & $\pm0.068$ &  $+0.043$ &  $0.064$  \\ 
        $\mathrm{T_{eff}}$ (K)  & $4777$& $\pm69.1$ &  $-49.9$ &  $51.77$  \\ 
        $\log(g)$ (dex)      & $2.64$&$\pm0.14$   &  $-0.034$ &  $0.091$\\
        $[\mathrm{C/Fe}]$ (dex) &$-0.014$&$\pm0.031$ &  $+0.016$ &  $0.043$ \\ 
        $[\mathrm{N/Fe}]$ (dex) & $0.20$&$\pm0.040$  &  $+0.036$ &  $0.052$  \\ 
        $[\alpha\mathrm{/M}]$ (dex) & $0.083$&$\pm0.029$ &  $-1.4\times10^{-3}$ &  $0.025$ 
    \end{tabular}
     \label{tab:APOGEE}
\end{table}

\subsection{APOGEE Comparisons}
\label{sec:Apogee_comparisons}

\cf{We compare the predictions of our approach to the spectroscopic data from APOGEE DR17. As the APOGEE data was used to train the NN, we expect the network to reproduce the APOGEE values well. These comparisons are a good measure of the quality of the NN's training, and its ability to accurately reproduce known `true' values.}

\begin{figure*}
	\centering
	\includegraphics[width=.49\textwidth, trim={0, 0, 80, 60}, clip]{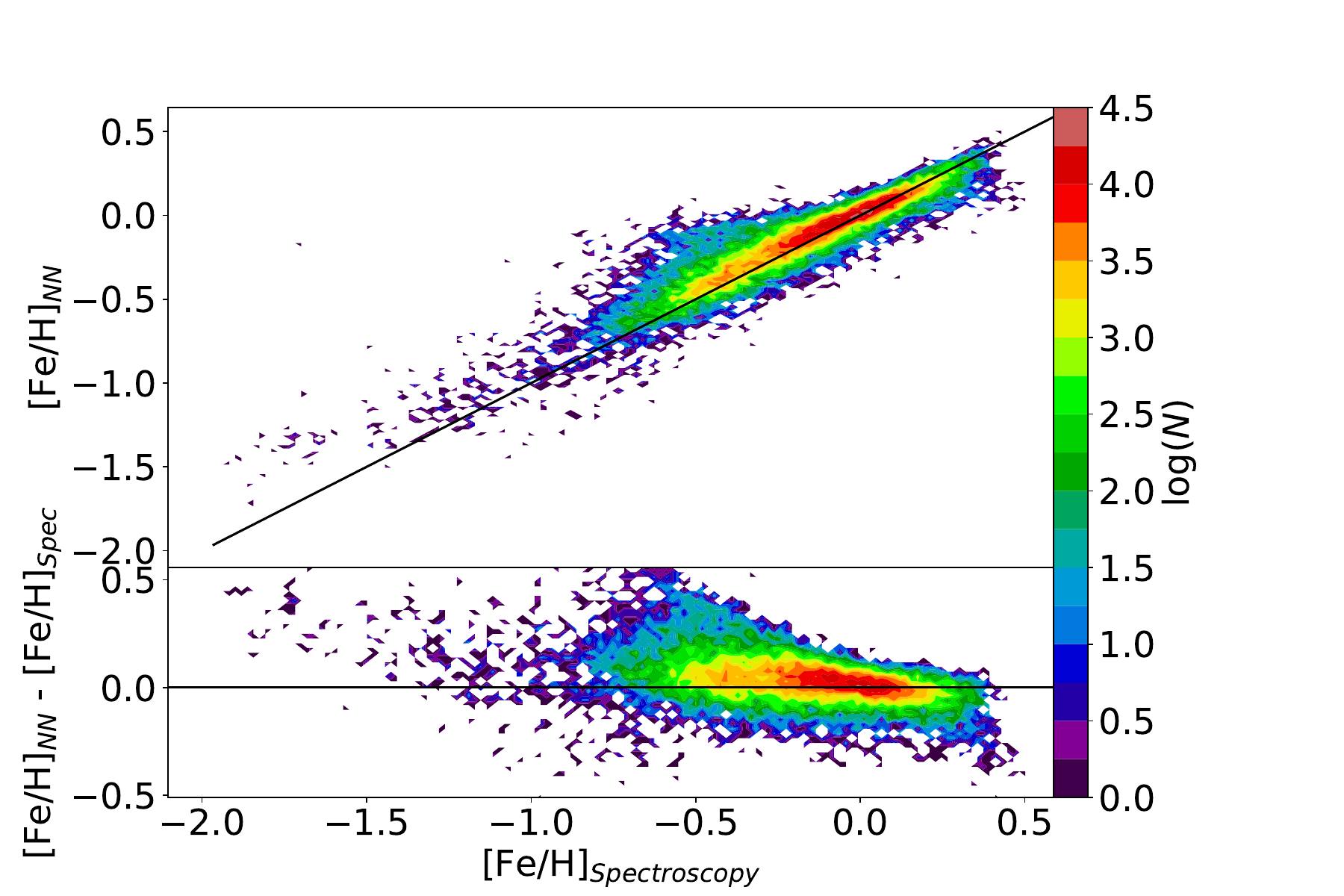}
	\includegraphics[width=.49\textwidth, trim={0, 0, 80, 60}, clip]{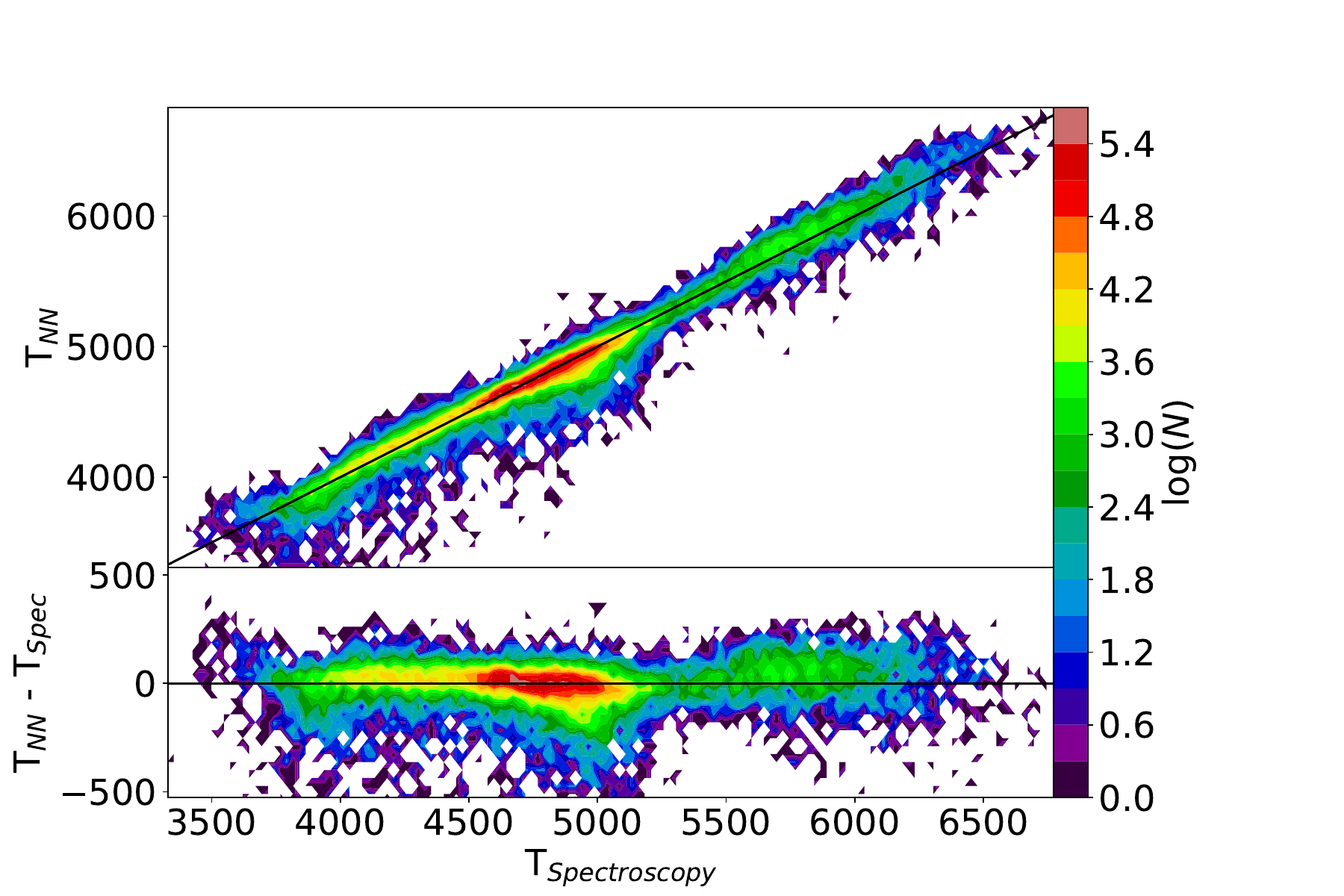}
    \includegraphics[width=.49\textwidth, trim={0, 0, 80, 60}, clip]{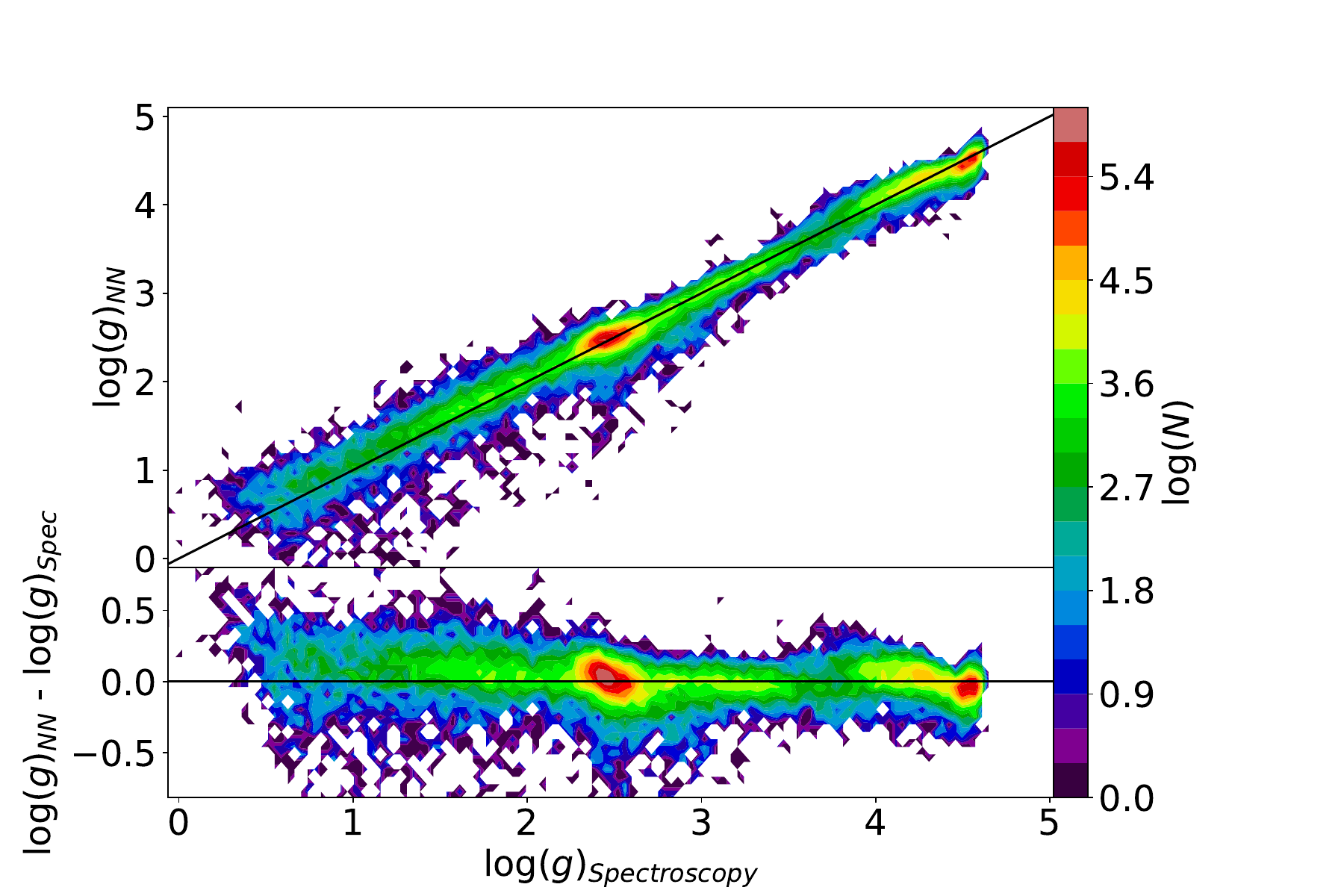}
    \includegraphics[width=.49\textwidth, trim={0, 0, 80, 60}, clip]{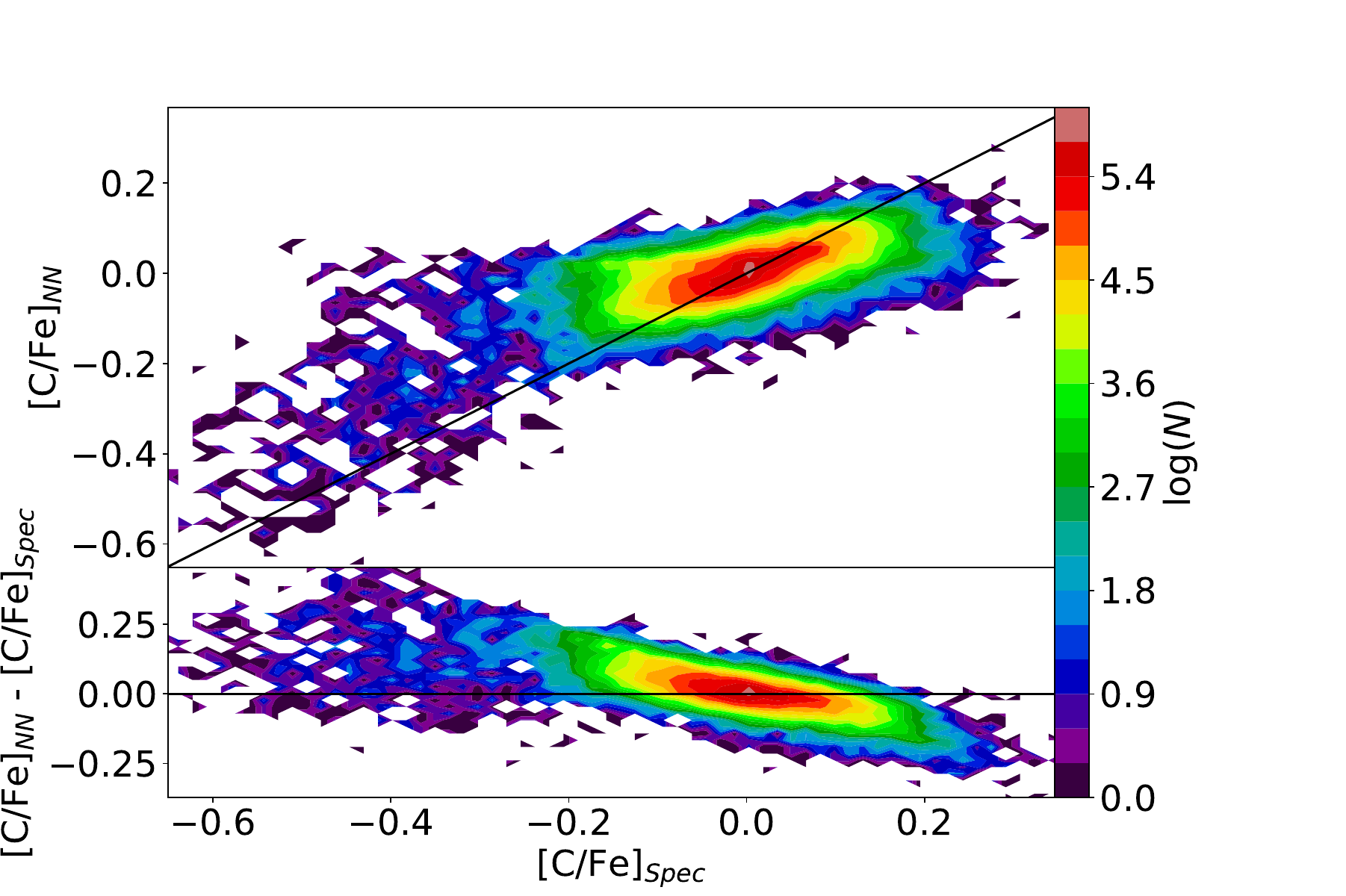}
    \includegraphics[width=.49\textwidth, trim={0, 0, 80, 60}, clip]{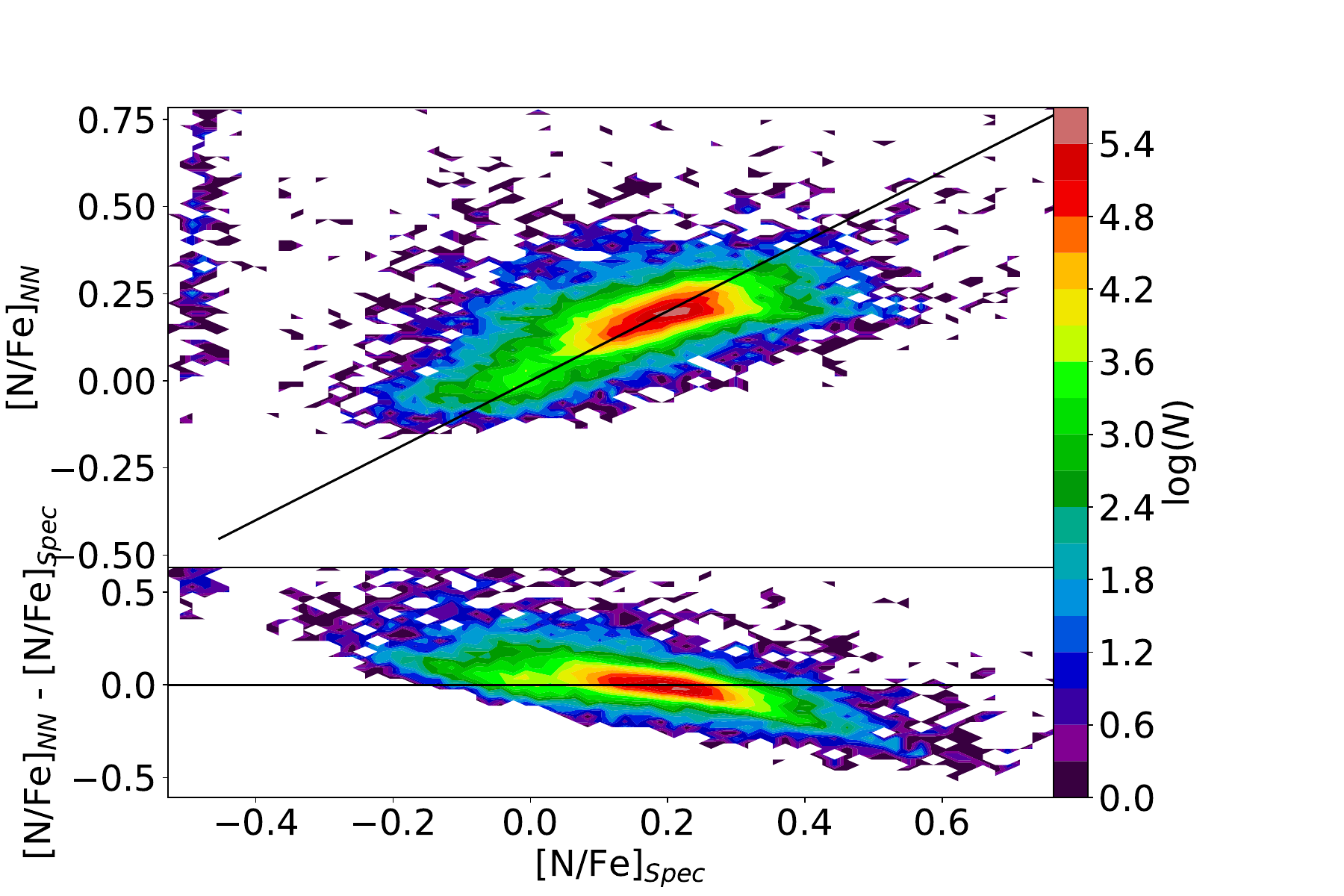}
    \includegraphics[width=.49\textwidth, trim={0, 0, 80, 60}, clip]{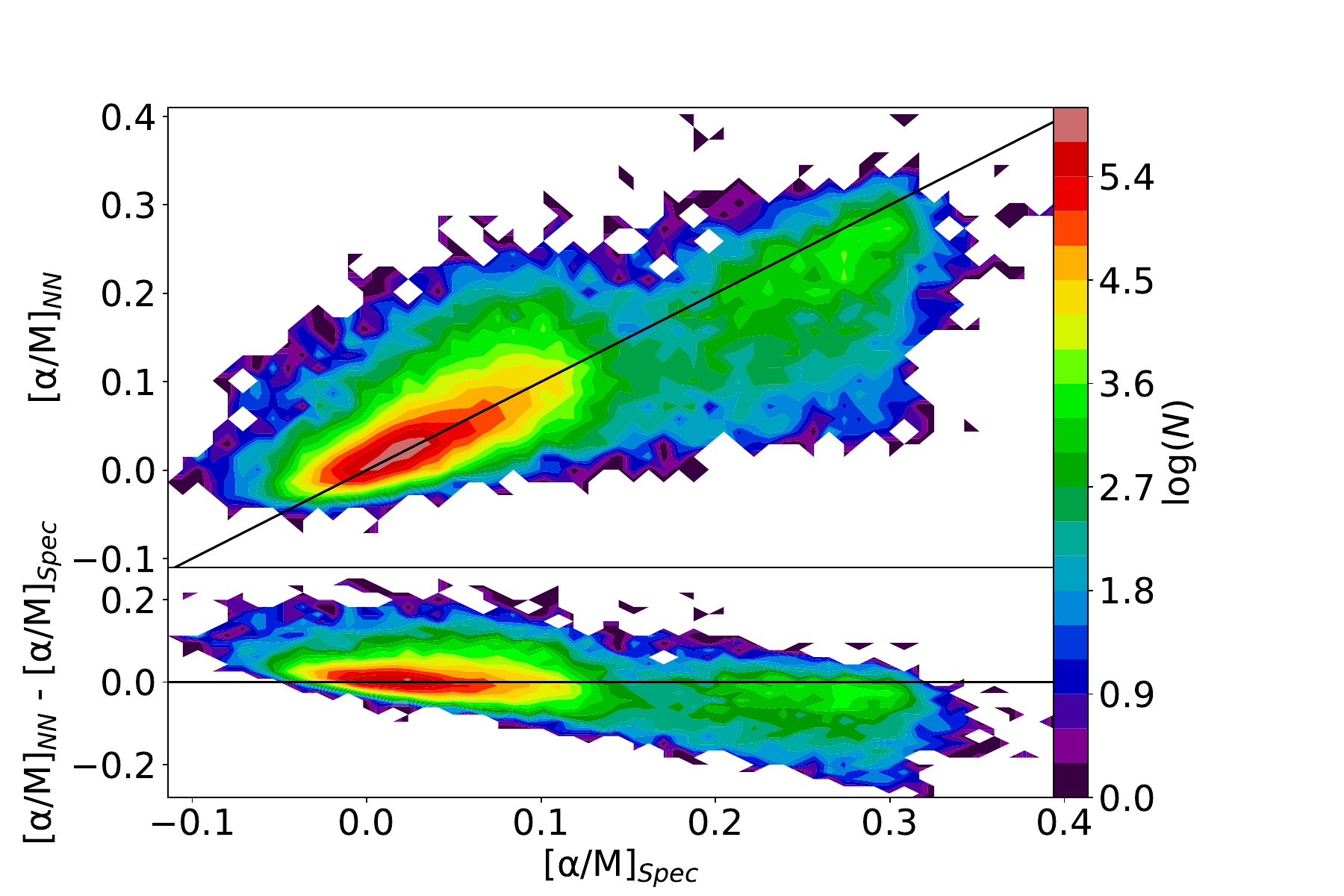}
    \caption{\rf{Comparison between the NN's predictions against values from APOGEE spectroscopy (`spec'), for metallicity, effective temperature, surface gravity, and carbon, nitrogen, and alpha abundances. Each figure shows a direct comparison in its upper panel, and the residual difference in its lower panel. Note that the contours on these plots are on a logarithmic scale.}}
    \label{fig:NN_vs_spec}
\end{figure*}

\cf{To ensure our validation identifies model biases caused by overfitting to the training dataset, we adopt an out-of-bag approach to separate our APOGEE cross-validation sample from our training sample. We remove 5\% of our APOGEE data to use for validation, while the remainder is used as our training sample. This ensures the predictions for the validation sample are unseen data for the model, rather than objects the NN has already seen from the training sample.}
We show these comparisons in Fig~\ref{fig:NN_vs_spec}, for the six output stellar parameters: [Fe/H], $\mathrm{T_{eff}}$, $\log(g)$, [C/Fe], [N/Fe], and [$\alpha$/M]. \cf{The mean scatter for these comparisons is shown in Table~\ref{tab:APOGEE}. }

Overall, our model seems to show good agreement with the APOGEE cross-validation sample, with all six of these plots showing regions of strong correlation. We also note generally small residuals for our [Fe/H], $\mathrm{T_{eff}}$, and $\log g$ predictions, compared to the median of the NN prediction for each parameter. This suggests our model is, on average across the validation sample, predicting stellar parameters that agree closely with the spectroscopic data. The mean RMSE values for the [C/Fe], [N/Fe], and [$\alpha$/M] predictions are larger compared to the mean prediction, but remain small on average across the sample.

We also note a slight negative bias in the residuals for [Fe/H], $\mathrm{T_{eff}}$, $\log g$, suggesting the model is systematically underestimating compared to the APOGEE observations. Similarly, our [N/Fe] predictions instead show a slight positive bias, suggesting systematic overestimation. However, these biases are consistently smaller than both the RMSE and the model's total uncertainty, suggesting any bias in the predictions is minor.

However, we do note some significant biasing present in Fig.~\ref{fig:NN_vs_spec} due to the model's tendency to `trend to the mean' when given poorer quality data. We find our model's training sample to have an overdensity of stars within a narrow range of each parameter. This bias ends up forming a biased underlying distribution learned by the model, from which it will draw predictions for poorer quality, high-uncertainty observations. We therefore find, for stars without `good' observations, the model will produce predictions closer to the mean of the training sample, producing a notable twist in the comparisons shown in Fig.~\ref{fig:NN_vs_spec}. As shown clearly in the [Fe/H], $\mathrm{T_{eff}}$, and [C/Fe] figures, there is a population of objects with APOGEE observations below the sample mean being overestimated by the model, while observations above the mean are underestimated.

The tendency of machine learning models (especially discriminative models, like ours) to predict values biased towards the mean values of the training data
is a major limitation of these algorithms. As we use real observed data to train our model, there will always be trends due to unbalanced stellar distributions in the galaxy or from survey selection functions. These underlying distributions therefore become embedded in the model's training, and present as the biases we observe in our predictions. 
However, biased outputs are typically accompanied by higher uncertainties allowing for their removal from any subsequent selections.

We finally note a significant population of unusual objects in our [N/Fe] predictions, which appear to have APOGEE observations below [N/Fe] $< -0.4$. Our model's predictions do not match these nitrogen abundances, and instead spread these objects over the much wider range of $-0.2 < \mathrm{[N/Fe]} < 0.6$. As these APOGEE measurements have generally higher uncertainties, we consider these anomalies to arise from issues with the APOGEE spectroscopic data or pipeline.

\cf{Overall, these comparisons show the general success of the model's predictions whilst highlighting the biases present in our NN's outputs. Of the six output stellar parameters, our [C/Fe] and [$\alpha$/M] predictions show the most significant divergences from the APOGEE data, suggesting a need for greater care when using these predictions in further analysis.}

\subsection{LAMOST comparisons} 
\label{sec:ap_vs_lam}

\begin{figure*}
	\centering
	\includegraphics[width=.49\textwidth, trim={0, 0, 70, 60}, clip]{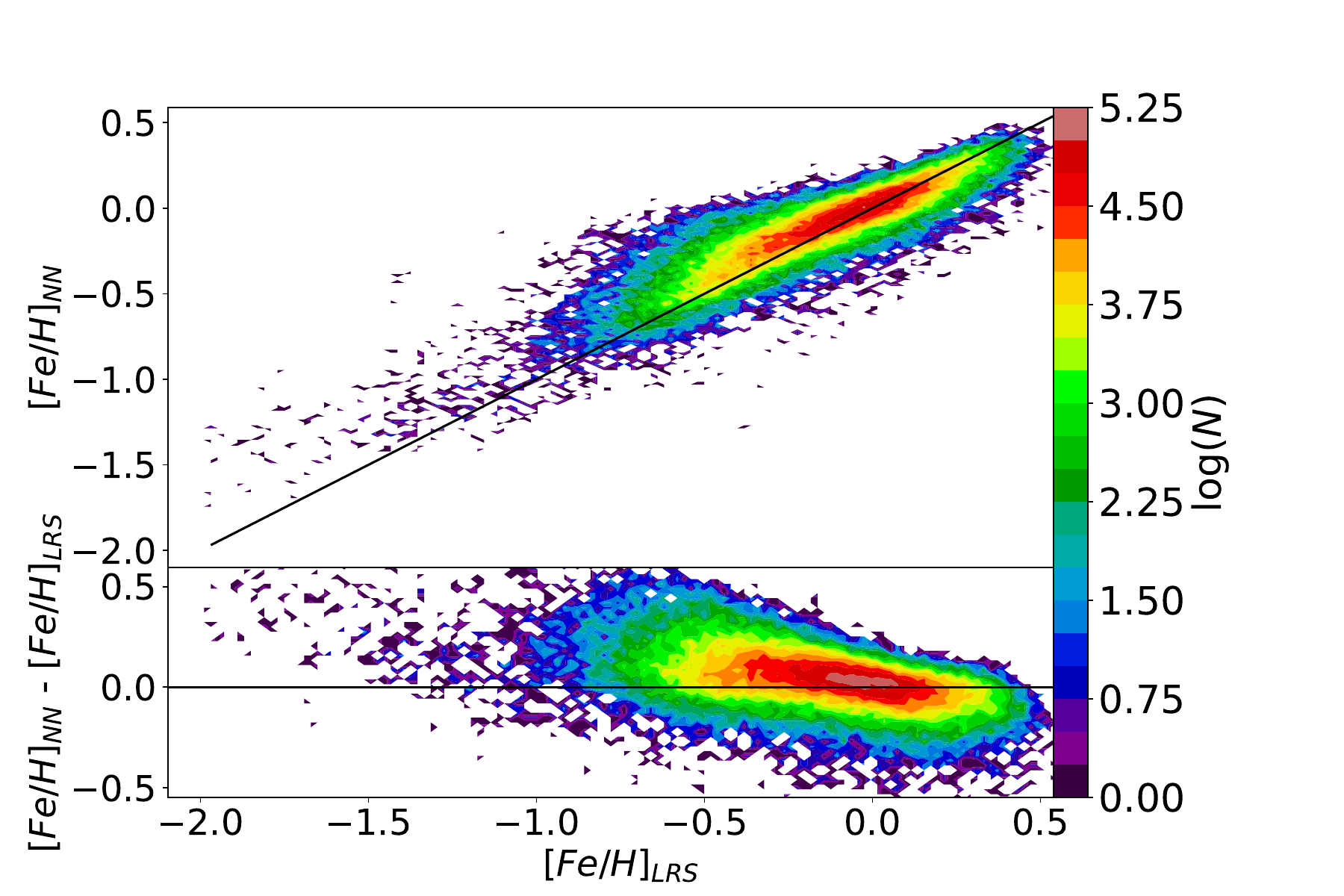}
    \includegraphics[width=.49\textwidth, trim={0, 0, 70, 60}, clip]{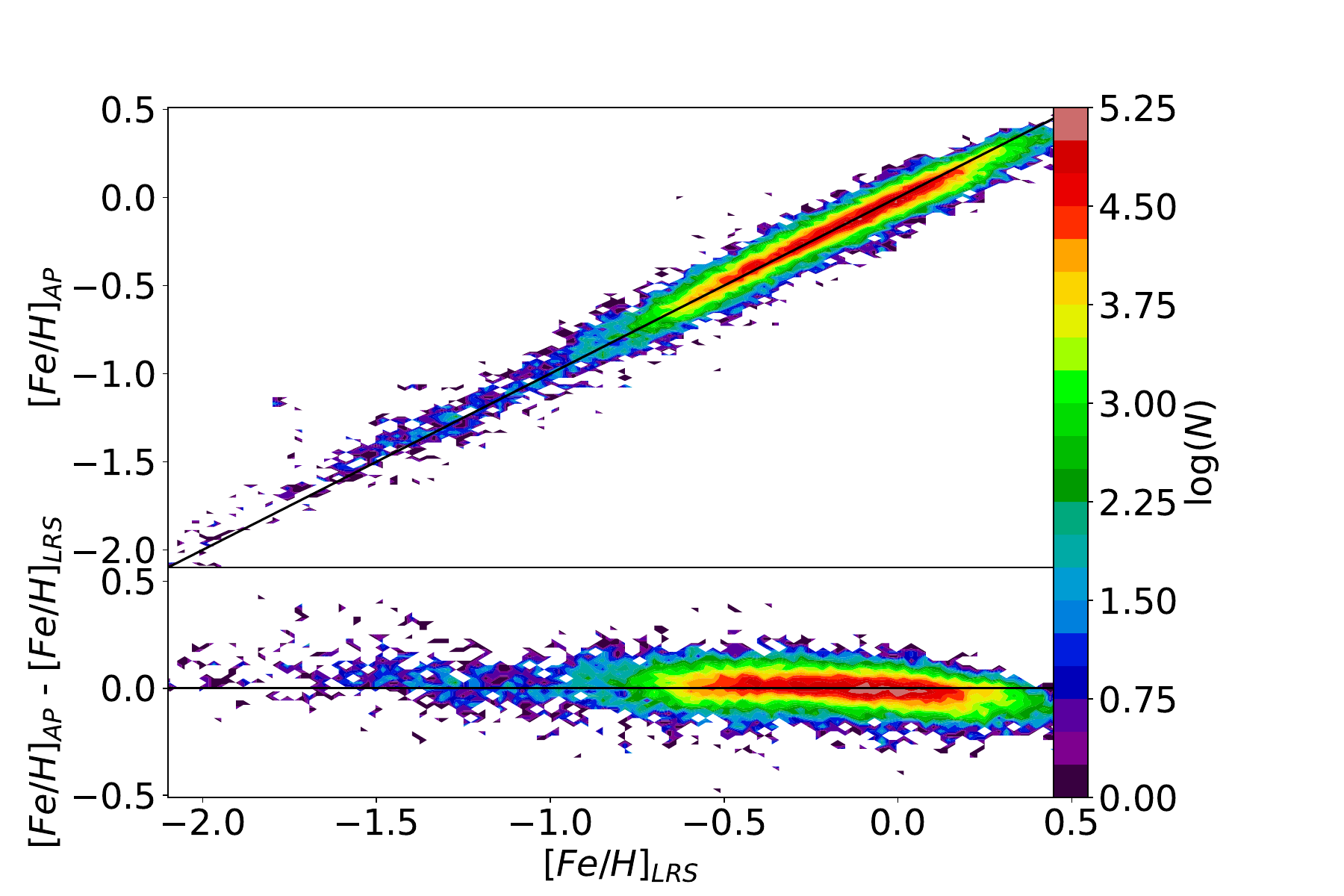}
	\includegraphics[width=.49\textwidth, trim={0, 0, 70, 60}, clip]{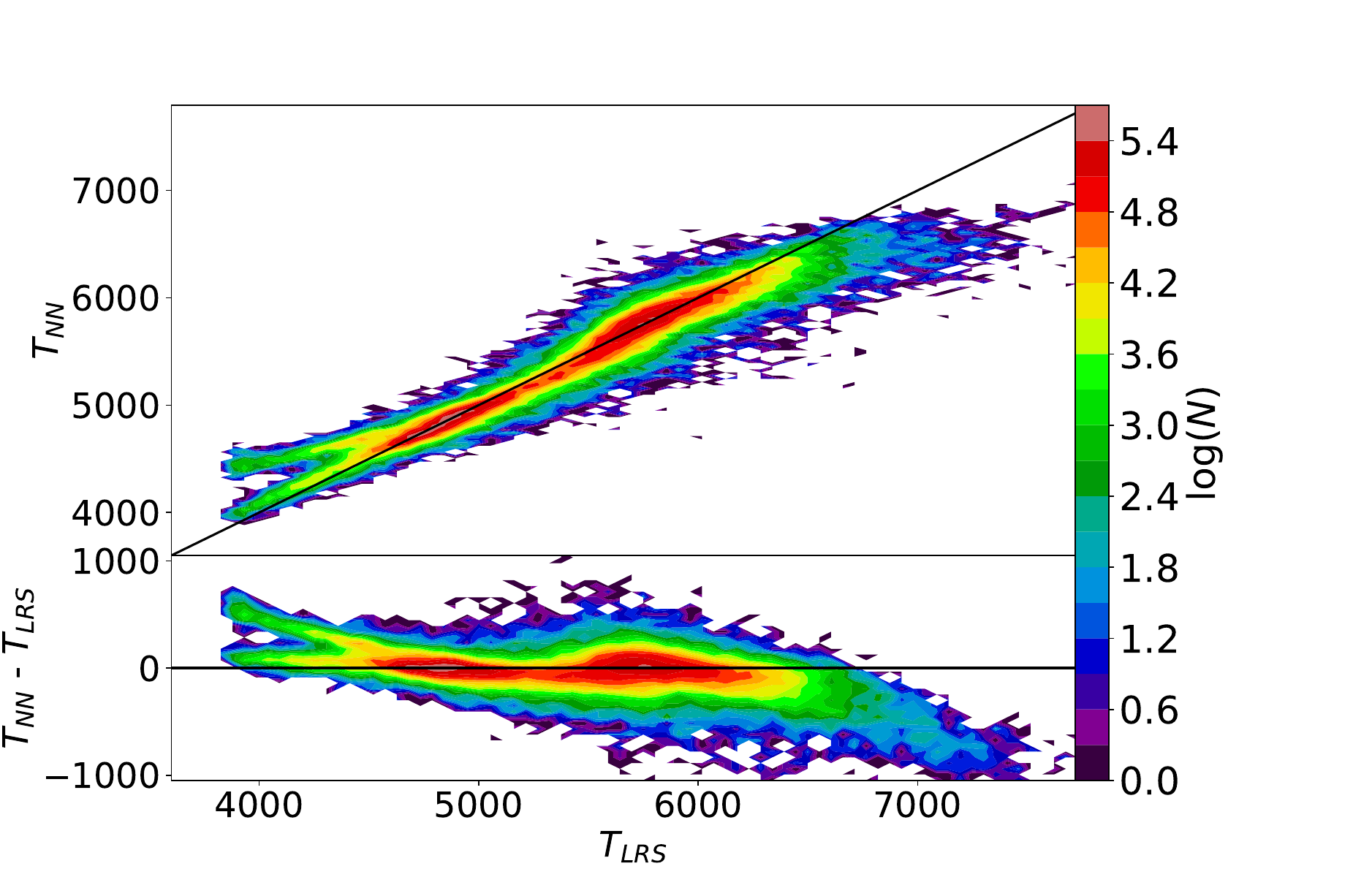}
    \includegraphics[width=.49\textwidth, trim={0, 0, 70, 60}, clip]{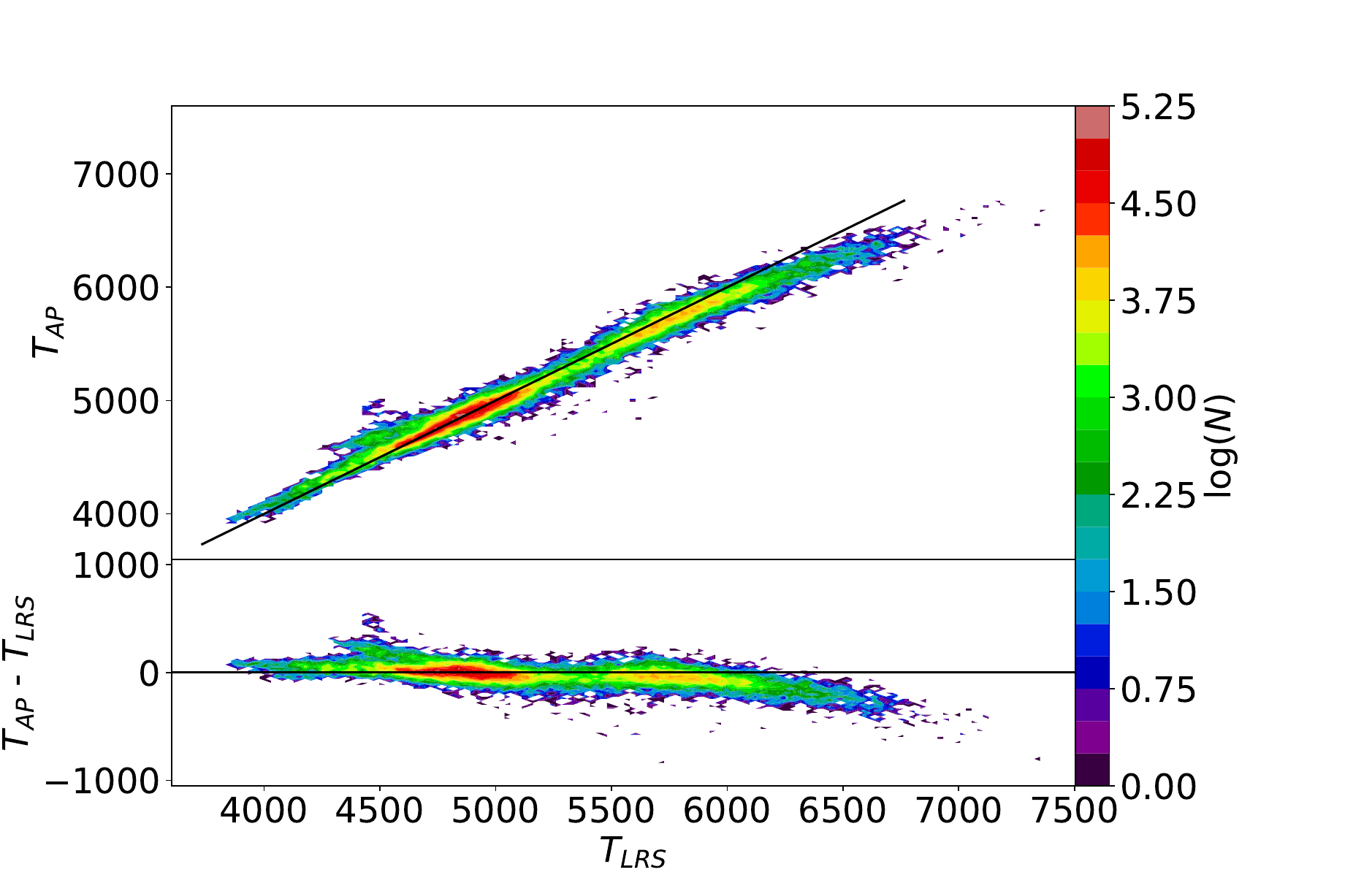}
    \includegraphics[width=.49\textwidth, trim={0, 0, 70, 60}, clip]{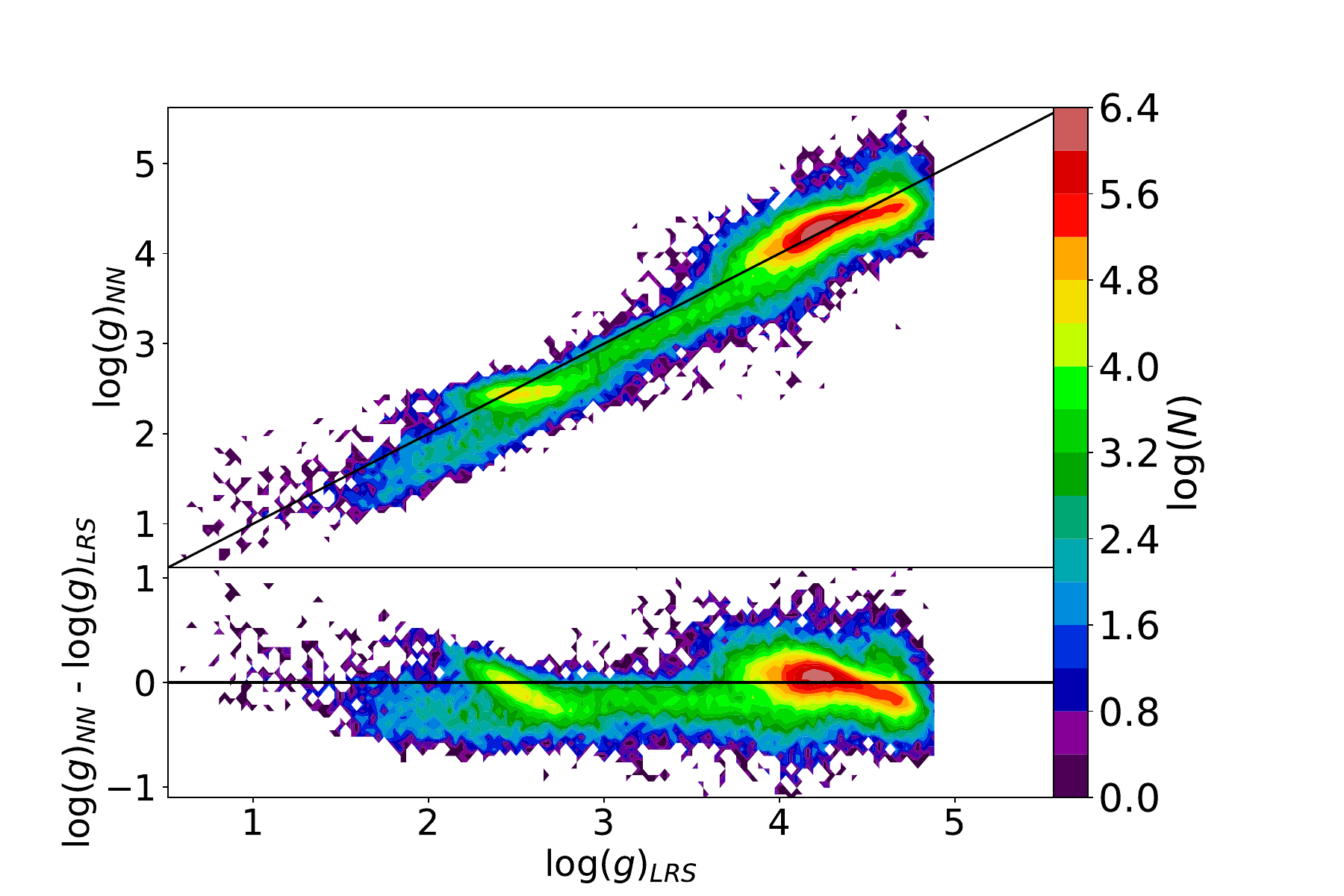}
    \includegraphics[width=.49\textwidth, trim={0, 0, 70, 60}, clip]{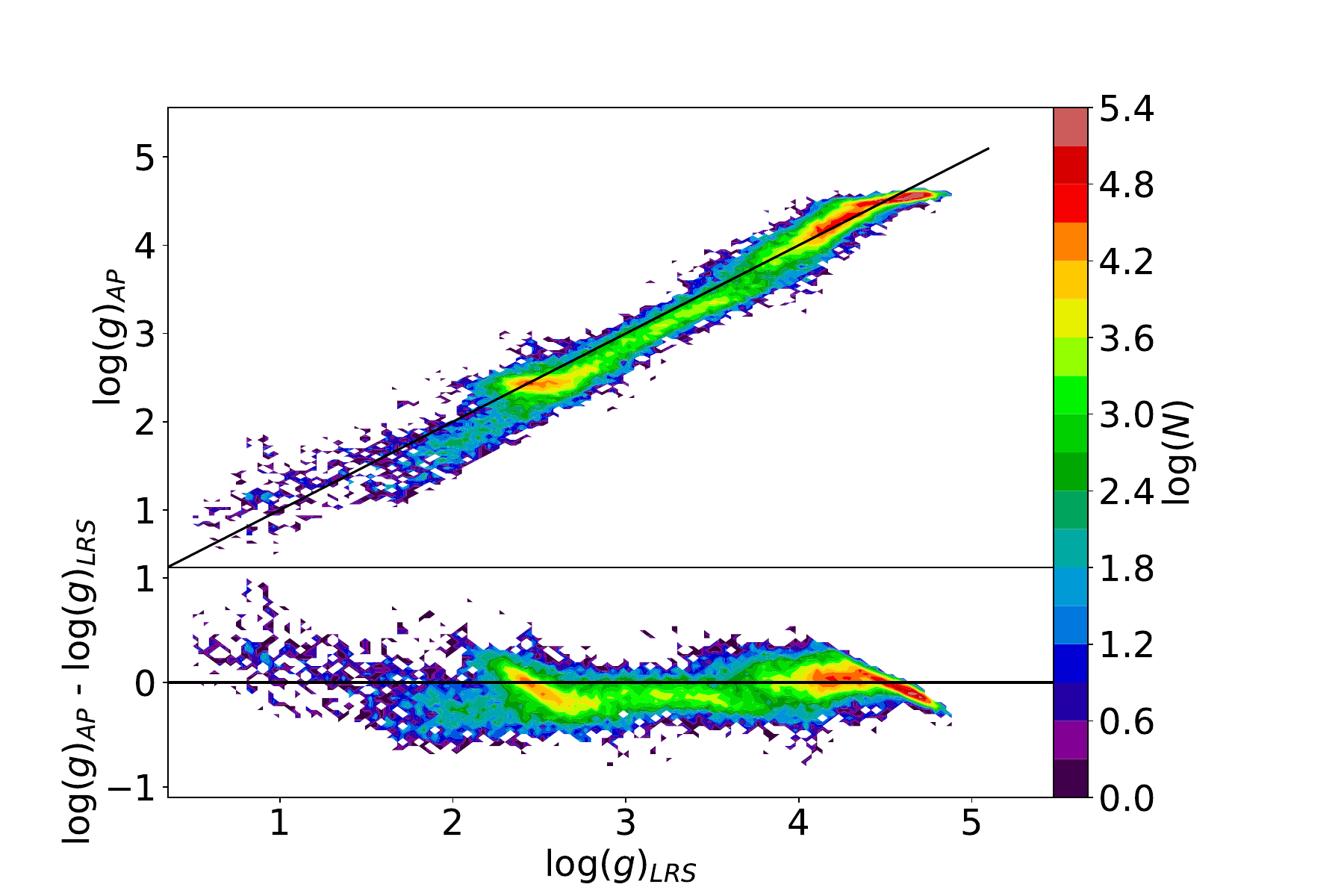}
    \caption{\rf{\emph{Left}, comparisons between our NN's predictions (NN) against LAMOST low-resolution spectra derived parameters (Spec) (sample size: $\sim$63,000). \emph{Right}, comparisons between LAMOST (LRS) and APOGEE (AP) parameters (sample size: $\sim$55,000)}}
    \label{fig:NN_vs_AP_vs_LAM}
\end{figure*}

\begin{figure*}
	\centering
    \includegraphics[width=.49\textwidth, trim={0, 0, 75, 60}, clip]{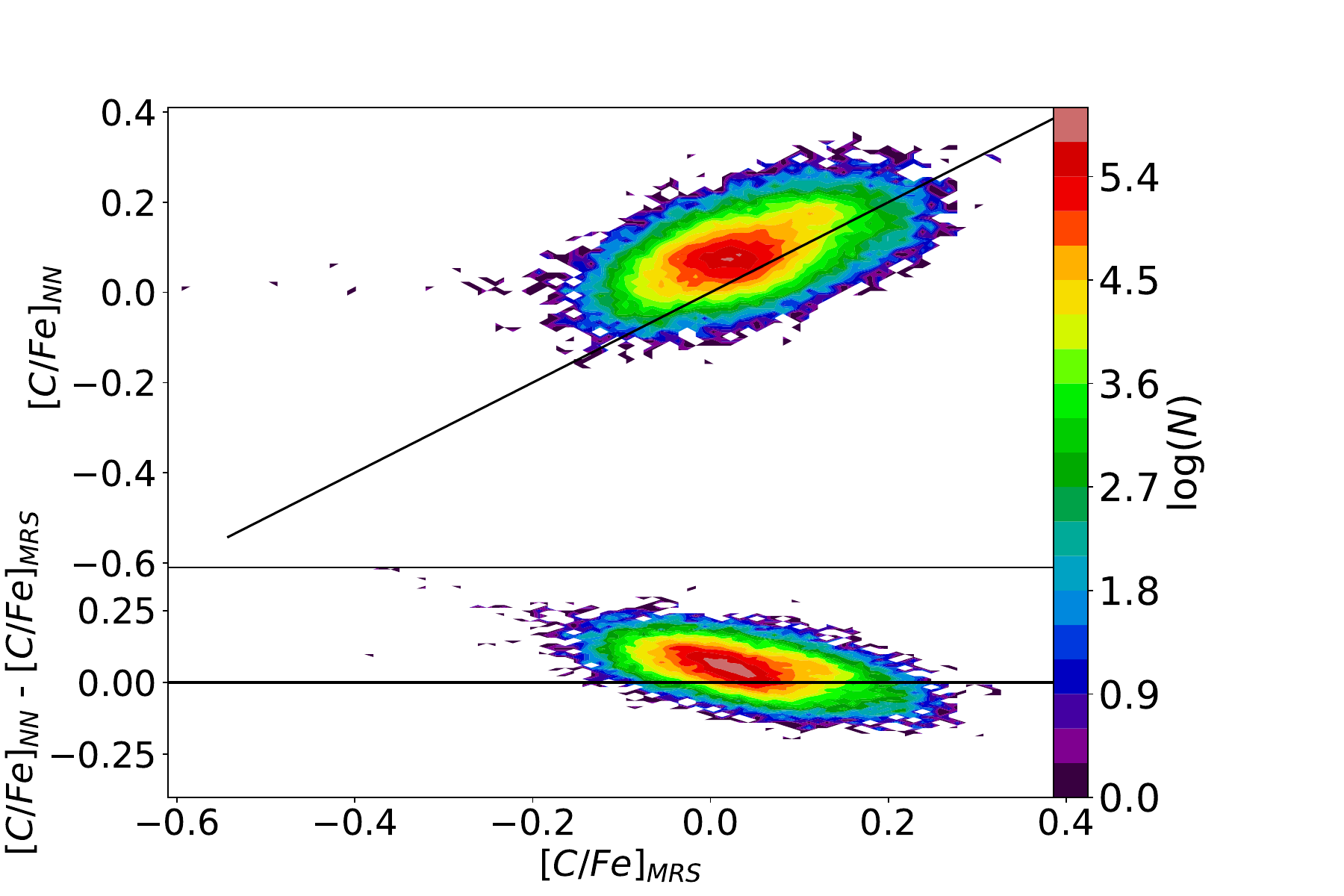}
    \includegraphics[width=.49\textwidth, trim={0, 0, 75, 60}, clip]{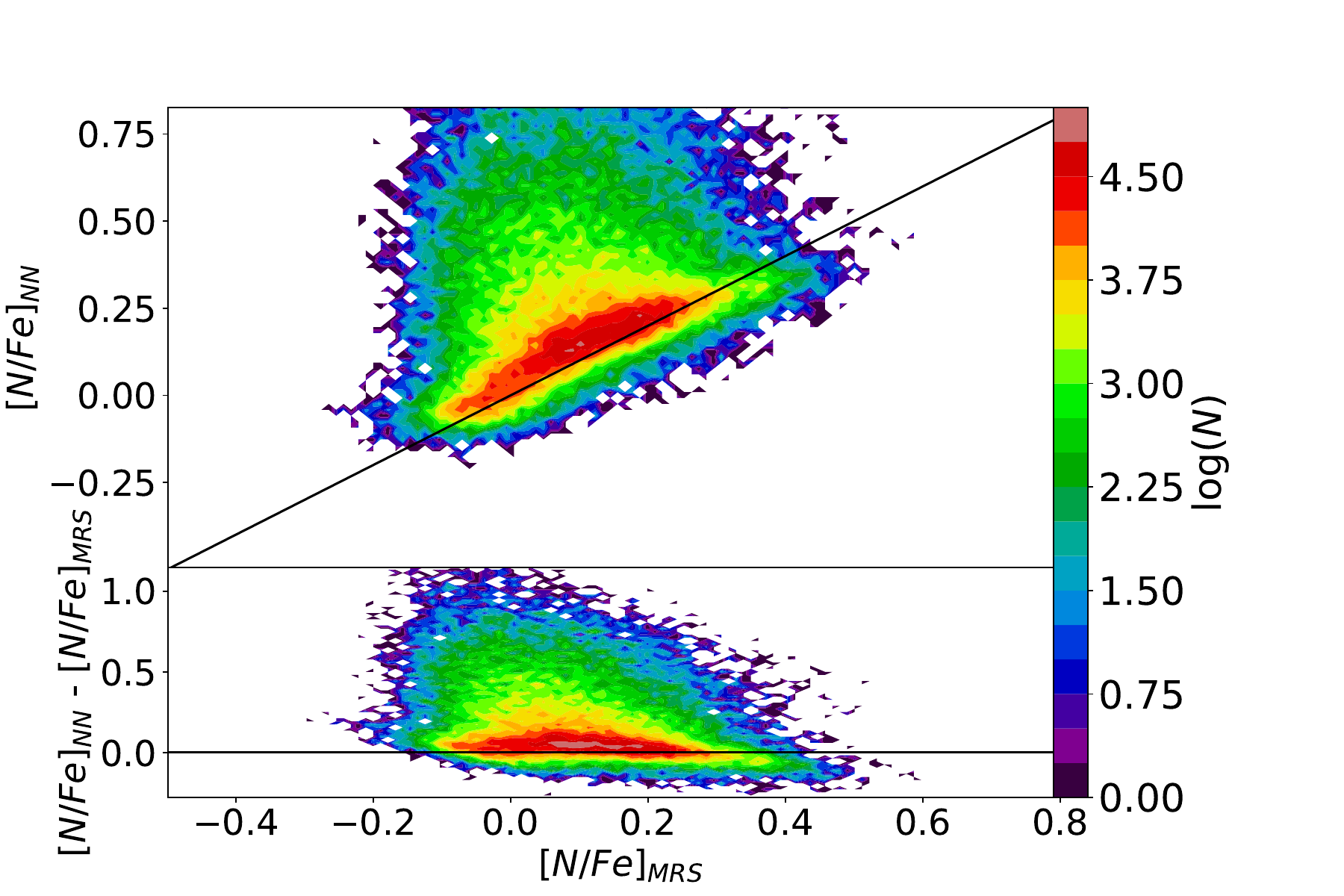}
    \includegraphics[width=.49\textwidth, trim={0, 0, 75, 60}, clip]{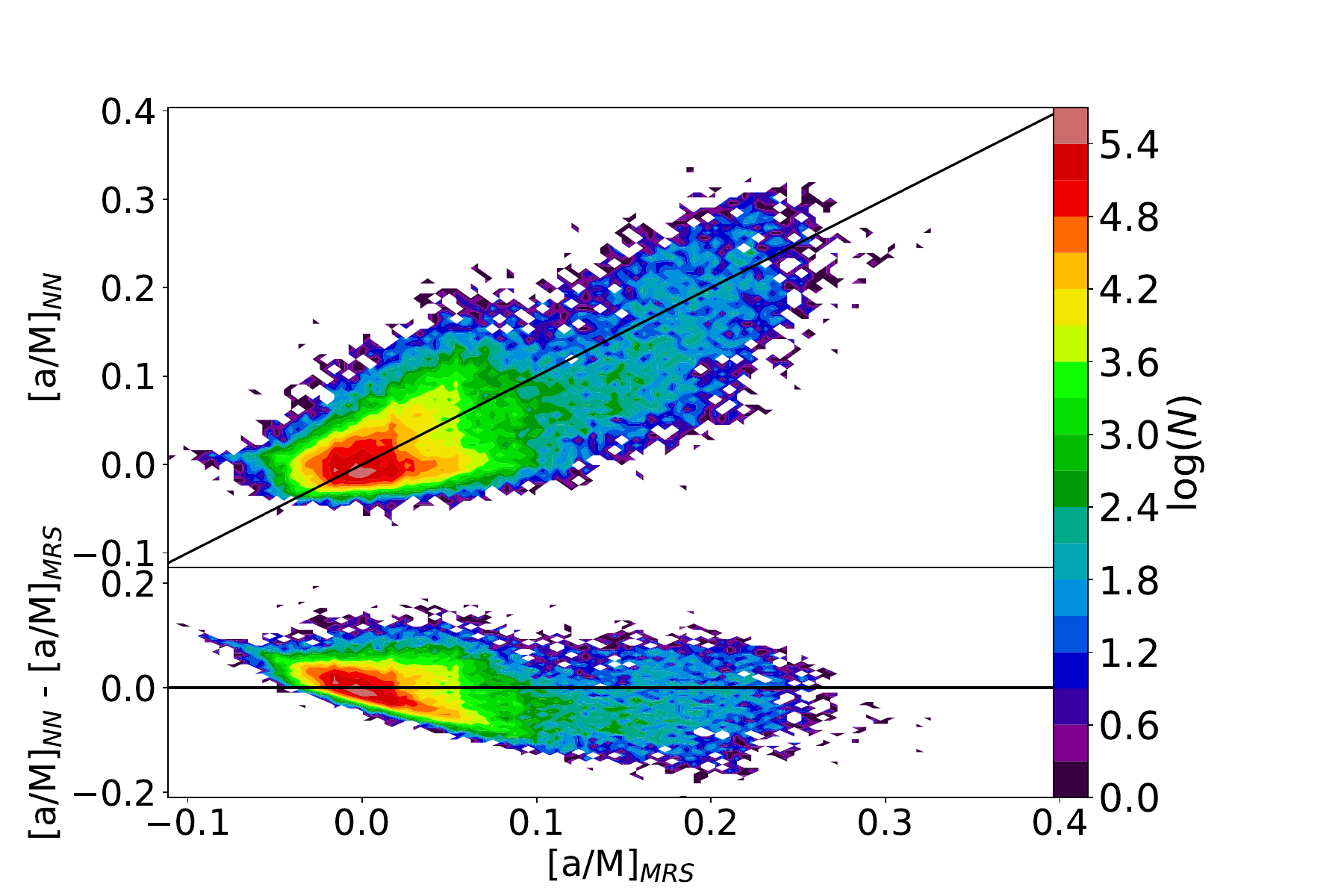}
    \caption{\rf{Comparisons between our NN's predictions (NN) against LAMOST medium-resolution spectra derived abundances (Spec).}}
    \label{fig:NN_vs_MRS}
\end{figure*}

\begin{table}
    \centering
    \caption{\rf{Statistics for our NN vs LAMOST LRS and MRS comparisons. We report the model's median prediction per parameter ($\mu_{\mathrm{NN}}$) and reported uncertainty ($\sigma_{\mathrm{total}}$), alongside the median residuals, and the root mean squared error (RMSE), $\sqrt{ \frac{1}{n} \Sigma^n_{i=1}  (y_{\mathrm{pred}, i}-y_i)^2}$, of the comparison. Our [Fe/H], $\mathrm{T_{eff}}$, and $\log(g)$ statistics are recovered from our LRS comparisons, while [C/Fe], [N/Fe], and [$\alpha$/M] are recovered from our MRS comparisons.}}
    \begin{tabular}{lll}
                        &Residuals & RMSE \\ \hline
        [Fe/H] (dex)         &$+0.096$ &  $0.104$  \\ 
        $T_\mathrm{eff}$ (K)  &$-34.3$ &  $81.5$  \\ 
        $\log g$ (dex)     & $+0.018$ &  $0.134$  \\ 
        $[\mathrm{C/Fe}]$ (dex) &$+0.052$ &  $0.059$  \\ 
        $[\mathrm{N/Fe}]$ (dex) &$+0.120$ &  $0.117$  \\ 
        $[\alpha\mathrm{/M}]$ (dex) &$+0.001$ &  $0.027$
    \end{tabular}
     \label{tab:LAMOST_LRS}
\end{table}

We further compare our NN's predicted atmospheric parameters with those derived by Data Release 7 (DR7) of the LAMOST survey. The Large Sky Area Multi-Object Fiber Spectroscopic Telescope (LAMOST) survey \citep{Cui2012} is a spectroscopic survey focussed on the northern sky, \cf{providing observations from both low and medium resolution instruments.}
For comparisons, we utilise both the low-resolution (LRS) and medium-resolution (MRS) spectra. We use the LRS sample due to its deeper depth and wider sky area than the MRS sample, while the MRS sample provides a shallower sample with measured [C/Fe], [N/Fe], and [$\alpha$/M] abundances. 
As described in \citet{Anguiano2018}, there is good agreement in atmospheric parameters between the APOGEE and LAMOST data, and thus we expect to see strong correlations between our APOGEE-trained NN outputs and the LAMOST atmospheric parameters.

\subsubsection{Low-Resolution Spectra}

We show comparisons between LAMOST LRS and our NN's predictions for [Fe/H], $T_\mathrm{{eff}}$, and $\log g$ in the left column of Fig.~\ref{fig:NN_vs_AP_vs_LAM} with corresponding scatter statistics given in Table~\ref{tab:LAMOST_LRS}. These comparisons show our predictions are generally well correlated with the LRS stellar parameters, with the median RMSE being small for [Fe/H], $T_\mathrm{{eff}}$, and $\log g$ compared to the magnitude of our parameters. However, there are clear biases in the comparison figures, especially for our $T_\mathrm{{eff}}$ and $\log g$ predictions. For $T_\mathrm{{eff}}$ we see a large tail of underestimated predictions above $T_\mathrm{{eff}}>6000\,\mathrm{K}$, and a similar tail of overestimated predictions below $T_\mathrm{{eff}}<5000\,\mathrm{K}$ suggesting a population of stars where the model is trending to the mean, as discussed with our APOGEE comparisons. Our $\log g$ biasing instead takes the form of similar overestimated/underestimated wings centred on the two sample overdensities at $\log g \sim 2.5$ and $\sim 4.5$. We find these wings are predominantly due to main sequence objects in our sample, specifically hot ($T_\mathrm{{eff}}>6000\,\mathrm{K}$) and very cool ($T_\mathrm{eff}<4500\,\mathrm{K}$) stars. 

We find these biases arise from differences between the LAMOST LRS parameters and the APOGEE sample we train the NN on. We show the direct comparison between APOGEE and LAMOST LRS in the right column of Fig.~\ref{fig:NN_vs_AP_vs_LAM}, which highlights similar biasing patterns as we see in our model comparisons. We therefore consider the model's predictions to agree well with the LAMOST LRS data with our noted divergences being primarily from the differences between the underlying spectroscopic data rather than poor quality predictions.

\subsubsection{Medium-Resolution Spectra}

We also compare our NN's predictions to parameters derived from the LAMOST Medium-Resolution spectra (MRS). Specifically, we select parameters derived by a convolutional neural-network label-transfer method applied to the LAMOST MRS data, as described by \citet{Wang2020}. As [C/Fe], [N/Fe], and [$\alpha$/M] were not present in the previous LRS comparison, we focus on validating our predictions for these parameters. These comparisons are shown in Fig.~\ref{fig:NN_vs_MRS}, with the statistics of the comparison shown in Table~\ref{tab:LAMOST_LRS}.

We find generally good agreement across the three abundances with the average RMSE similar to that of our APOGEE comparisons. There are notable deviations in the [N/Fe] comparisons, with a population of stars being overestimated by the model compared to the LAMOST data. These objects are predicted by the model to have [N/Fe] $>0.4$, which is uncommon in the LAMOST data but available in our APOGEE sample. We suggest this divergence in the NN's predictions is due to this difference between the APOGEE and LAMOST parameters, which is then reflected in the model's training and predictions.
Our [$\alpha$/M] comparisons seem to show a more symmetric scatter than our APOGEE comparisons, with lesser biasing due to over- and under-estimation by the NN. We also see distinct high and low alpha populations in these comparisons, as we saw in the APOGEE data, although the LAMOST data contains a larger proportion of low-alpha stars. From these comparisons, we further confirm our model's predictive accuracy against this medium-resolution spectroscopic data, alongside our prior APOGEE and LAMOST LRS comparisons. 

\begin{figure*}
	\centering
    \includegraphics[width=.49\textwidth, trim={0, 0, 80, 60}, clip]{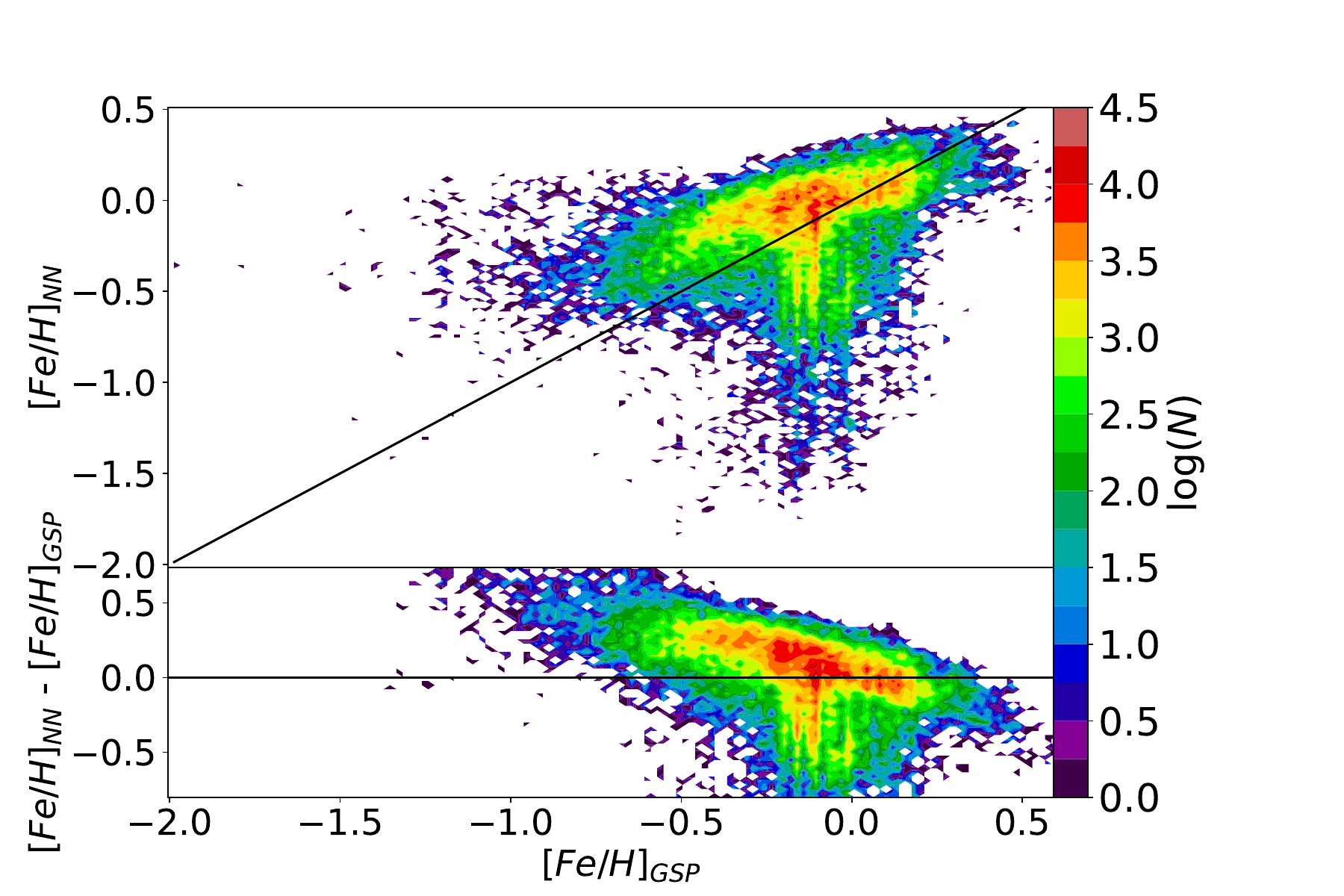}
    \includegraphics[width=.49\textwidth, trim={0, 0, 80, 60}, clip]{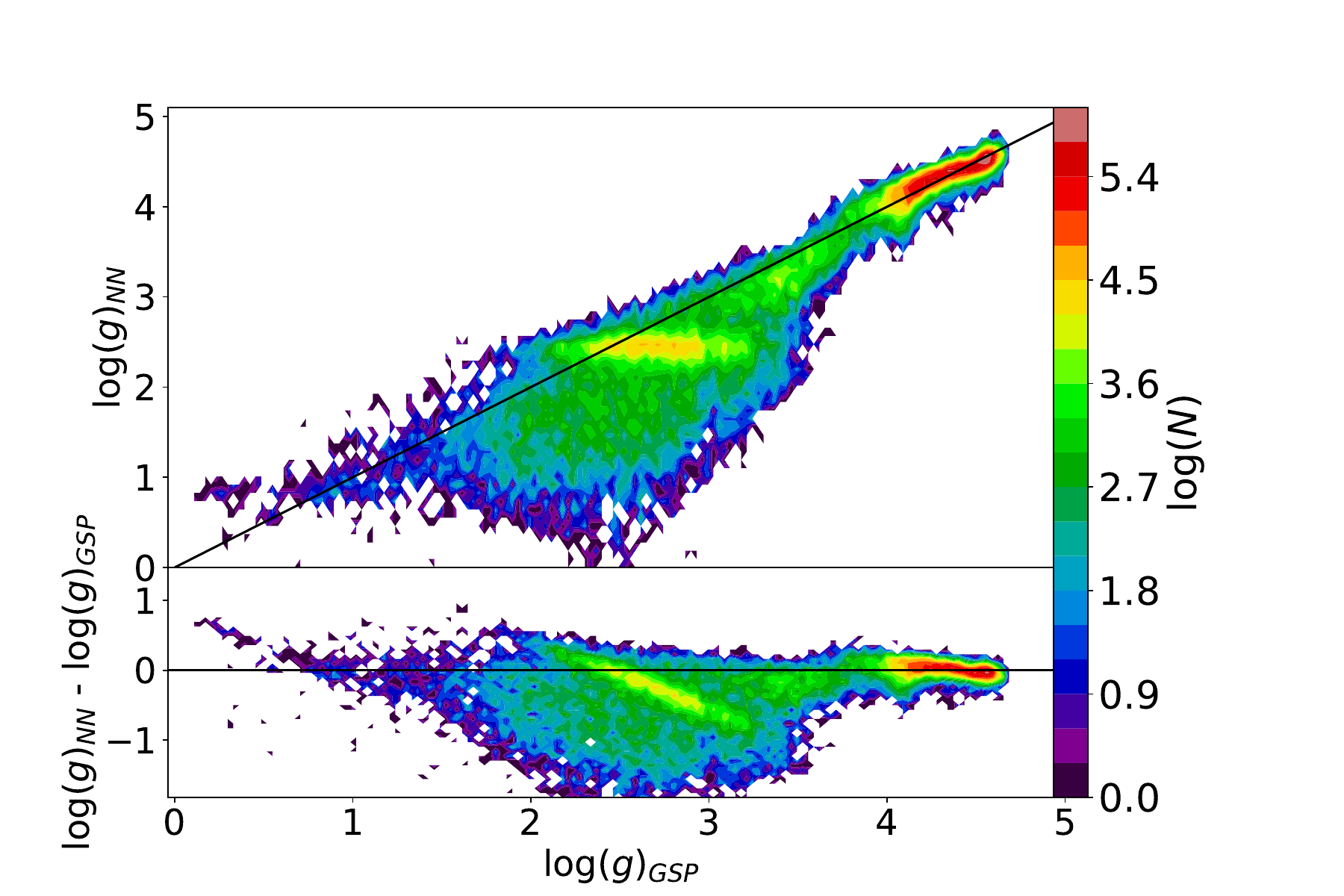}
    \includegraphics[width=.49\textwidth, trim={0, 0, 80, 60}, clip]{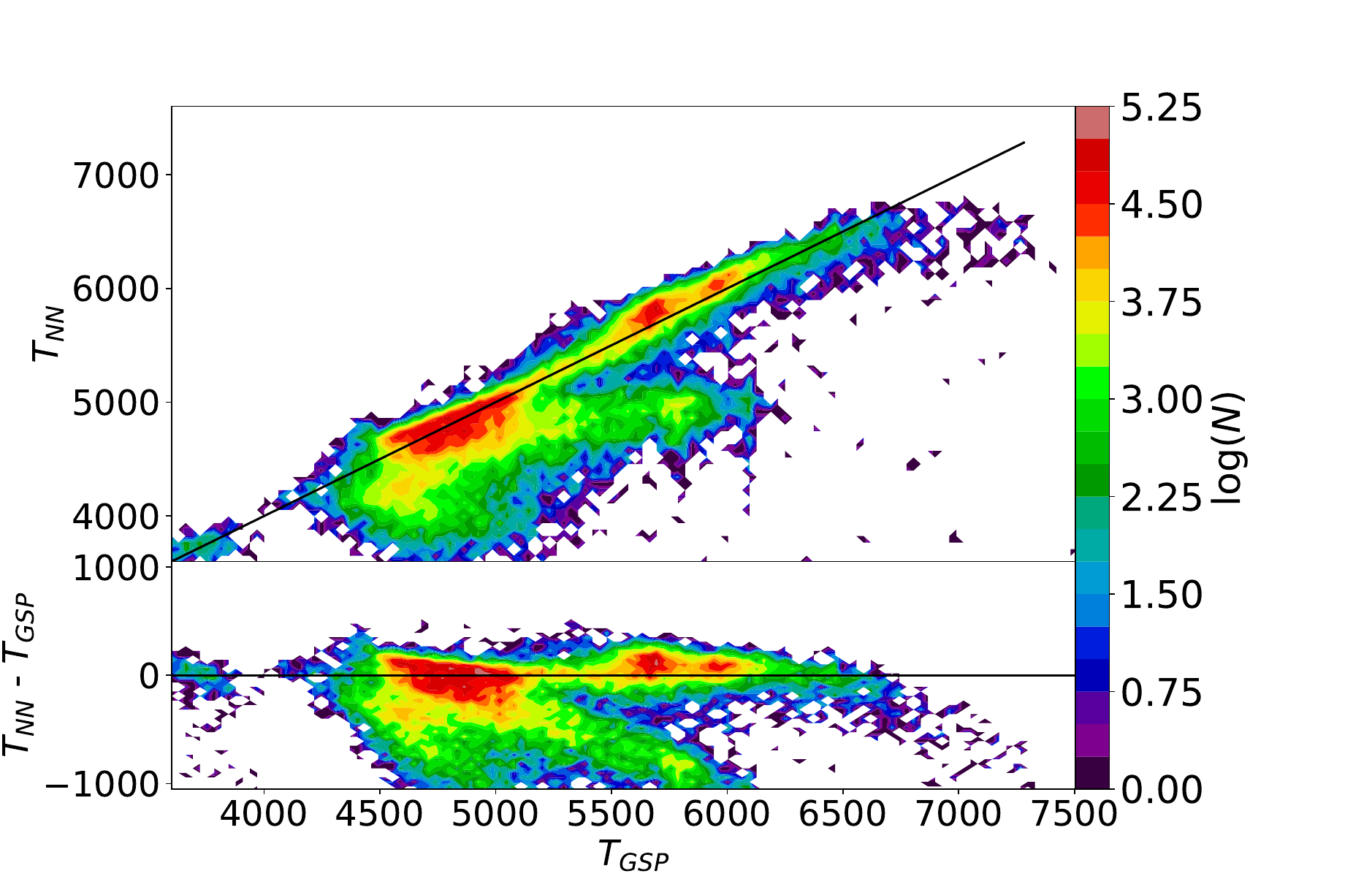}
    \caption{\rf{Comparison between our NN predictions and atmospheric parameters from the Gaia `GSP-Phot' algorithms.}}
    \label{fig:NN_vs_GSPPhot}
\end{figure*}

\subsection{Gaia GSP-Phot comparisons}

We also compare our NN's predictions against the General Stellar Parametizer from Photometry (GSP-Phot) data available as part of the Gaia survey \citep{Andrae2023a}. The GSP-Phot data are generally known to be good for higher quality observations, but have significant limitations for objects with high extinction or low-quality observations. We are therefore interested in how our model compares with this Gaia-only pipeline, and where our model makes notable improvements.

Figure~\ref{fig:NN_vs_GSPPhot} shows comparisons between our model's predictions and the Gaia GSP-Phot data. We note that we compare our NN's [Fe/H] predictions to the GSP-Phot [M/H] parameters. \cf{While we expect [Fe/H] and [M/H] to agree well for the model's predictions, small differences between these two abundances (e.g: due to alpha element abundance) will likely add additional scatter to our comparison figures.}

For the metallicity, we see the NN predicts a wide range of values for stars the GSP method considers solar-like. Similarly, there is a trend for objects the NN considers solar-like to be assigned lower metallicities in the GSP-Phot sample. When we filter for these divergent objects, we find these differences are most prominent in luminous giant stars. Further, when we analyse the objects that also diverge in $T_\mathrm{{eff}}$, we again find these same luminous giant populations. As we expect luminous objects to be more visible at larger distances, we expect observations of these objects to be more strongly affected by extinction effects. 

Similar divergences are present in the surface gravity comparisons, becoming prominent for giant stars with $\log g < 3.5$. This is clearest for red clump stars ($\log g \sim 2.5$) in our sample, which make up a significant proportion of our NN's APOGEE-based training sample. We see good agreement between NN and GSP-Phot for main sequence stars, which are generally fainter, and thus observed at smaller distances, and thus will have much lower extinction effects.  Therefore, we can again note extinction as a major cause for divergences between our model and the GSP-Phot observations.

Overall, we therefore agree with previous studies the general limitation of the GSP-Phot method is being unable to accurately model extinction \citep{Andrae2023a, Li2023_extinction}, and thus the method returns biased stellar parameters for high extinction objects. Further, this comparison appears to show further strengths in our model's predictions, which are seemingly less biased by interstellar extinction than the GSP-Phot observations. 

\subsection{Uncertainty Distribution}

\begin{table}
    \centering
    \caption{Our model's reported uncertainties compared with those from \citet{Andrae2023a} and \citet{Leung2023}.}
    \begin{tabular}{|lccc|}
    &[Fe/H] (dex)&$T_\mathrm{{eff}}$ (K) &$\log g$ (dex)\\\hline
    This work&$0.068$&$69.1$&$0.142$\\
    \citet{Leung2023}&$0.070$&$47.0$&$0.110$\\
    \citet{Andrae2023b}&$0.039$&$21.5$&$0.028$
    \end{tabular}
     \label{tab:ext_paper_comp}
\end{table}

In Table~\ref{tab:ext_paper_comp} we report our model's average uncertainties for each of the predicted parameters compared to the estimates from other works using BP/RP spectra: \citet{Leung2023} and \citet{Andrae2023a}. Note that, due to known underestimations in the \citet{Andrae2023a} uncertainties, we apply the suggested `inflation factors' for this comparison. We find our model compares well to predictions by \citet{Leung2023}, and has somewhat larger uncertainties than those reported by \citet{Andrae2023a}. 

\begin{figure*}
	\centering
	\includegraphics[width=.49\textwidth, trim={40, 0, 10, 40}, clip]{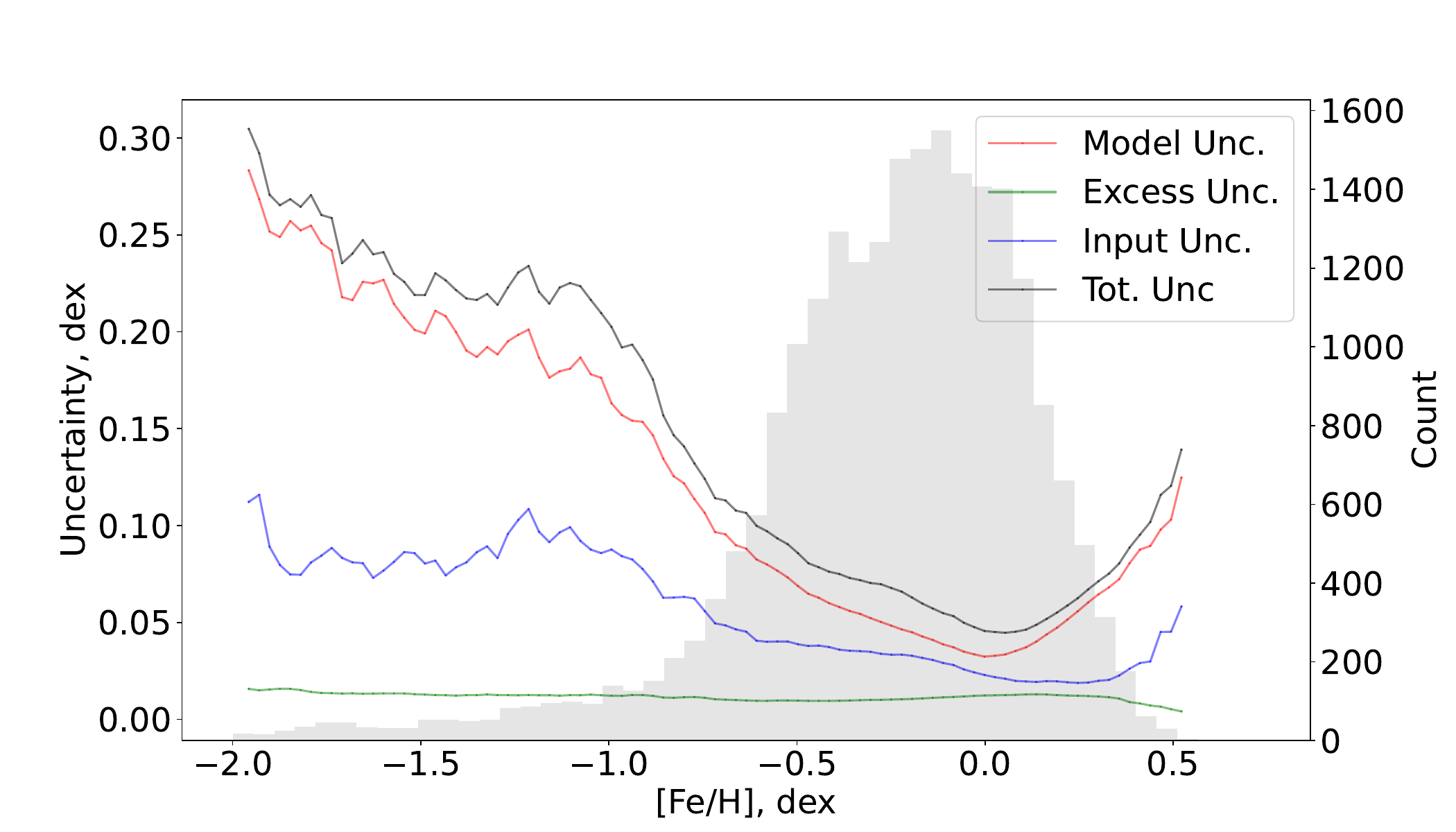}
	\includegraphics[width=.49\textwidth, trim={40, 0, 10, 40}, clip]{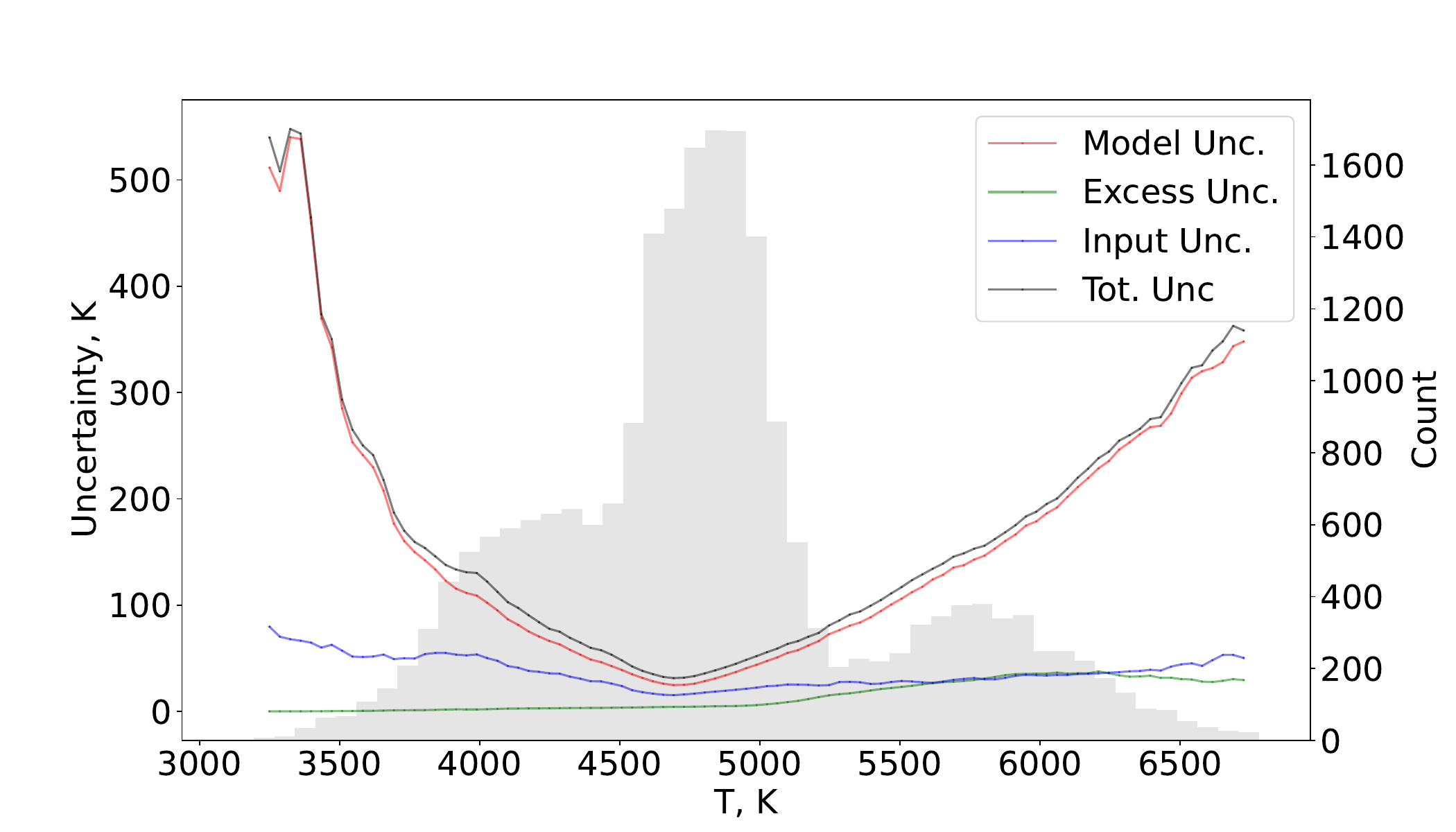}
    \includegraphics[width=.49\textwidth, trim={40, 0, 10, 40}, clip]{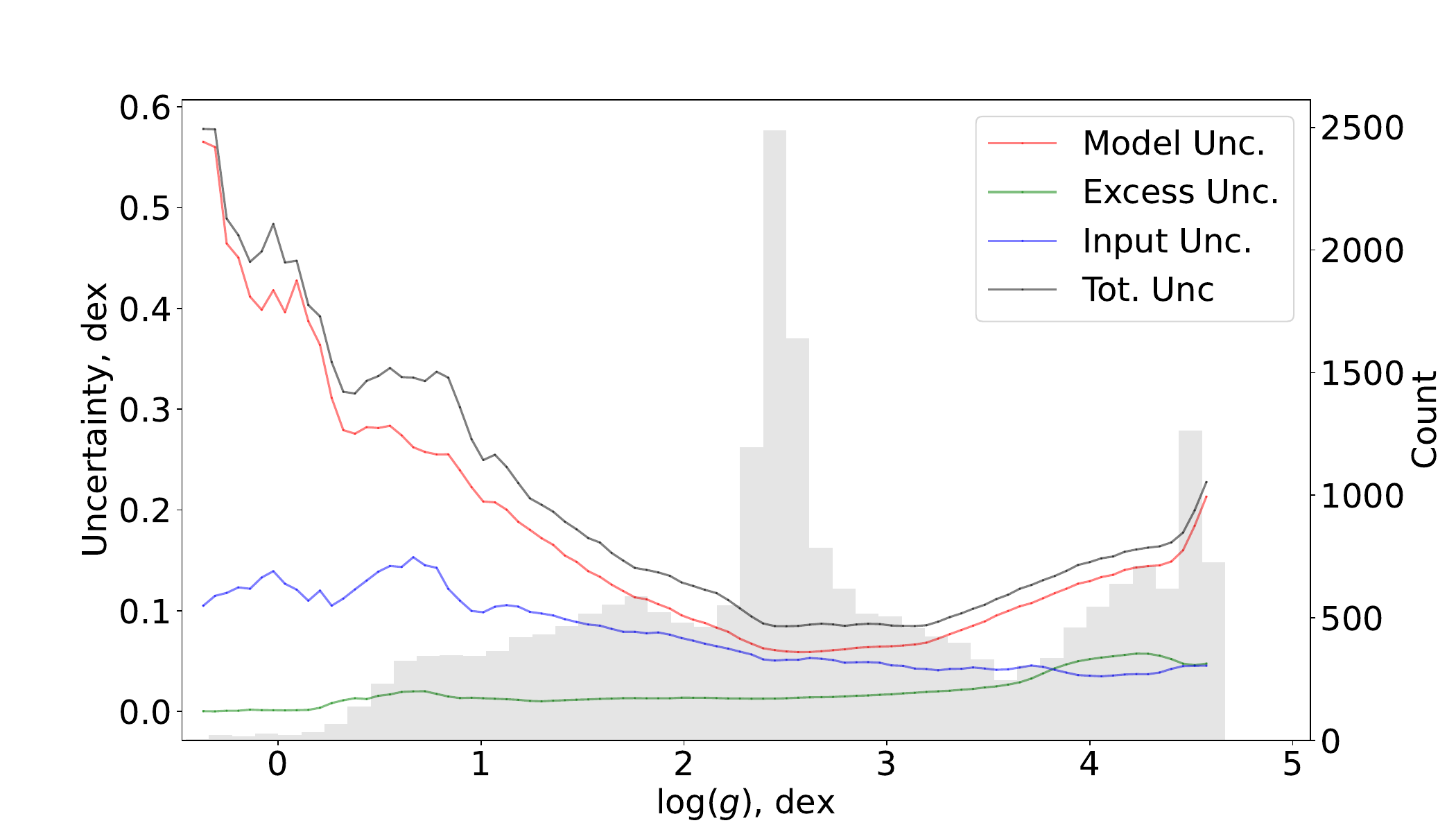}
    \includegraphics[width=.49\textwidth, trim={40, 0, 10, 40}, clip]{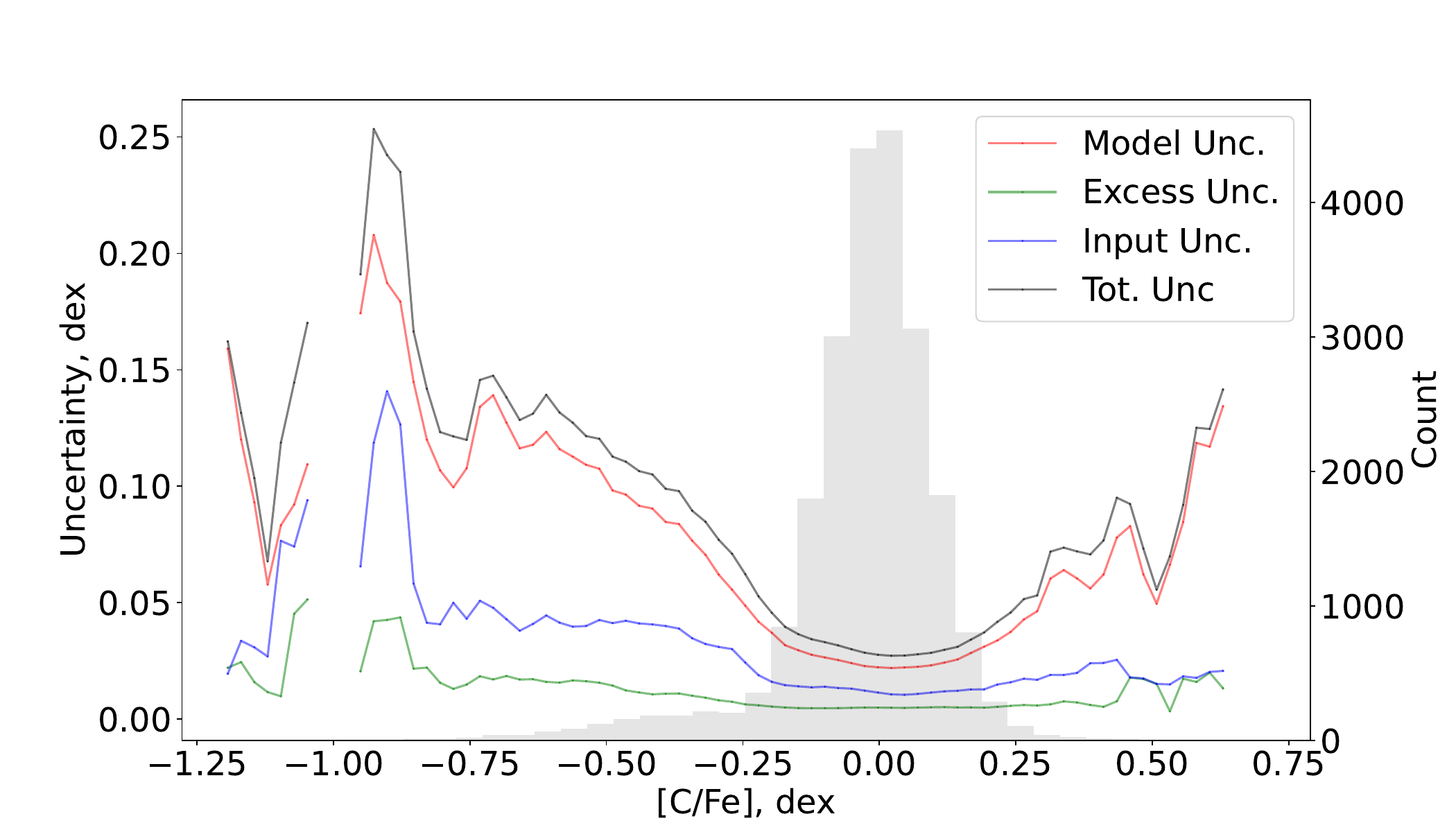}
    \includegraphics[width=.49\textwidth, trim={40, 0, 10, 40}, clip]{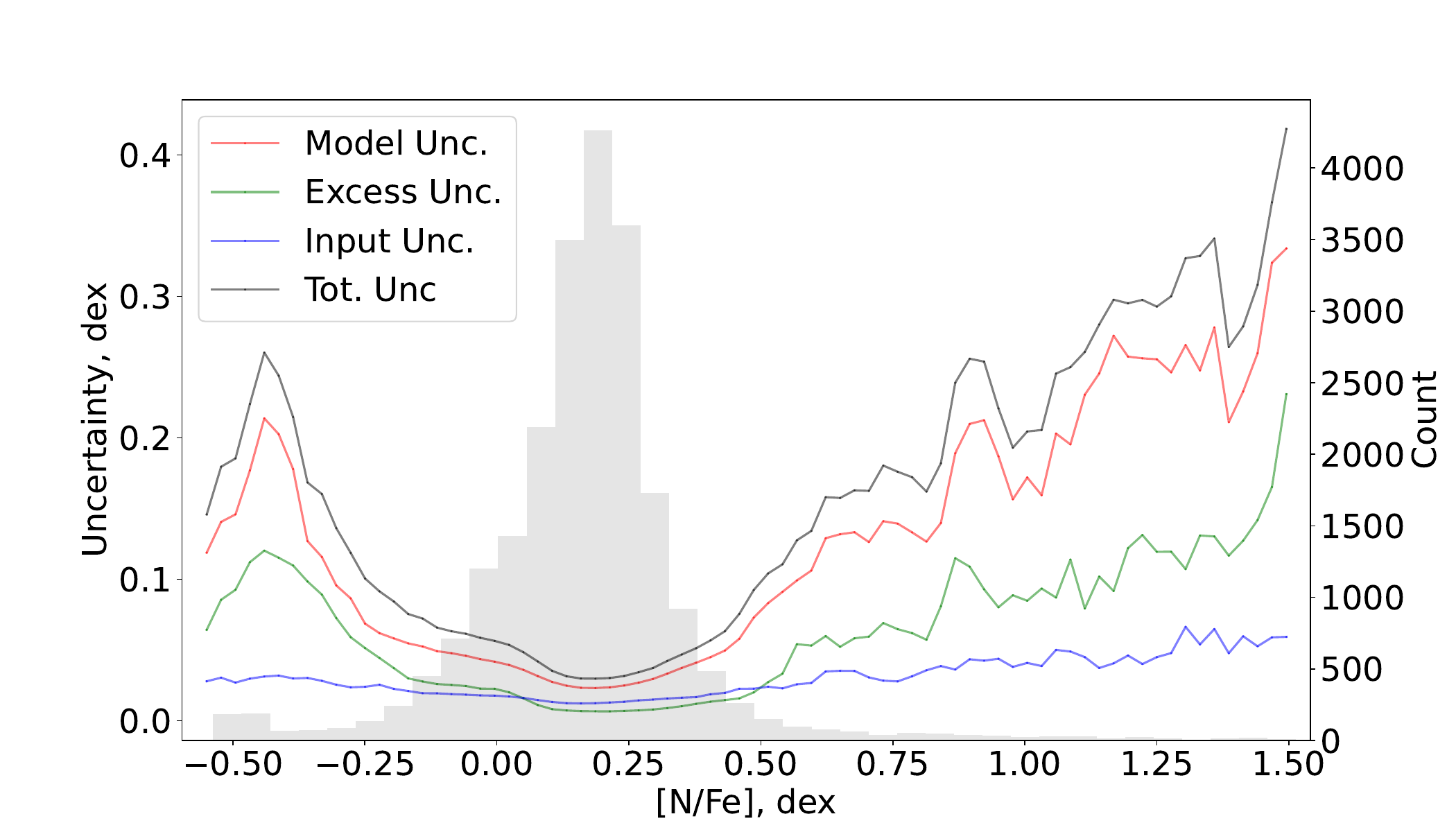}
    \includegraphics[width=.49\textwidth, trim={40, 0, 0, 40}, clip]{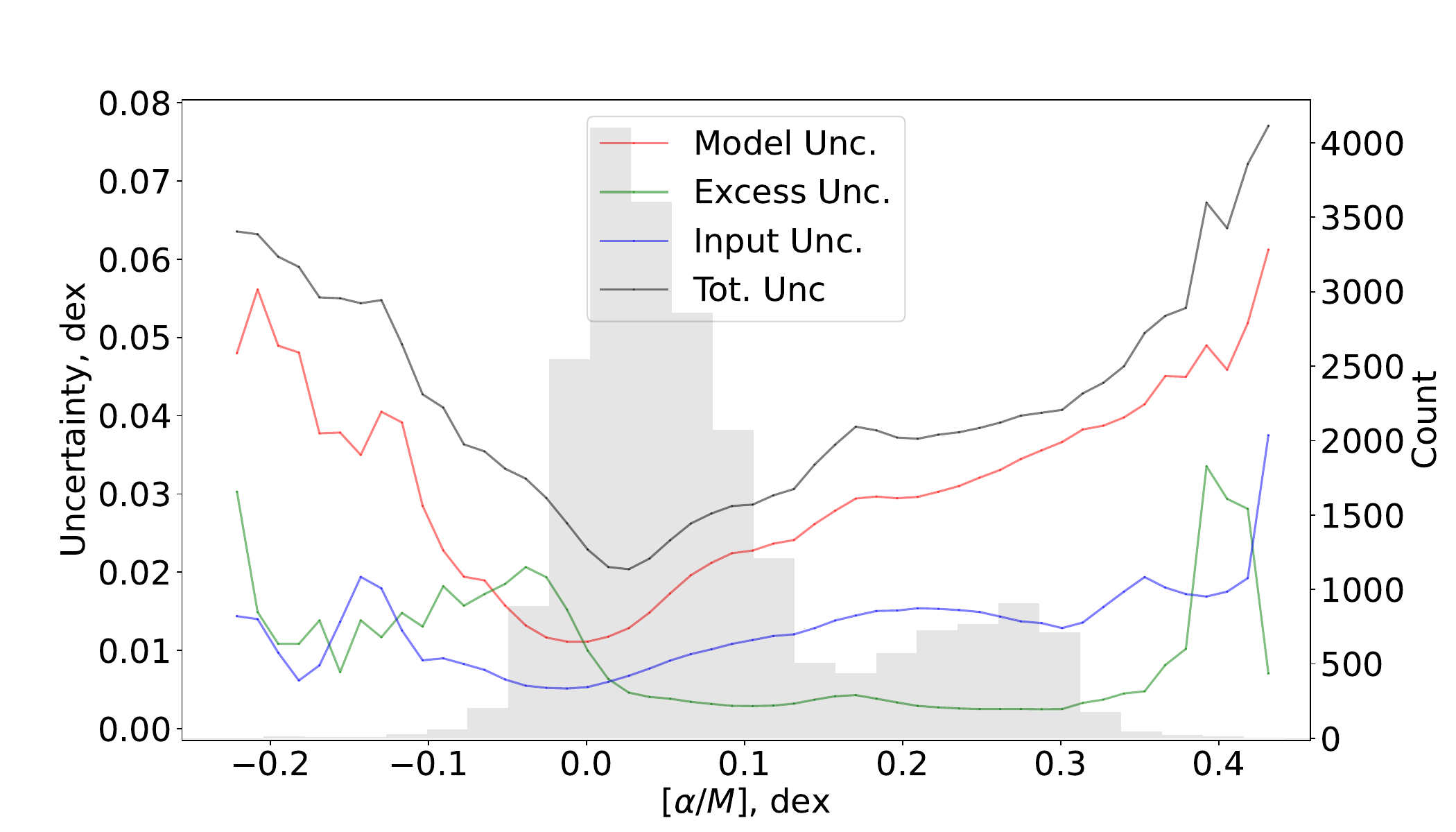}
    \caption{\rf{Our NN's reported uncertainties against the predicted stellar parameters. Figures are shown for [Fe/H], $\mathrm{T_{\mathrm{eff}}}$, $\log{g}$, [C/Fe], [N/Fe], and [$\alpha$/M]. The total uncertainty returned by the NN (\emph{black}) is further split into the three components described in Section~\ref{sec:unc_metrics}: model uncertainty (\emph{red}), input uncertainty (\emph{blue}) and excess uncertainty (\emph{green}). We include a histogram (\emph{grey}) to show the distribution of objects in each parameter.}}
    \label{fig:NN_vs_NNunc}
\end{figure*}

\cf{Average uncertainties do not \rf{robustly} describe how the model's \rf{precision} changes with \rf{respect} to each of our parameters. We therefore analyse the distribution of our uncertainties across the parameter space to identify any regions of unexpectedly high or biased uncertainties. We show comparisons between our model's uncertainties and the APOGEE parameters in Fig~\ref{fig:NN_vs_NNunc}, alongside the distribution of objects in the APOGEE-based training sample.}
\cf{We find our uncertainty distributions tend to be biased in correlation with the distribution of the training sample, where regions with large object counts have lower uncertainty predictions. This is an expected feature of using our NN method, as the model can be more certain in its predictions when it has more data available to it. Conversely, for the parameter ranges where the model has limited information (such as at the minimum and maximum edges of its training sample), it will struggle to predict with high certainty. }

\cf{Our three uncertainty metrics provide further insight into how the model's performance changes across the parameter ranges. We note that our model uncertainty (the uncertainty arising from the NN's architecture, estimated by applying dropout to the network) tends to be the largest of the three metrics for our [Fe/H], $T_\mathrm{{eff}}$, $\log g$, and [C/Fe] predictions. This suggests, that for these parameters, the predictions are highly dependent on the architecture of the model, and less so on scatter from input data or unaccounted-for errors. This trend is notably broken by our [N/Fe] and [$\alpha$/M] predictions, which both show regions of enhanced excess uncertainty. For [$\alpha$/M], this excess uncertainty appears for the most alpha-enhanced objects in our sample [$\alpha$/M]$>0.35$, suggesting the model is identifying some anomalous scatter being present in this small fraction of objects. However, for our [N/Fe] predictions, the excess uncertainty appears to dominate across the parameter range. This suggests there is some source of uncertainty present in these predictions that is being systematically missed: either our model is inaccurately estimating its own internal scatter, or the uncertainties in the input data are underestimated. As noted in Section~\ref{sec:Apogee_comparisons}, we have already shown some systematic issues in the observed APOGEE [N/Fe] values, which the model has identified as enhanced excess uncertainty. We finally note that, while this excess uncertainty describes uncertainty that isn't captured in our other metrics, it is incorporated in the reported uncertainties for the prediction - and thus we do not consider the NN's [N/Fe] uncertainties to be underestimated.}

\begin{figure}
	\includegraphics[width=0.5\textwidth, trim={50, 0, 0, 60}, clip]{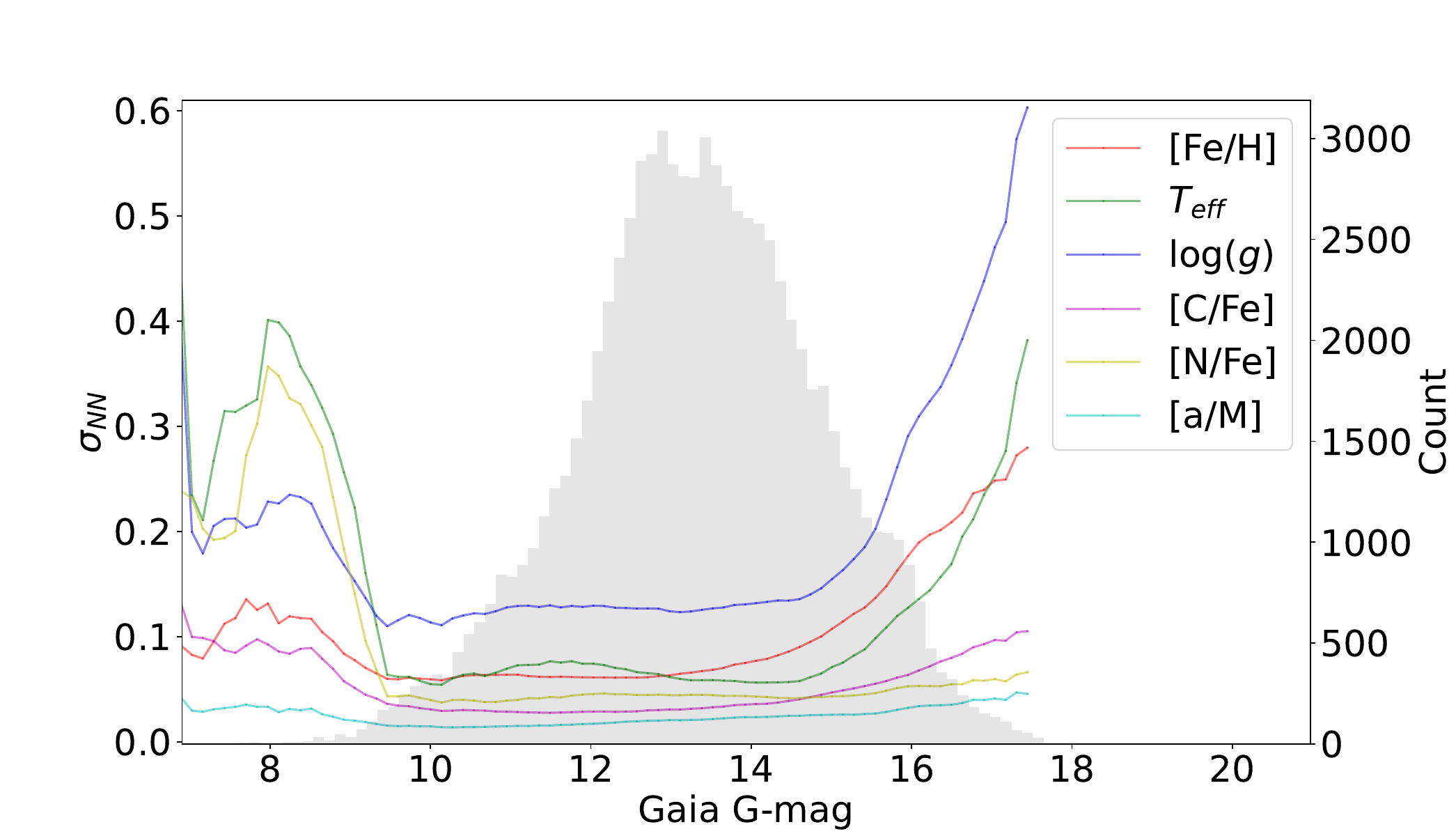}
    \caption{\rf{Our model's reported uncertainty against Gaia $G$ magnitude. Note we plot $\sigma_\mathrm{T_{eff}}/1000\,\mathrm{K}$ for the $T_\mathrm{{eff}}$ uncertainties.}}
    \label{fig:G_over_unc}
\end{figure}

As our model is working with the signal-to-noise of the BP/RP spectra, we expect the model's predictions to be higher uncertainty for fainter objects. We therefore compare our model's uncertainty against apparent Gaia G-magnitude in Fig.~\ref{fig:G_over_unc}. \cf{We see a reasonably flat uncertainty distribution across the $G$-magnitude range, with the distribution peaking for objects at the edges of our sample distribution ($G<9$ and $G>14.5$). This is an expected increase due to the model's predictions becoming less \rf{precise} when less information is available, as discussed previously. However, we also see a general trend to larger uncertainty for fainter $G$-magnitudes across the parameter range, albeit this change is small compared to the magnitude of the uncertainties. This suggests our model's predictions become more uncertain as objects get fainter (and thus have lower signal-to-noise ratios), and thus our model is behaving as expected.}

One final concern with the uncertainties returned by the NN is whether they are accurate for each prediction. To this end, we compare the model's uncertainties against the residuals between the NN's predictions and the APOGEE training data, following:
\begin{equation}
    A = \frac{\hat{m} - m}{\sigma_{\mathrm{NN}}},
	\label{eq:resid_vs_unc}
\end{equation}
where $m$ is the prediction from the NN, $\hat{m}$ is the APOGEE training value, and $\sigma_{\mathrm{NN}}$ is the uncertainty returned by the NN. For a sample where the scatter in the residuals is entirely accounted for by the model's uncertainties, $A$ will be distributed as a unit Gaussian. Deviations from a unit Gaussian imply our NN's uncertainties are either underestimated or overestimated, being over-broadened or over-narrowed respectively. 

\begin{figure*}
	\includegraphics[width=\textwidth, trim={90, 0, 100, 30}, clip]{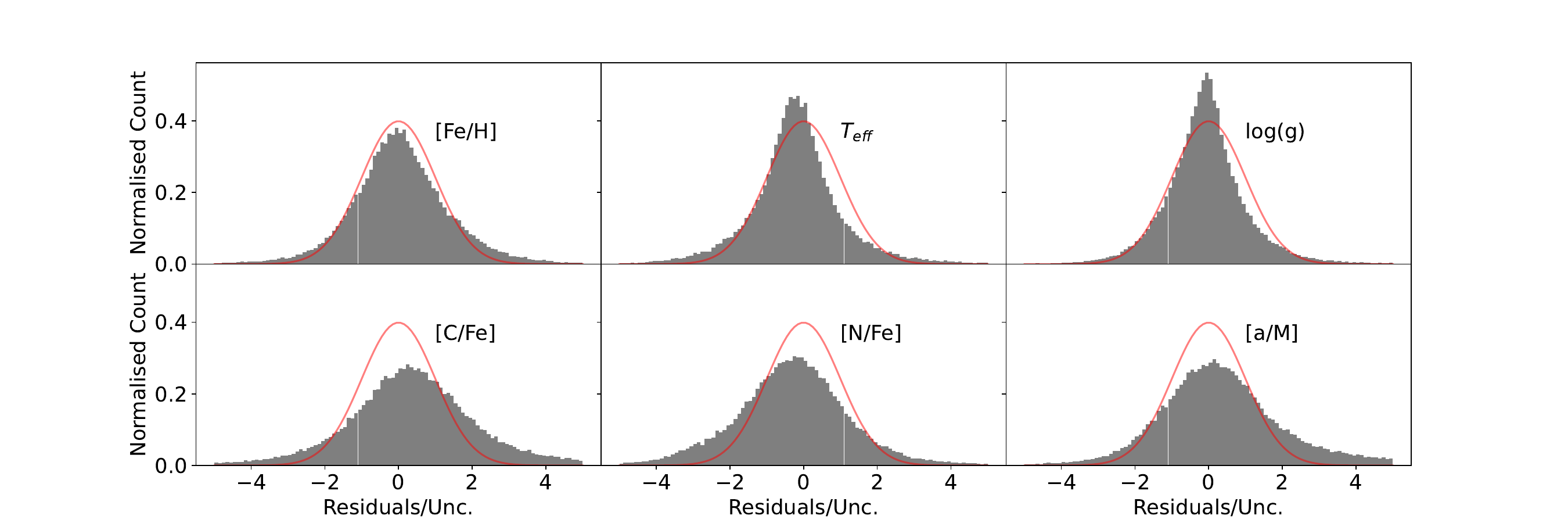}
    \caption{Prediction residuals divided by reported uncertainties, with over-plotted unit Gaussian (\emph{red}) for each of our predicted parameters.}
    \label{fig:residuals_over_unc}
\end{figure*}

We plot the distribution of $A$ in Fig.~\ref{fig:residuals_over_unc} for the NN's predicted atmospheric parameters. The six distributions all appear to match reasonably well to the expected unit Gaussian. For [Fe/H], $\mathrm{T_eff}$, and $\log(g)$, the distribution is slightly narrower than the unit Gaussian, suggesting our model's uncertainties are somewhat overestimated for these parameters. Conversely, the carbon, nitrogen, and alpha abundances show a very minor broadening relative to the unit Gaussian, suggesting our model's uncertainties may show minor underestimation for some objects. However, the overall good correlation between our distributions and the unit Gaussian further confirms our model is returning \rf{precise} uncertainty estimations.

\subsection{Per-Coefficient Attention}

\begin{figure*}
	\centering
	\includegraphics[width=.99\textwidth, trim={100, 0, 130, 40}, clip]{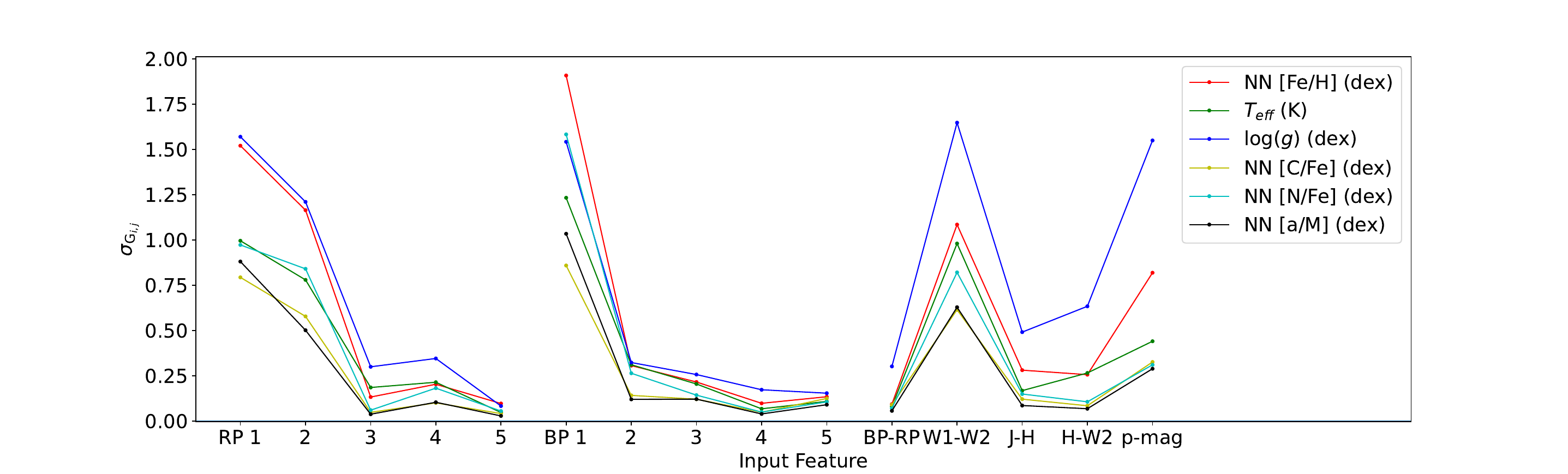}
    \caption{Standard deviation model attention, $\sigma_{\mathrm{G_{i,j}}}$, to our input features, over all objects in our sample. We show the gradients for the first 5 coefficients in the BP and RP spectra, photometric colours, and pseudo-absolute magnitude (shortened to `p-mag' here). Note that, for illustrative purposes, we plot attention for $\mathrm{T_{eff}}$ as $\sigma_{\mathrm{G_{i,j}}}/1000\mathrm{K}$.}
    \label{fig:per_coeff_derivs}
\end{figure*}

\cf{Alongside the metrics we have already discussed, we also analyse the importance of each input feature to the success of the model's predictions. To achieve this, we utilise the gradients across the NN used in Section~\ref{sec:unc_propagation} as a metric for the model's attention to each input feature. Previously we used the product of the model gradient and the input uncertainties to estimate the impact of those uncertainties on our predictions. Therefore, to measure the importance of each input feature to the NN's predictions, we instead calculate the product of the NN gradient, $\frac{\partial \boldsymbol{y}_{\mathrm{pred}}}{\partial \boldsymbol{x}}$, and our input features, $\boldsymbol{x}$, by}
\begin{equation}
    G_{i,j} = \frac{\partial {y}_{\mathrm{pred}, j}}{\partial {x}_i} {x}_{i},
	\label{eq:gradient_normalised}
\end{equation}
where $G_{i,j}$ is the model's attention to each $i$th input feature for a each $j$th stellar parameter output by the NN. We then take the standard deviation of $G_{i,j}$ over all stars in our sample, where a larger value represents a higher attention to a chosen feature. 

We plot the per-coefficient attention in Fig.~\ref{fig:per_coeff_derivs}. As the BP/RP spectra are constructed such that lower-order coefficients contain a larger proportion of the spectral information \citep{Carrasco2021}, we expect our model's attention to focus more on these lower-order terms. We therefore simplify Fig.~\ref{fig:per_coeff_derivs} to only the first five BP and RP input features, alongside the photometric colour inputs.

Primarily, we see our model appears to draw significant information from both the BP/RP spectra and the photometric colours. While the lowest-order coefficients in BP and RP show the highest attention from the model, following our expectations, we see some of the photometric colours are considered as important as our BP/RP terms. Specifically, we see the model draw significant information from the WISE W1-W2 colour which, as discussed by \citet{Grady2021} and \citet{Fallows2022}, is known to be sensitive to changes in metallicity.
Furthermore, we see all of our predicted stellar parameters draw information from the pseudo-absolute magnitude input. As pseudo-absolute magnitudes are designed to include luminosity information in our model, they provide information on stellar mass and evolutionary stage. These properties are important in predicting all of our stellar parameters - especially in separating giant and main sequence objects via $\log g$.

\subsection{Abundance Distributions}
\label{sec:abundance_dbns}
One additional metric to validate our model's predictions is whether the abundance distributions we observe in our data match the distributions we expect from existing models and observations. \cf{While this analysis alone does not prove the success of a model's predictions (as a model suffering large biasing from overfitting may still recover the sample distribution, while predicting poorly per individual object), we can use anomalies in these predicted distributions to understand where the NN is under-performing.}

\begin{figure*}
	\centering
	\includegraphics[width=.49\textwidth, trim={30, 10, 70, 60}, clip]{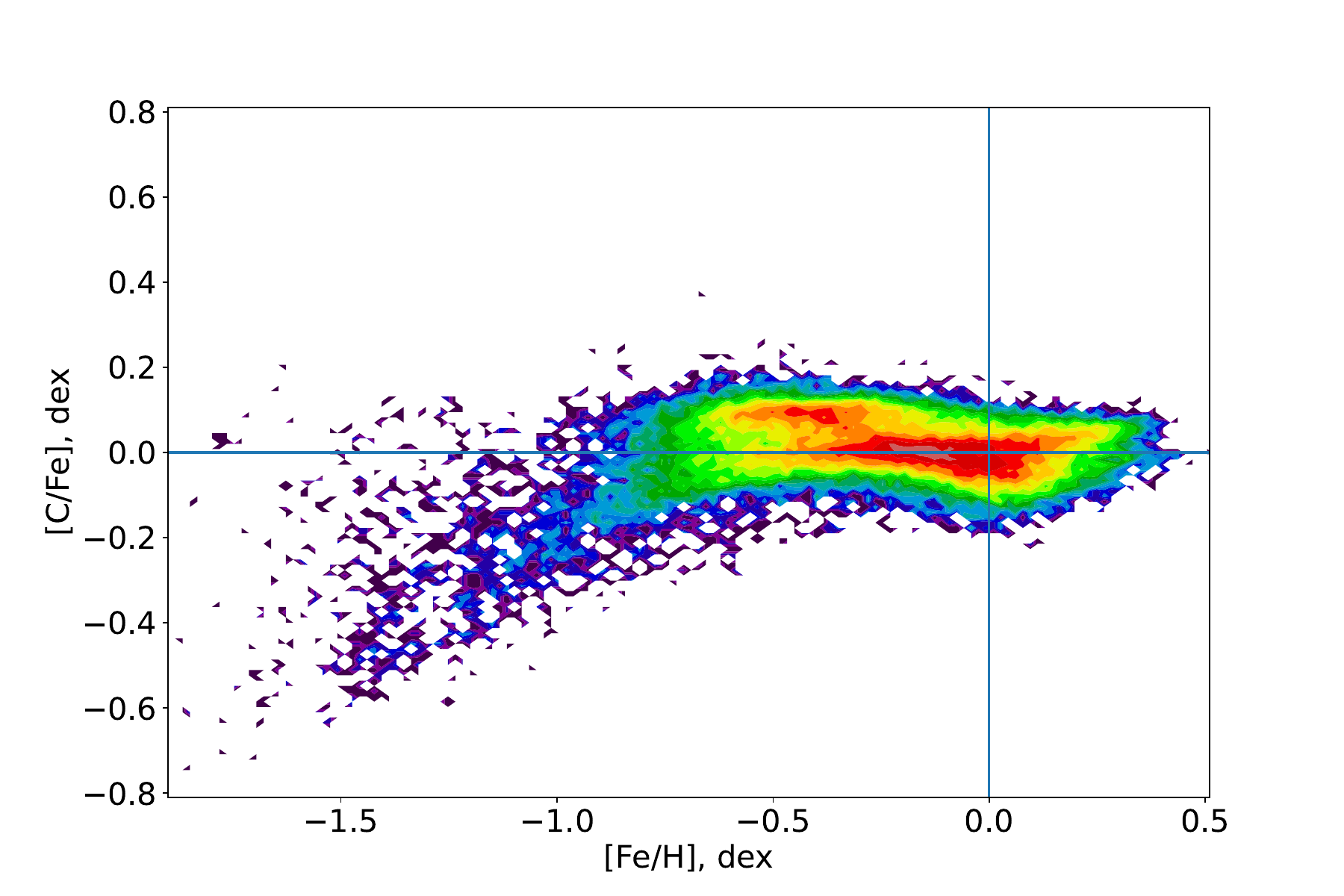}
    \includegraphics[width=.49\textwidth, trim={30, 10, 70, 60}, clip]{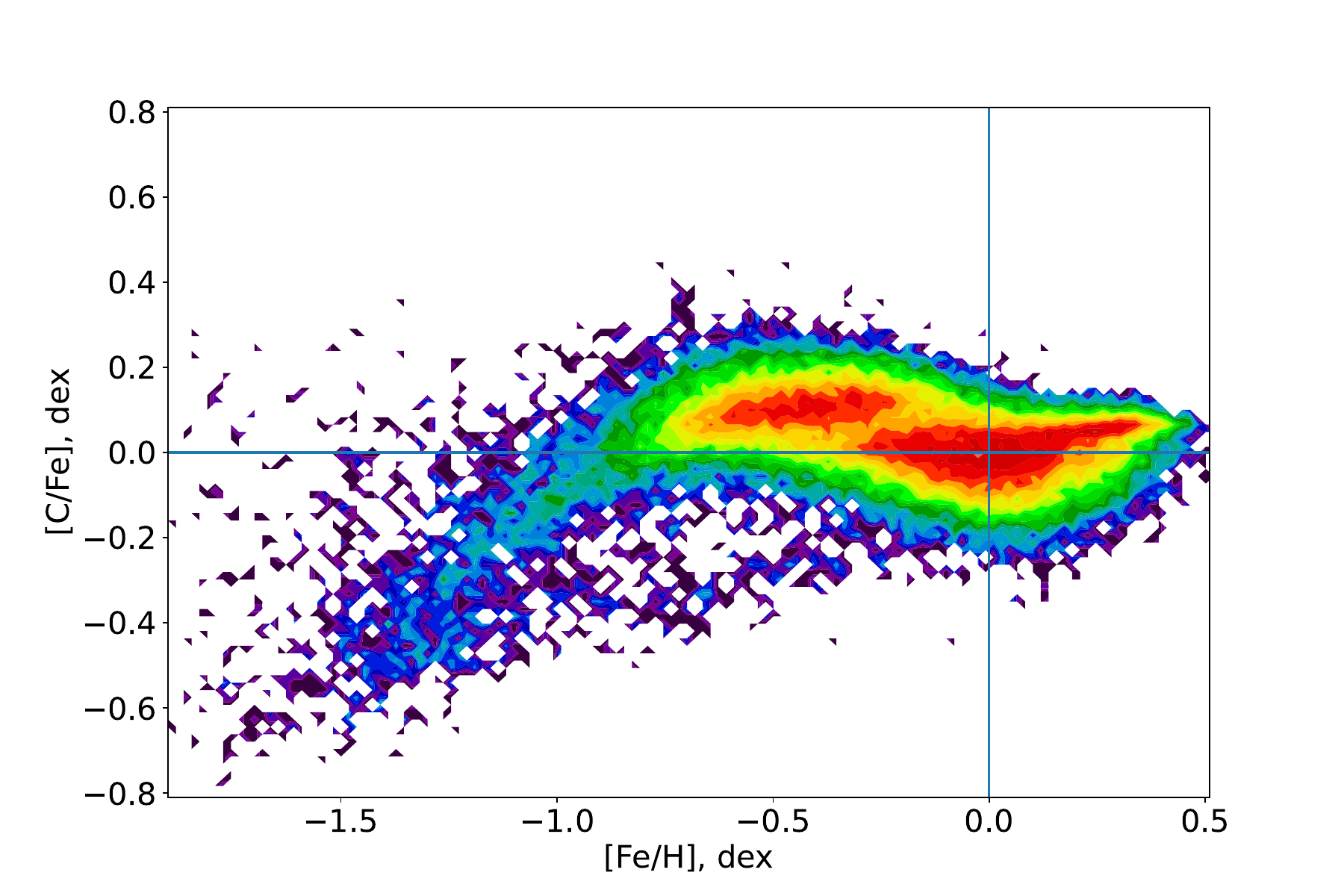}
	\includegraphics[width=.49\textwidth, trim={30, 10, 70, 60}, clip]{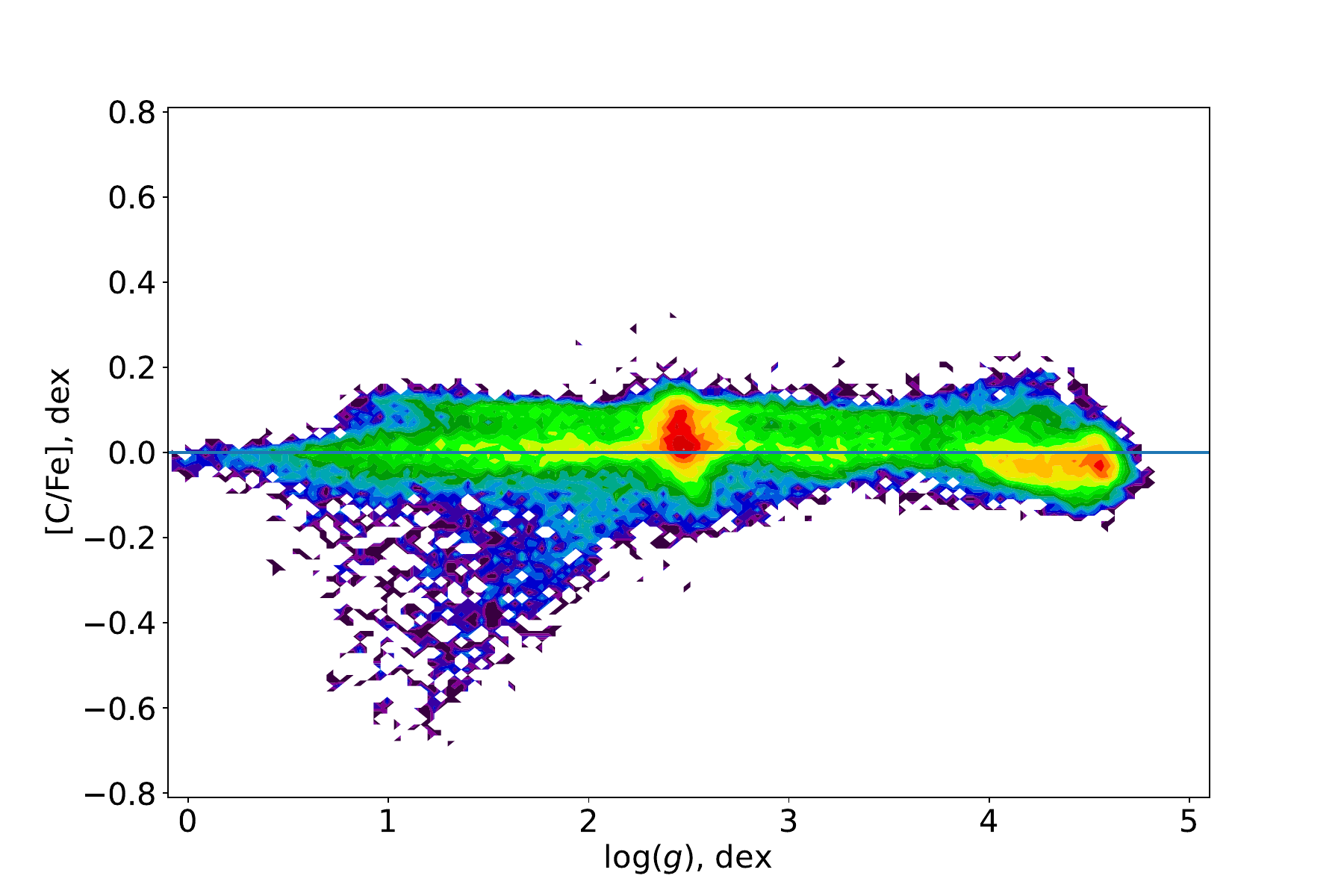}
    \includegraphics[width=.49\textwidth, trim={30, 10, 70, 60}, clip]{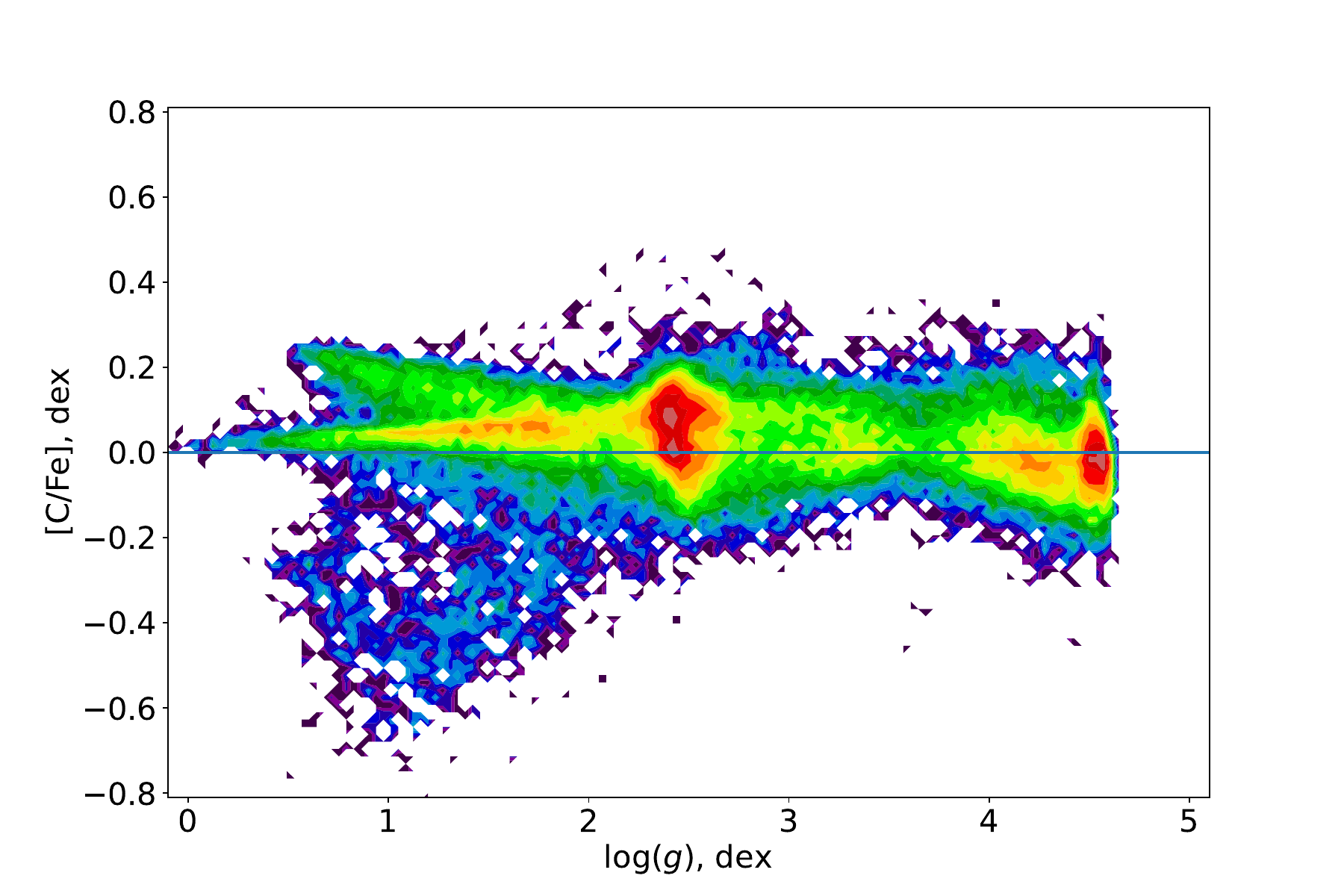}
    \caption{Carbon abundances against [Fe/H] and $\log(g)$, as estimated by the NN (left) and as observed by APOGEE (right).}
    \label{fig:carbon_iron_comps}
\end{figure*}

\begin{figure*}
	\centering
	\includegraphics[width=.49\textwidth, trim={10, 10, 70, 60}, clip]{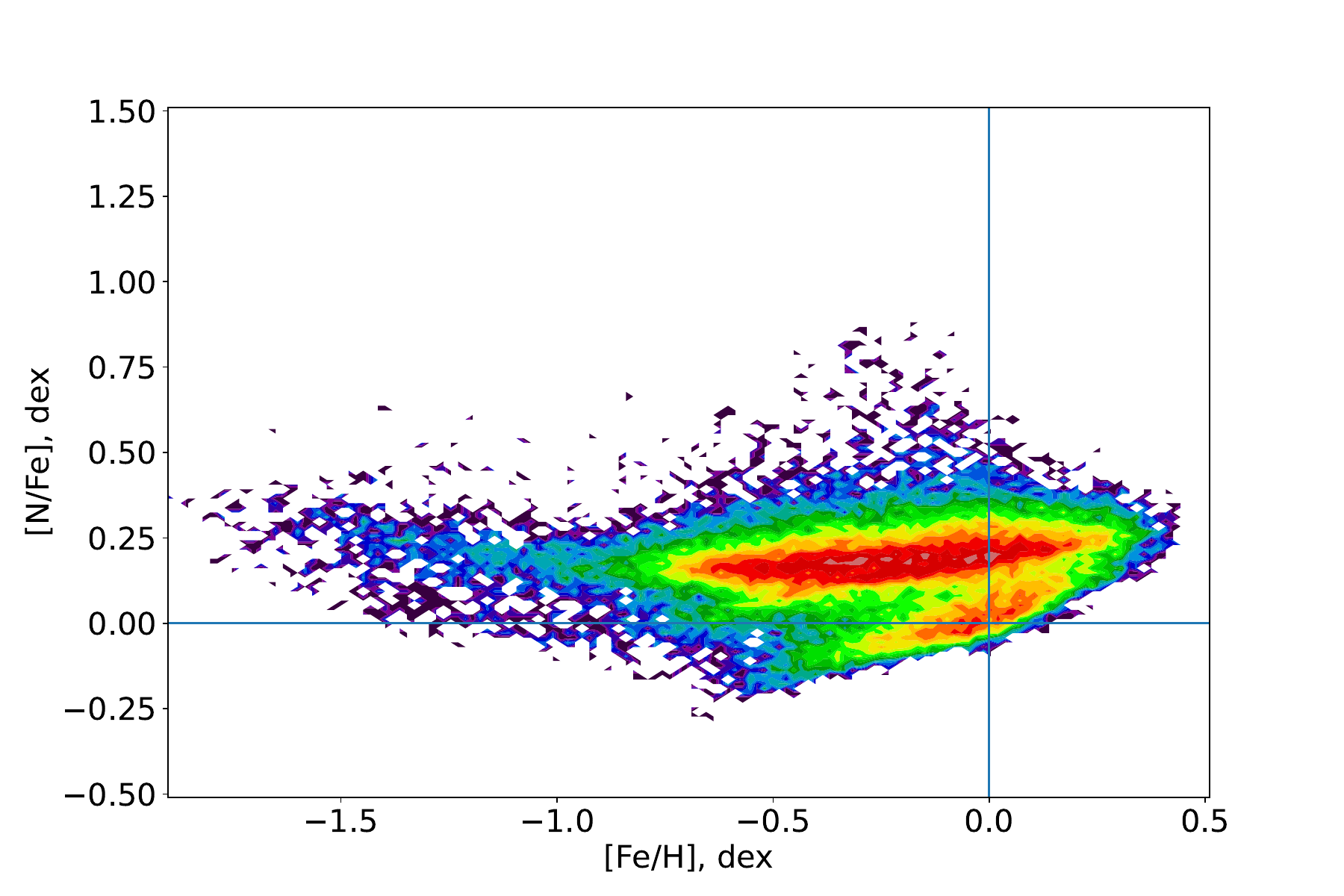}
    \includegraphics[width=.49\textwidth, trim={10, 10, 70, 60}, clip]{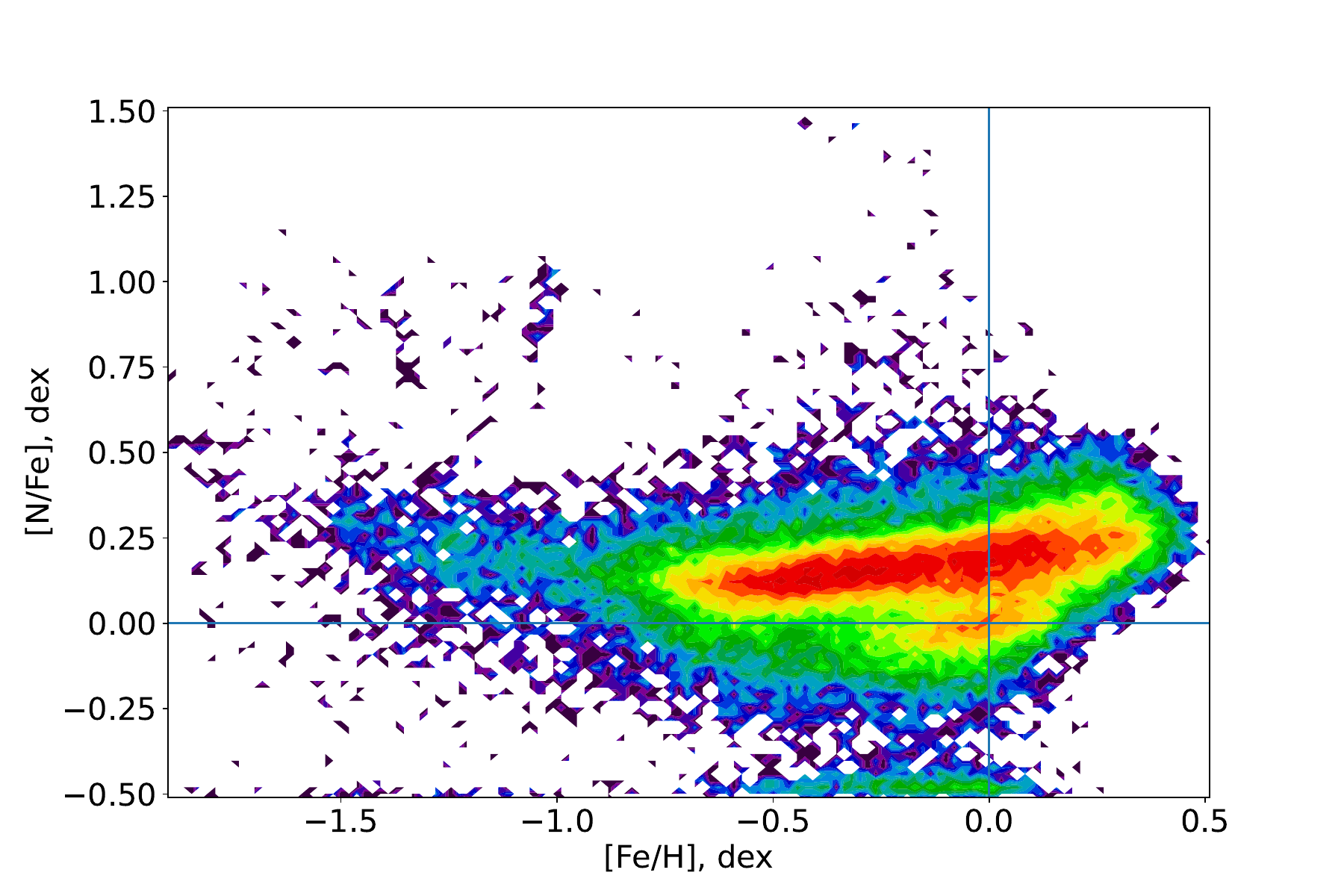}
    \includegraphics[width=.49\textwidth, trim={10, 10, 70, 60}, clip]{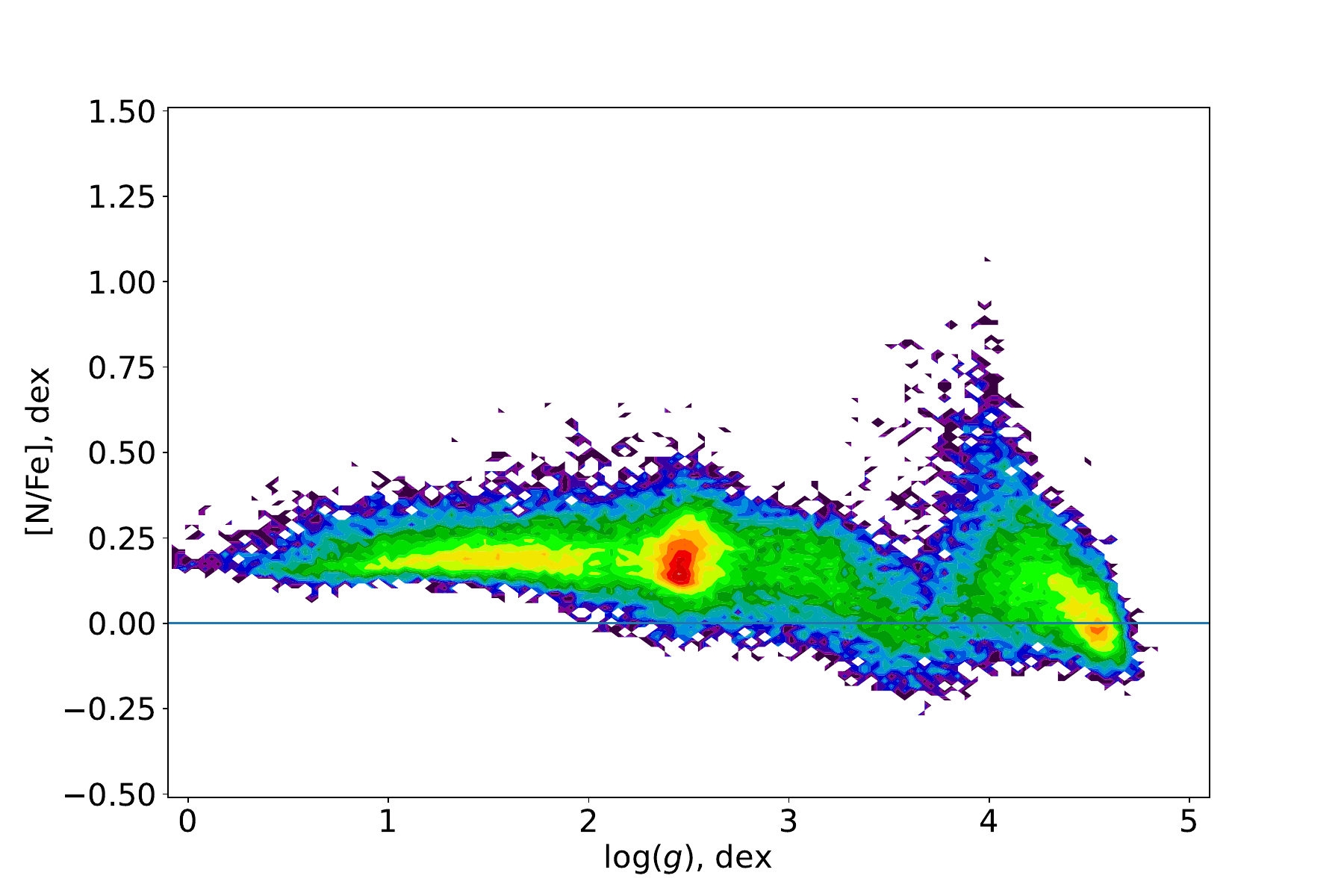}
    \includegraphics[width=.49\textwidth, trim={10, 10, 70, 60}, clip]{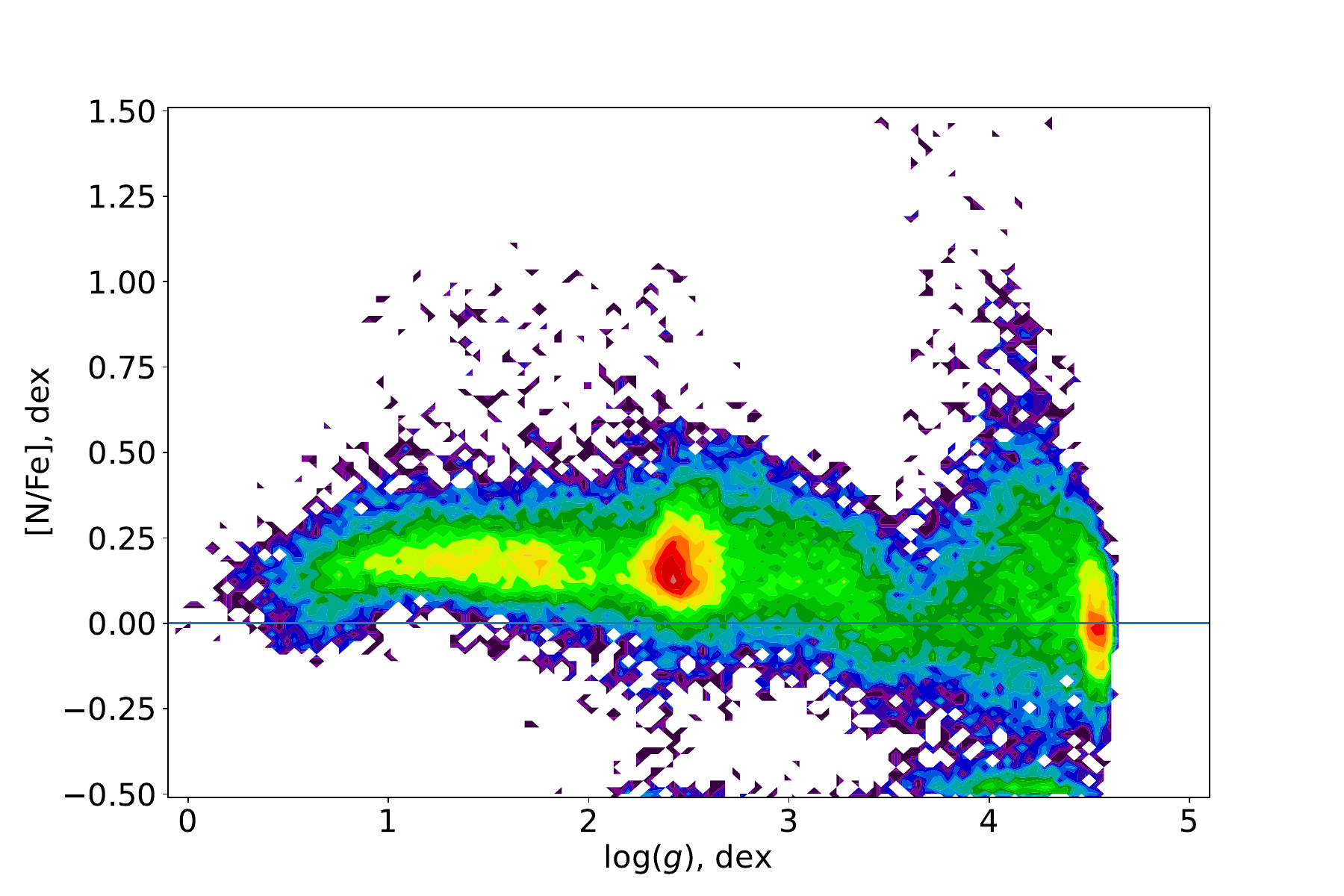}
    \includegraphics[width=.49\textwidth, trim={10, 10, 70, 60}, clip]{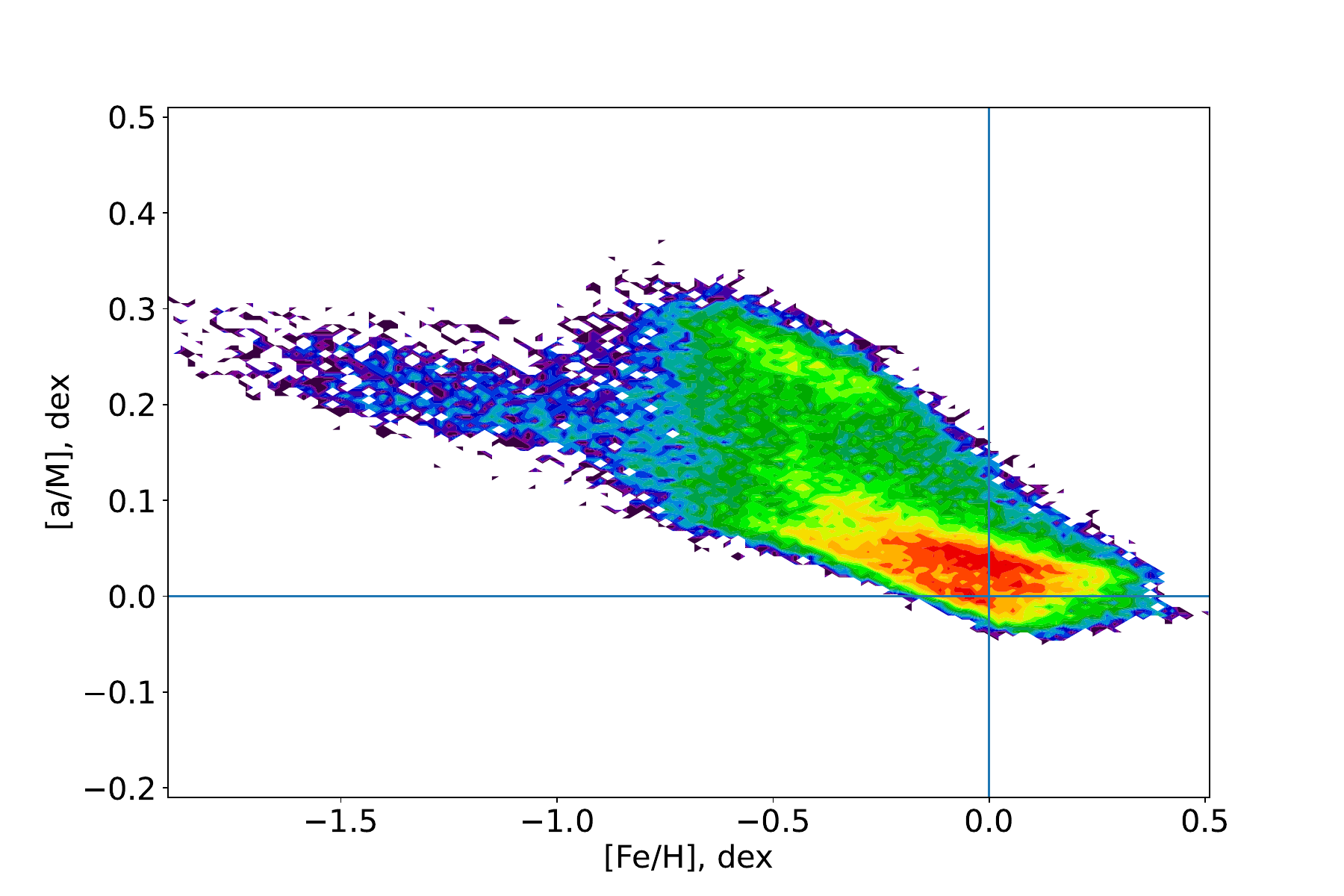}
    \includegraphics[width=.49\textwidth, trim={10, 10, 70, 60}, clip]{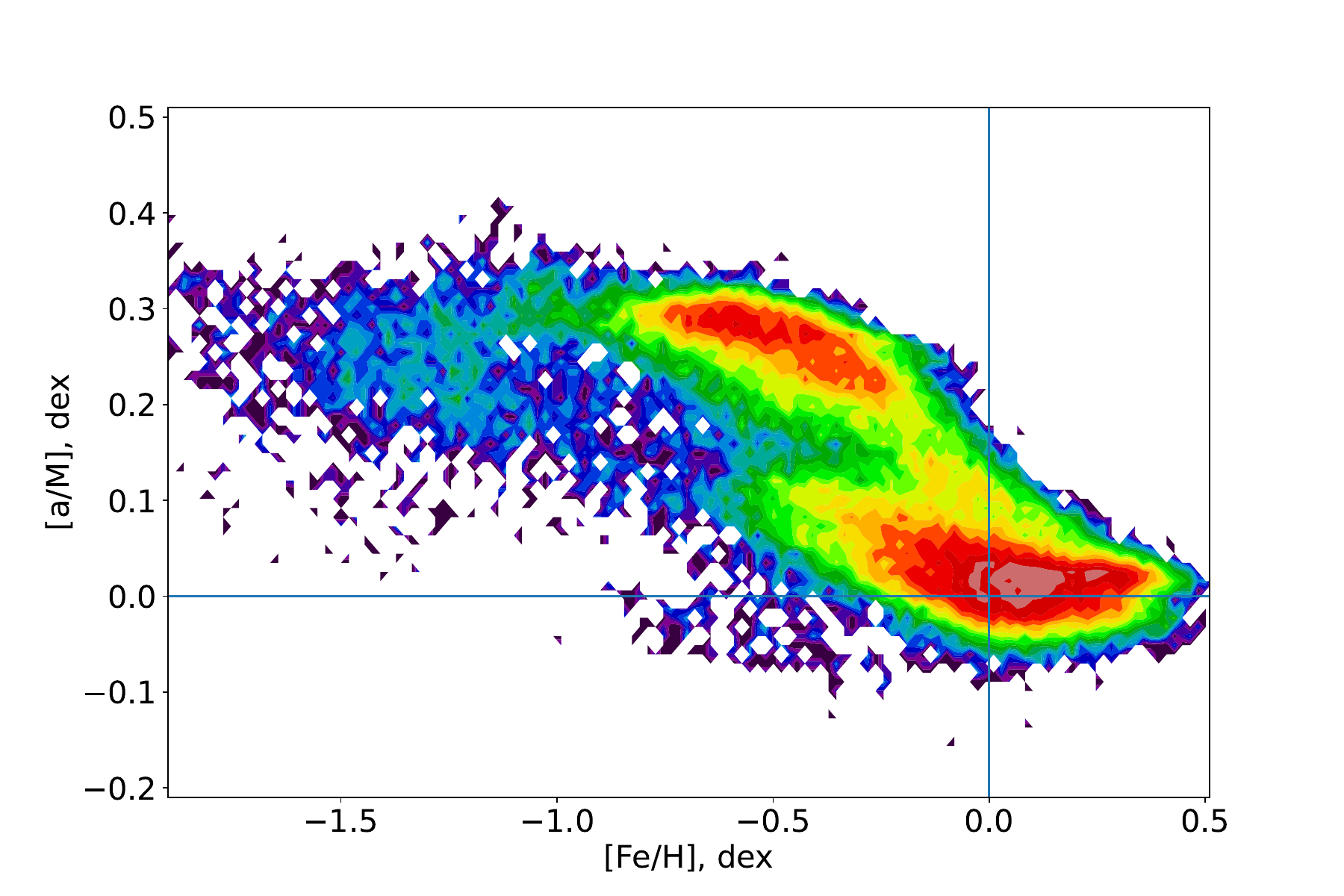}
    \caption{\rf{Nitrogen and alpha abundances against [Fe/H] and $\log g$, as estimated by the NN (left) and as observed by APOGEE (right).}}
    \label{fig:nitrogen_iron_comps}
\end{figure*}

We focus on five comparisons: [C/Fe], [N/Fe], and [$\alpha$/M] against [Fe/H]; and [C/Fe] and [N/Fe] against $\log(g)$. These abundance distributions allow us to identify several important populations in our data, due to how abundances trend along stellar evolutionary pathways (e.g. how carbon and nitrogen abundances evolve along the giant branch).  We show these distributions in Fig.~\ref{fig:carbon_iron_comps} and \ref{fig:nitrogen_iron_comps}, alongside comparable distributions in the APOGEE survey data. 

For our model's [C/Fe] vs [Fe/H] distributions, we see a generally good correlation with the APOGEE parameters. We note a similar overall shape, showing distinct thin and thick disk populations and a tail of iron and carbon-poor giant stars. However, our model's [C/Fe] predictions do not separate the two components of the Milky Way disk as cleanly as the APOGEE data, with a larger spur of metal-poor objects with solar carbon abundances being visible in our predictions. These discrepancies are largely reduced when filtering for only objects with small predicted uncertainties, suggesting our model is \rf{precisely} recovering these populations in its higher quality predictions. 

Similarly for the comparisons between [C/Fe] and surface gravity, we again see our predictions generally mimic the APOGEE data. The primary difference of note is the overall lower scatter in the carbon abundance distributions. Our model shows a slightly narrower profile in [C/Fe], with fewer objects predicted above [C/Fe] $>0.2$ or below [C/Fe] $<-0.2$ than we would expect from the APOGEE sample. We also do not see the carbon-enhanced giant-branch spur present in the APOGEE data. This further suggests a trend to the mean in the carbon predictions, with the model struggling to classify objects with larger deviations in carbon abundance, and thus falling back to the sample mean ([C/Fe] $\sim$ 0). However, the stronger correlations between our predictions and the spectroscopy visible around the major stellar populations in these comparisons (red clump and main sequence) do suggest the model is better equipped to recover \rf{precise} predictions where data is abundant.

Our [N/Fe] vs [Fe/H] comparisons generally compare well to the spectroscopic parameters, showing a clear separation of giant and main sequences in the predictions. We do however note a small `spur' around [Fe/H] $\sim$ $-0.5$ and [N/Fe] $\sim$ $0.1$. These objects are not associated with any unusual or anomalous population, suggesting this is likely an artefact of the model's training. Furthermore, we see a similarly good correlation in the [N/Fe] vs surface gravity comparisons, with the overall shape of the plot matching well. We again see a slight decrease in scatter in the abundance compared to APOGEE, as we noted in the carbon abundances. This again suggests a trend for the model towards the sample mean for particularly unusual or uncertain objects, thus compressing the distribution away from extreme deviations.

Finally, we show our comparisons between [Fe/H] and our predicted alpha abundances. This comparison seems to agree well with APOGEE for metal-rich objects, and we see a \rf{weak bimodality of high- and low-alpha sequences, associated with thin and thick disk populations \citep{Navarro2011}. While we do not see the strong separation we may expect from the APOGEE data, prior Gaia-based methods have identified similar bimodal sequences using Gaia RVS spectra \citep{Guiglion2024} and BP/RP spectra \citep{Li2024}. Thus, the appearance of this pattern in our NN's predictions suggests the model is extracting this sequence separation from our input BP/RP spectra and photometry. We note that, as mentioned above, machine learning models like ours tend to make predictions following the distribution of its training data. Therefore, the presence of bimodality is expected, as it is difficult to disentangle the performance of the model from this trend.}

Furthermpre, the NN's predictions diverge considerably from APOGEE for more metal-poor objects ([Fe/H] < -0.75), reducing the reliability of predictions in the metal- and alpha-poor regime. As the low-[Fe/H], low-[$\alpha$/M] regime is associated with material accreted onto the Milky Way \citep{Helmi2018}, this biasing makes separation of in-situ and accreted material difficult without additional data.

This difficulty in estimating alpha abundance for metal-poor stars is in agreement with \citet{Li2023_AspGap}, who note the weakened metal features in low-[Fe/H] stars make [$\alpha$/M] estimations difficult from low-resolution BP/RP data.

\section{Spectral Information}
\label{sec:spectral_info}

\begin{figure*}
	\centering
	\includegraphics[width=.95\textwidth, trim={110, 0, 120, 50},clip]{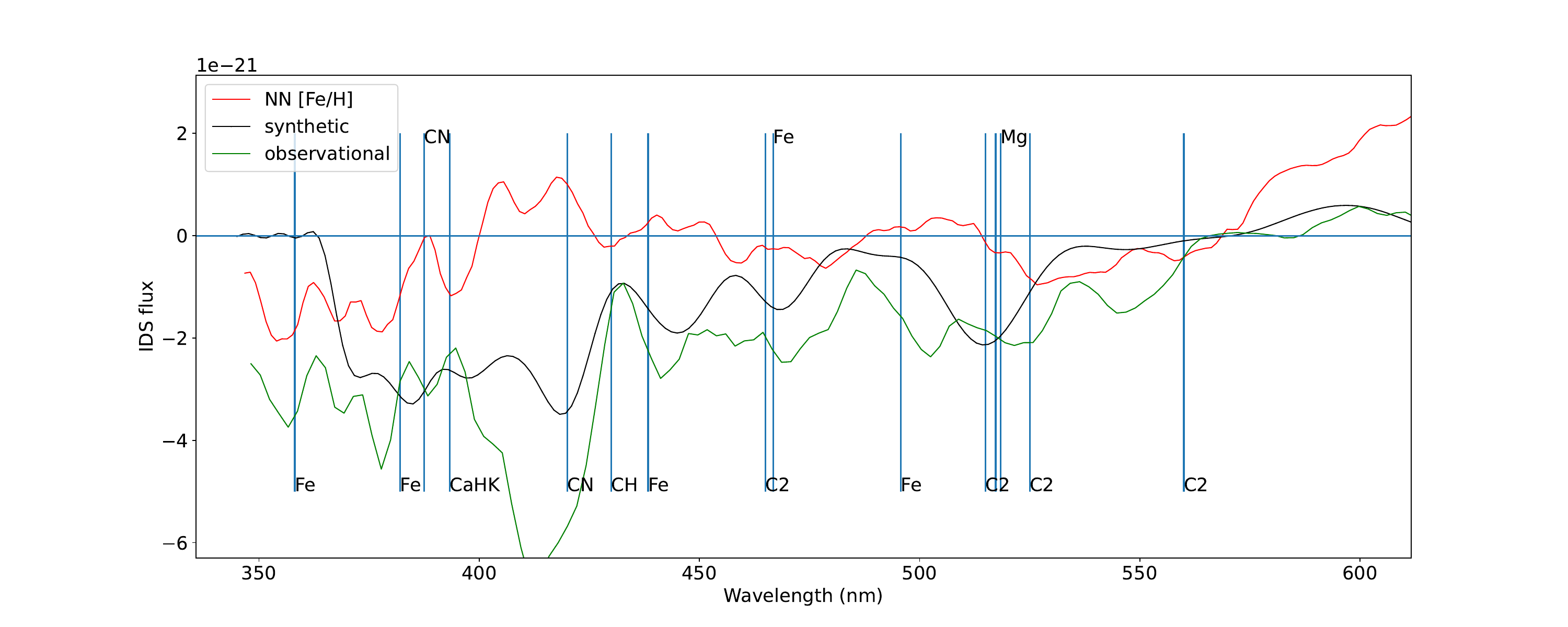}
    \includegraphics[width=.95\textwidth, trim={110, 0, 120, 50},clip]{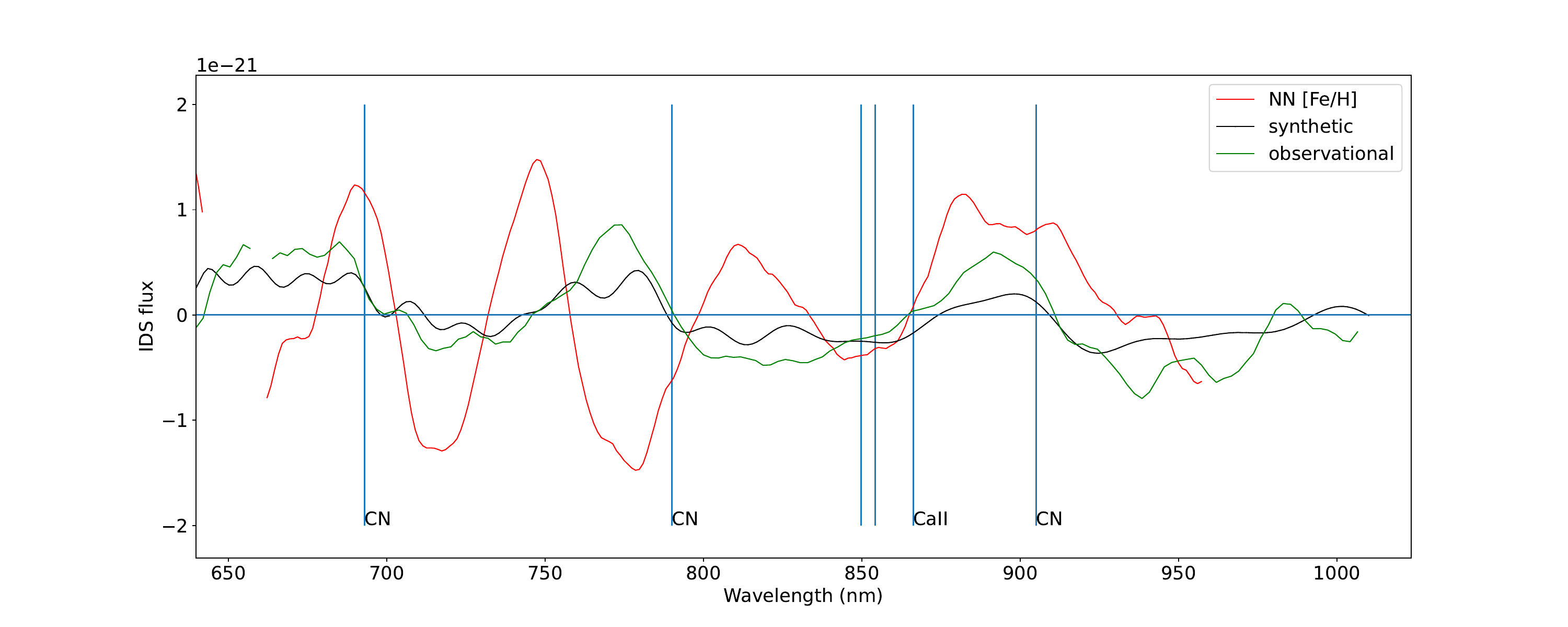}
    \caption{[Fe/H] comparisons for red clump stars ([Fe/H] $\sim$ 0.0, T $\sim$ 4750, $\log g \sim$ 2.5) between our NN's DS, synthetic spectral differences, and APOGEE observational differences.}
    \label{fig:synth_vs_NN_feh}
\end{figure*}

\begin{figure*}
	\centering
	\includegraphics[width=.95\textwidth, trim={110, 0, 120, 50},clip]{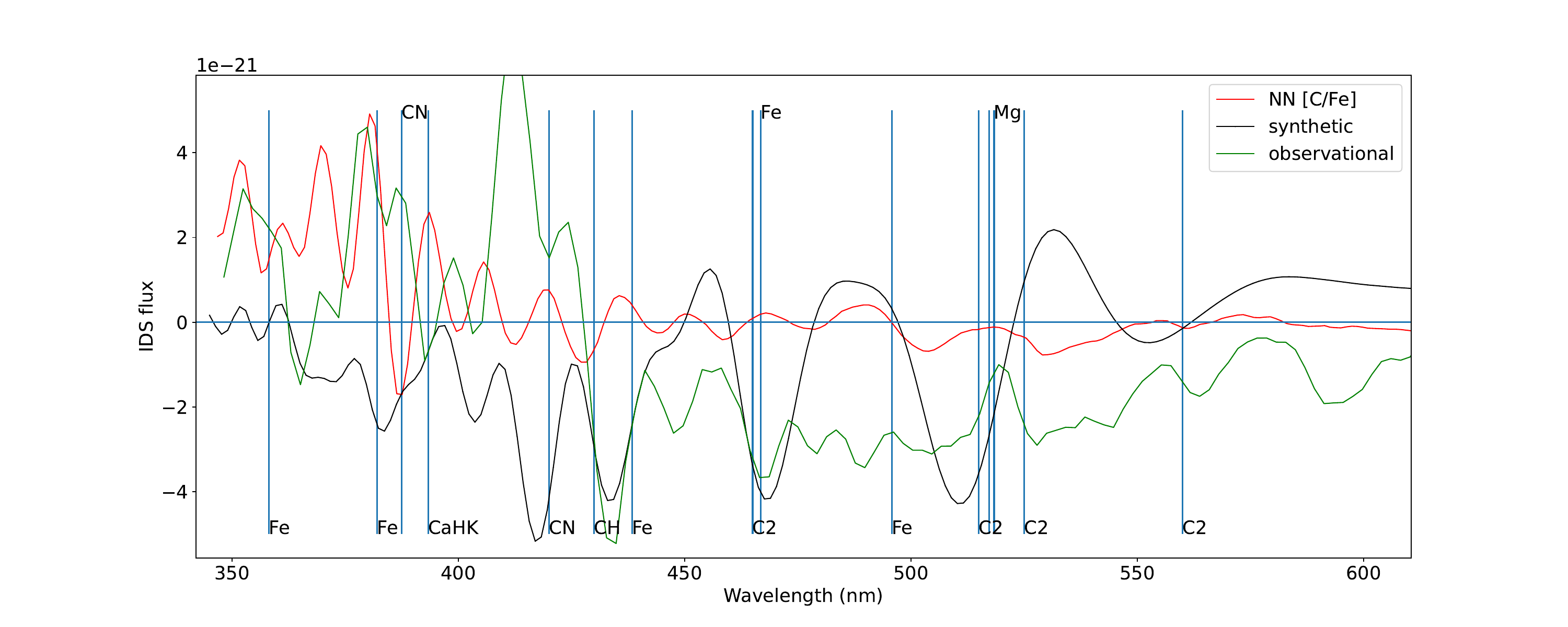}
    \includegraphics[width=.95\textwidth, trim={110, 0, 120, 50},clip]{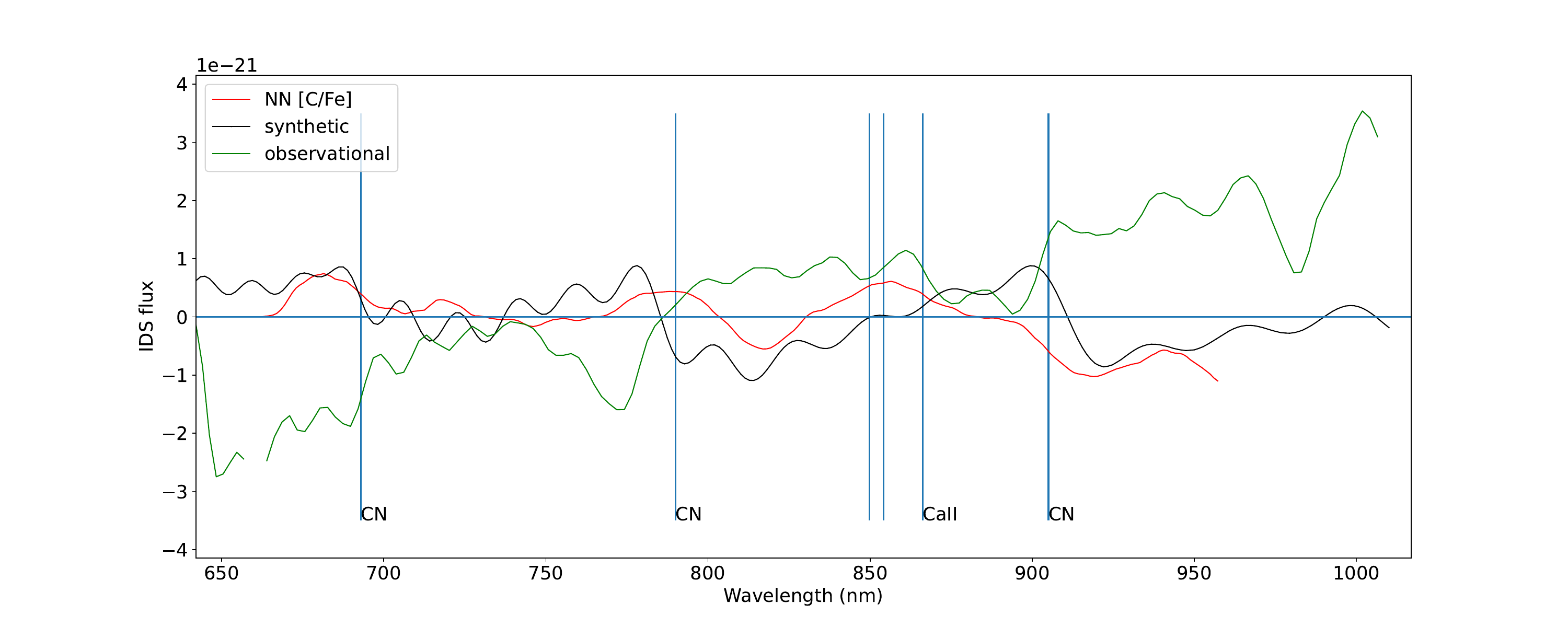}
    \caption{[C/Fe] comparisons for red clump stars ([Fe/H] $\sim$ 0.0, T $\sim$ 4750, $\log g \sim$ 2.5) between our NN's DS, synthetic spectral differences, and APOGEE observational differences.}
    \label{fig:synth_vs_NN_cfe}
\end{figure*}

\begin{figure*}
	\centering
	\includegraphics[width=.95\textwidth, trim={110, 0, 120, 50},clip]{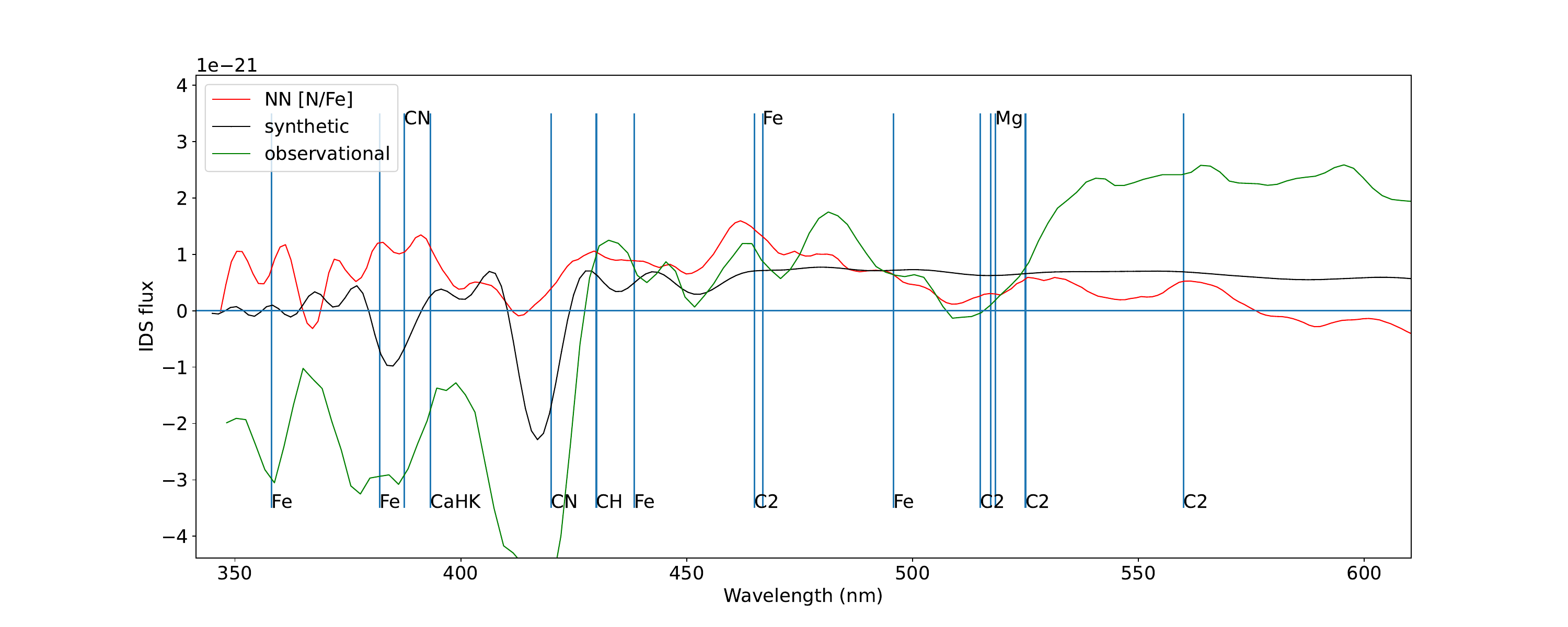}
    \includegraphics[width=.95\textwidth, trim={110, 0, 120, 50},clip]{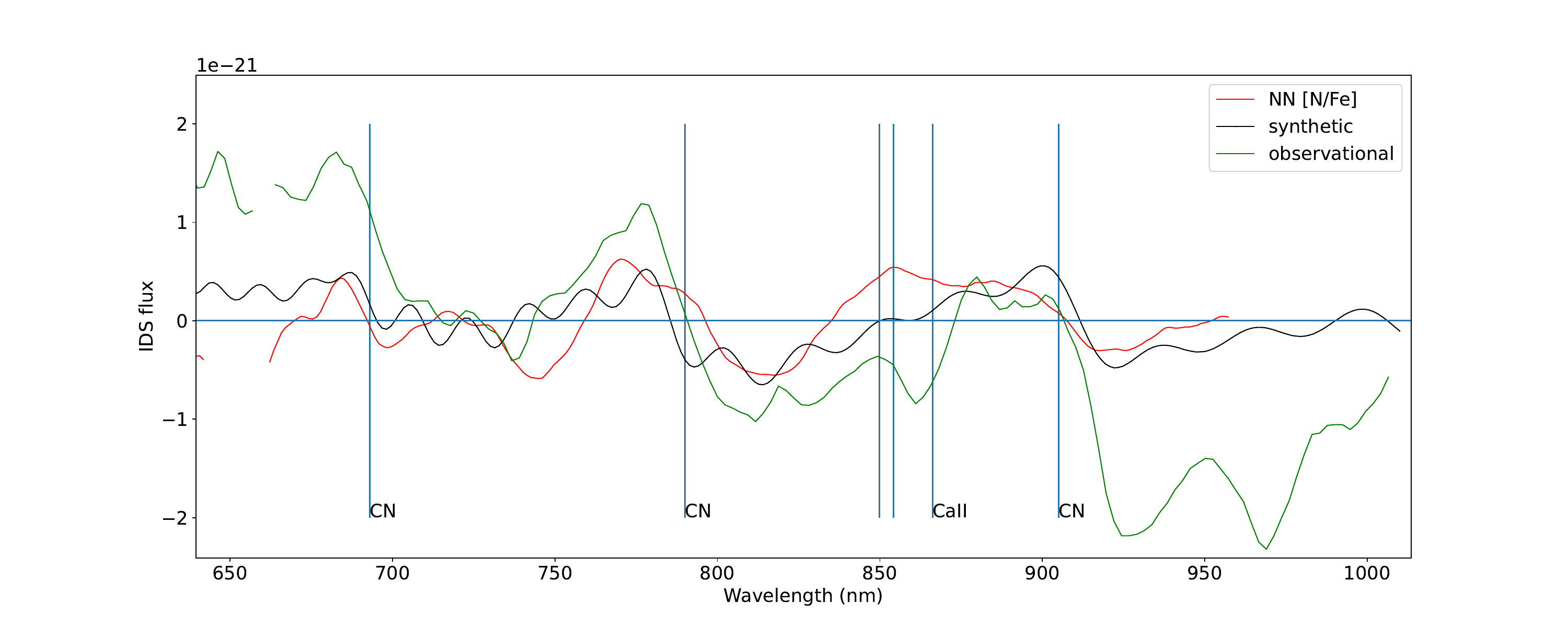}
    \caption{[N/Fe] comparisons for red clump stars ([Fe/H] $\sim$ 0.0, T $\sim$ 4750, $\log g \sim$ 2.5) between our NN's DS, synthetic spectral differences, and APOGEE observational differences.}
    \label{fig:synth_vs_NN_nfe}
\end{figure*}

\begin{figure*}
	\centering
	\includegraphics[width=.95\textwidth, trim={110, 0, 120, 50},clip]{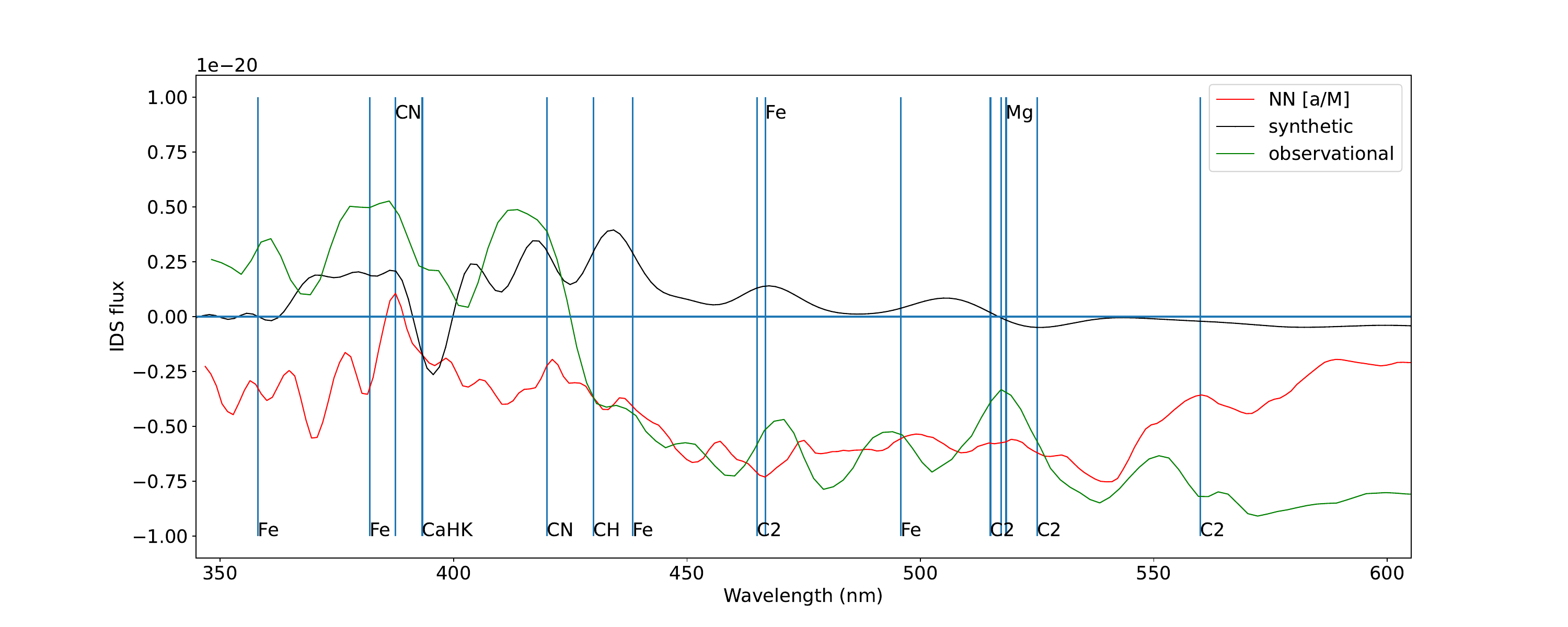}
    \includegraphics[width=.95\textwidth, trim={110, 0, 120, 50},clip]{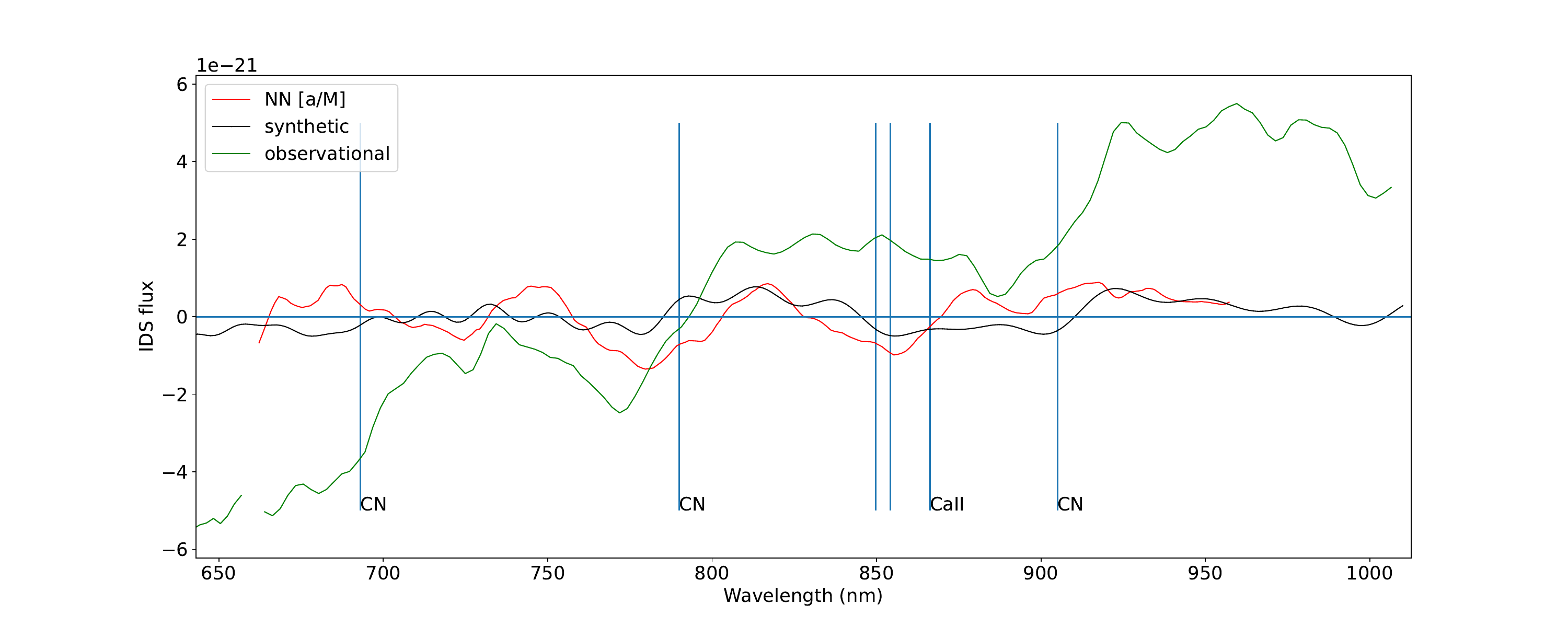}
    \caption{[$\alpha$/M] comparisons for red clump stars ([Fe/H] $\sim$ 0.0, T $\sim$ 4750, $\log g \sim$ 2.5) between our NN's DS, synthetic spectral differences, and APOGEE observational differences.}
    \label{fig:synth_vs_NN_am}
\end{figure*}

\subsection{NN Derivative Spectra}
An important question when using discriminative models is ``where is our information coming from?'' While such methods are generally powerful when applied to the predictive problems we propose, it can be particularly problematic to parse whether the model is predicting physical patterns in the data or has instead learned broader correlations in the sample. \cf{While general correlations are not necessarily a failure of the model (and still allow accurate predictions to be made), reliance on non-physical parameters can stop the model from adapting to new data. Furthermore, additional inference cannot reasonably be done for predictions dominated by non-physical trends, as it becomes difficult to disentangle useful analyses from the NN's learned prior distributions.}

If, for example, a model learns some parameter as generally correlated with a chosen input feature, the model would \rf{make precise predictions} so long as all our observed stars follow this trend. However, if some secondary stellar population is included in the data (which does not follow this broad correlation) the model's predictions will be unable to adapt to these new objects. If instead the model can learn physically meaningful features during training, then its predictions come from correlations in physical parameters rather than general trends. New populations can then be learned as a different region of physical parameter space, allowing the model to identify trends that are not present in general correlations.

\cf{To probe our model's ability to learn physically meaningful features, we investigate our NN's attention to spectral features in the BP/RP spectra. Utilising the NN gradients we have already calculated to estimate our input uncertainties in Section~\ref{sec:unc_propagation}, we can analyse the relationship between some prediction and the input features that produced it. Previously we have used the derivative $\partial y_{\mathrm{pred}}/\partial {x}$ to calculate how changes to our input features cause changes in our NN's predictions. However, to probe the model's attention to the BP/RP spectra we instead require $\partial {x}/\partial y_{\mathrm{pred}}$.
As $\partial y_{\mathrm{pred}}/\partial {x}$ calculated across the NN is a non-square matrix (due to the model having more input features than output parameters), this inversion is ill-specified. Thus, to invert $\partial y_{\mathrm{pred}}/\partial {x}$, we utilise a pseudoinverse function \citep{Penrose1955} to calculate our inverse gradient.}

\cf{This approach does require a small modification to the pseudoinverse method, due to the function computing the best fit of the inverted matrix by considering all elements of the matrix equally. As our gradients have the spectral information more heavily weighted towards lower-order coefficients \citep{Carrasco2021}, we rescale the coefficients to account for this. We select the standard deviation of each coefficient averaged over the sample as our normalisation factor. As we expect greater variation in lower-order coefficients, due to their higher sensitivity to changes in the spectra, this value can be used to scale our inverse gradients as needed.}

For a set of gradients, $\partial y_{\mathrm{pred}}/\partial {x}$, we apply a weighting term, $w^2$ equal to the standard deviation of each of our input BP/RP coefficients over all stars in our sample. This weighting term is applied along the axis of $\partial y_{\mathrm{pred}}/\partial {x}$ aligning with the input features. We apply this weighting and calculate the pseudo-inverse, $P(x)$, by
\begin{equation}
    \frac{\partial x}{\partial y_{\mathrm{pred}}} = P\left(\frac{\partial y_{\mathrm{pred}}}{\partial {x}} w^2\right)^\mathrm{T} w^2,  
	\label{eq:coeff_bias}
\end{equation}
where the transpose ensures matrix axes remain matched. Note that Eq.~\ref{eq:coeff_bias} is an approximation of our model's attention due to our choice of normalisation factor. There are a multitude of other equally valid factors that could be chosen to weight this calculation.

We therefore use $\partial {x}/\partial y_{\mathrm{pred}}$ to estimate the model's `attention' to each of the NN's inputs. This coefficient gradient is mapped into a BP/RP derivative spectrum via the \texttt{GaiaXPy} module \citep{GaiaXPy}, with the inverse gradients used as coefficients to produce our `derivative spectra' (DS). As we expect stronger correlations between our abundance predictions and elemental/molecular spectral features, we focus only on [Fe/H], [C/Fe], [N/Fe], and [$\alpha$/M] for this analysis.

\subsection{Simulated Difference}

\begin{table}
    \centering
    \caption{Table of parameter step sizes used to compute numerical derivatives for our simulated spectra. Note that we simulate enhanced alpha abundance by increasing O, Ne, Mg, Si, S, Ar, and Ca abundances by this step value.}
    \begin{tabular}{|lc|}
        \hline
                        & Step Used \\ \hline
        [Fe/H] (dex)    & $0.5$ \\ \hline
        [C/Fe] (dex)    & $0.2$ \\ \hline
        [N/Fe] (dex)    & $0.2$ \\ \hline
        [$\alpha$/M] (dex)    & $0.2$ \\ \hline
    \end{tabular}
     \label{tab:sim_diff}
\end{table}

Abundance variations cause changes to both the formation of individual line features and the overall structure and opacity of the stellar atmospheres. Model atmospheres can therefore provide a probe of these combined effects, and provide a good comparison to our NN's DS.
Using the ATLAS12 and SYNTHE modules described in \citet{Kurucz2005}, we build atmospheric models comparable to our object sample. From these simulated atmospheres, we construct spectra for stars with various properties and abundances. We then apply the \texttt{simulator} module of \texttt{GaiaXPy} \citep{GaiaXPy} to 
incorporate BP/RP spectra's line-spread function, and produce an `ideal' BP/RP spectra which we can compare to our NN's DS

We estimate the information content of our simulated spectra (and thus where we expect our model to pay attention to) as the regions where the spectra change the most for a change in each parameter. We therefore shift our simulated atmospheres by the small steps noted in Table~\ref{tab:sim_diff}. 

Additionally, we only choose to plot our derivative spectra for red clump stars, as they are the most common object in our NN's training sample. We use the [Fe/H], $\mathrm{T_{eff}}$, and $\log g$ values in Table~\ref{tab:obs_diff_bins} to define our red clump spectra, as these objects tend to have the lowest uncertainty predictions from the model.

\subsection{APOGEE Observational Difference}
\label{sec:obs_diff}
\begin{table}
    \centering
    \caption{Table of parameter bounds used to select red clump objects for our difference calculations.}
    \begin{tabular}{|lll|}
                        & Mean & Range \\ \hline
        [Fe/H] (dex)    & $0.0$ & $\pm0.5$ \\ 
        $T_\mathrm{{eff}}$ (K)    & $4750$ & $\pm50$ \\ 
        $\log g$ (dex)    & $2.5$ & $\pm0.5$ \\
        $[\mathrm{C/Fe}]$    & $0.0$ & $\pm0.2$ \\ 
        $[\mathrm{N/Fe}]$    & $0.2$ & $\pm0.2$ \\ 
    \end{tabular}
     \label{tab:obs_diff_bins}
\end{table}

For a change in observed APOGEE abundance, we can also directly compute how the BP/RP spectra change, 

We split our sample of red clump objects (defined in Table~\ref{tab:obs_diff_bins}) into `high' and `low' bins in our APOGEE abundances, with the median value of the chosen abundance being the threshold between them. The median BP/RP spectrum within the high and low bins is calculated, with the difference between describing how the spectra change with abundance. We repeat this for each of our chosen APOGEE abundances, and plot the APOGEE difference spectra alongside our DS and simulated difference spectra.  
We also limit to objects with Gaia $G$-band extinction below $0.5$ magnitudes, following the approach described in Section~\ref{ap:pseudo_mag}.

\subsection{Analysis}

We plot our NN DS, synthetic differences, and observed differences in Fig.s~\ref{fig:synth_vs_NN_feh} to \ref{fig:synth_vs_NN_am}. Note that we apply the normalisation process described in Section~\ref{sec:coeff_norm} to the BP/RP coefficients used to calculate our simulated and observed spectra so that they can be directly compared to the NN's DS.
We plot our three spectral information measures for the blue and red ends of the spectra respectively. 
The BP/RP spectra have a resolution of less than 100 \citep{Carrasco2021}, resulting in a resolving power of between 4nm and 23nm across the wavelength range. Therefore, while we expect the peaks in the model's DS to coincide with absorption features in the spectra, offsets can occur due to this limited resolving power. 

We find, across all inspected abundances, that the model's DS has smaller magnitude peaks than the APOGEE differences, as well as often diverging significantly from the synthetic spectra.
Furthermore, peaks in the DS tend to be narrower around spectral features, while our APOGEE difference spectra tend to show broader features. This suggests our model focuses its attention on a smaller number of features within the spectra, rather than drawing information from every available elemental or molecular peak. As we have previously detailed the high \rf{precision} of the model's predictions, this appears to suggest the NN does not require all of the available spectral information to \rf{successfully predict stellar properties}. Either the model is extracting all necessary information from the identified spectral regions, or is drawing information from elsewhere in the input data.
For example, abundances of carbon and nitrogen are known to evolve along the giant branch (as seen in Fig.~\ref{fig:nitrogen_iron_comps}), meaning the model may be identifying correlations between these abundances and $\log g$. 

For our [Fe/H] predictions, we see the model generally focuses on the Fe features towards the blue end of the spectra, as well as slightly smaller peaks at the CaHK and CN features between 380nm and 430nm. The NN's DS does not seem to show strong attention to the Mg features ($\sim$520nm), which is apparent in the synthetic spectra. \cf{This is a known reliable indicator of metallicity \citep{Lyubimkov2005}, and so we expect it to be well correlated with [Fe/H] abundance}. Therefore, the model's lack of attention to this region is unexpected. 

Instead, the DS appears to show large peaks at the redder end of the spectra. These peaks are likely associated with the CN molecular features in this region, although these peaks do not fully correlate with the comparison synthetic spectra. Our DS seems to focus its attention on the $\sim$690nm feature, while this feature appears more weakly in the synthetic spectra. Similarly, the model's attention to the $\sim$790nm feature appears to be opposite in magnitude to the CN feature in the synthetic comparison. The only general agreement occurs between the CaII triplet \citep[a known tracer of metallicity,][]{Cenarro2002} and CN features above 830nm. We note that, while the model's DS appears to be offset from the associated CN lines, these molecular features are the band-heads of broader features that extend blueward of annotated lines. Thus, the DS peaks are still likely associated with the CN features, even if they do not line up with the band head wavelengths. While not directly associated with iron, we do note the potent correlations between iron, carbon, and nitrogen in Sec.~\ref{sec:abundance_dbns}. We can suggest, therefore, that the NN is successfully leveraging these correlations, and thus the model's attention focuses on the CN features while predicting [Fe/H].

For our model's [C/Fe] predictions, we see the NN's DS diverge quite considerably from the synthetic spectra. While the synthetic and observational spectra appear to show strong negative peaks associated with the CN, CH, and C$_{2}$ features between $\sim$420nm and $\sim$550nm, the model's DS instead seems to focus its attention on the Fe, CN, and CaHK features below 400nm. The model correlates better with the comparison spectra at redder wavelengths, with peaks associated with the CN features here. However, the NN's DS does not show notable peaks associated with strong carbon-based features in the spectra, suggesting the model is likely deriving its [C/Fe] predictions from either carbon-nitrogen-iron correlations (as noted above), or from other input data (such as the photometric data focus seen in Fig.~\ref{fig:per_coeff_derivs}). We note that the model's attention to the $W1-W2$ colour is likely due to the CO feature present in the $W2$ wavelength range.
Therefore, while the model is not seemingly identifying the expected carbon features, the NN does appear to be drawing useful information from important elemental and molecular features in the spectra. 

The NN's [N/Fe] predictions also show strong physical features in the spectra. While the three spectra are reasonably flat at blue wavelengths, the synthetic and observational spectra show a significant negative peak associated with the CN feature at $\sim$425nm. Our model's DS does show a minor negative peak associated with this feature, but it doesn't appear as strong as the difference spectra would suggest. We do, however, note significant correlation between our three spectra at red wavelengths. The DS appears to show significant attention to the CN bands in this region, which are expected to correlate strongly to nitrogen abundance.

We finally plot our DS for the [$\alpha$/M] predictions.
We see strong features correlated with CN and CaHK in both our DS and the simulated spectra, suggesting the model has successfully identified the importance of these regions. However, between $\sim$425nm and $\sim$550nm, we start to see the DS follow a broader feature present in the observational difference (but not present in the simulated spectra). This seems to show the BP/RP spectra change unexpectedly with changes in alpha elements, such that the model begins to pay attention to a broader region of the spectrum. We also note the NN's DS shows larger peaks at redder wavelengths, that can broadly be associated with CN and Ca features. These features are notably similar to the DS for our [Fe/H] predictions at redder wavelengths suggesting the model may be using broader correlations with [Fe/H] in its alpha abundance predictions. Such correlations provide information to the model, and thus the NN's attention will be drawn to spectral features (despite them not being directly related to alpha elements).

Overall, we see our model appears to identify physically significant elemental and molecular features in the BP/RP spectra, especially for our iron and nitrogen abundance predictions. While we note there are discrepancies between the model's DS and the expected information in the BP/RP spectra (from both synthetic calculations and directly inspecting the data), the model's attention always appears focused on elemental and molecular features. We therefore suggest our model's predictions are generally based on the detection of physically meaningful regions of the spectra. It is clear that, while this approach does allow us to determine the model's attention within the BP/RP spectral data, it does not describe \emph{all} of the information the model is using for its predictions, as both non-BP/RP input features and internally identified correlations are missing from the DS method.

\subsection{A Note on BP/RP Information Content}

Previous works \citep[e.g.][]{Witten2022} have aimed to identify the potential sensitivity of the BP/RP spectra to changes in elemental abundance. From this prior work, there is an expectation of good metallicity and carbon abundance sensitivity for brighter, cooler stars, with scope for identifying carbon-enhanced metal-poor objects from BP/RP data alone \citep{Lucey2023}. \cite{Witten2022} also note that, while alpha abundance should be identifiable from the BP/RP data, this is only true for metal-rich, cool objects. For higher-temperature (T > 5000K), metal-poor ([Fe/H] $\leq$ -2.0) objects they show expected alpha abundance uncertainty to be too high for meaningful use or analysis. 

From our APOGEE difference spectra, we see generally large peaks associated with Fe, CN, and $\mathrm{C_2}$ features. These features are prominent for changes in all four of our elemental abundances, suggesting the BP/RP spectra are sensitive to Fe, C, and N elements. As described by \citet{Ting2018}, this sensitivity introduces further correlations with oxygen abundance due to the CNO equilibrium in stellar atmospheres. As oxygen presence within the atmosphere causes the production of CO molecules, there is an altered abundance of CH, CN, and $\mathrm{C_2}$ molecules. Therefore, while extracting oxygen abundance directly from the BP/RP isn't possible, information is encoded in the CN and $\mathrm{C_2}$ peaks we do identify in our spectra.

We also note these spectra do not appear to show prominent peaks for Ca features for any of our chosen abundance changes. The BP/RP spectra therefore likely do not have the necessary sensitivity to extract these features, which are found to be correlated with stellar metallicity \citep{Cenarro2002}. We also see a weaker peak associated with Mg for our [Fe/H] variation spectra, but find it generally missing on our other abundance differences. As noted by \citet{Ting2022}, there are very strong correlations between [Fe/H] \& [Mg/Fe] abundances and abundances of O and Si, suggesting the BP/RP spectra are also generally insensitive to directly extracting alpha abundance.

We therefore find general agreement with \citet{Witten2022} that, for our sample of objects, the BP/RP spectra appear sensitive to iron, carbon, and nitrogen abundances. We also agree with the general limited ability of the BP/RP data to extract alpha elements, excluding the correlations between C, N, and O discussed above.

\section{Conclusions}\label{sec:conclusions}

Our machine learning method has successfully derived \rf{precise} stellar atmospheric parameters using Gaia BP/RP spectra and selected broadband photometric colours from Gaia, 2MASS and WISE. This approach has leveraged statistical methods to produce reliable prediction uncertainties arising from random input uncertainties (aleatoric) and systematic (epistemic) uncertainties. The median uncertainties in the model's predictions, for our six predicted stellar parameters, are provided in Table~\ref{tab:APOGEE}. We have also provided a catalogue of our model's predictions applied to the full Gaia BP/RP dataset at \url{https://zenodo.org/doi/10.5281/zenodo.10471095}.

These predicted stellar atmospheric parameters have been validated against results from the APOGEE and LAMOST spectroscopic surveys, the Gaia GSP-Phot parameters, and through analysis of the physical trends observed in the model's predictions. From these comparisons, we have found generally very good agreement between our method and these higher-resolution surveys, with any notable biases following the observed divergences between different survey results. We have demonstrated that the provided uncertainties are well-calibrated and typically smaller uncertainties are predicted where more training data is available. 

Finally, we investigated the model's ability to extract physical information from the training data by mapping its `attention' across the BP/RP spectra. Despite only an approximate method being employed, we found significant correlations between the model's attention and prominent elemental and molecular features in the spectra. We further identified weaknesses in this approach, as we were unable to fully account for all information available to the trained model. These additional information sources, such as photometric colours or the model's internally identified parameter correlations, move the model's attention away from the BP/RP spectral features, causing attention to appear lower than may be expected when compared to simulated data.

\subsection{Method Improvements}

While we are confident in our method's ability to \rf{precise} recover traditionally spectroscopically derived parameters from photometry and low-resolution Gaia spectra, there are some weaknesses we wish to discuss.

Primarily, as we are utilising machine learning algorithms, our model's success is highly dependent on our chosen training data. 
We focus our training on recovering APOGEE-derived parameters, due to the large available sample size and high accuracy measurements. 
For a well-trained model, the predictions end up strongly modelling the training dataset, including any systematic or methodological quirks present. Future applications of this method may wish to incorporate a broader sample of surveys for use as training data, sacrificing the model's \rf{high precision compared to the APOGEE sample} to allow greater applicability across a wider range of stellar types.

Additionally, due to our model being discriminative, we also face the limitations of such algorithms. Namely, that during training, the model incorporates the prior distribution of the training data into its predictions. Thus, when the model is dealing with biased or high uncertainty input data, it can become over-reliant on this prior when making its predictions. This leads to the pattern of predictions trending to the sample mean we have discussed in previous sections. The primary solution to limit the impact of this effect is to train the model with unbiased training data. This is, however, difficult to apply to models using observational data, as all survey data is biased by the selection effects of the chosen instruments. Instead, using alternative model architectures, especially generative algorithms like Normalising Flows \citep{NormFlow}, can somewhat alleviate these issues by directly incorporating prior distributions into its predictions - greatly limiting their effect on the final predictions.

Alternative model architectures can also be used to improve the efficiency of our DS calculations. As generative architectures (such as autoencoders from \citet{Kramer1991}) are symmetric, they effectively train both a `forward' and an `inverse' model simultaneously. Gradient calculations using such models would be significantly simpler than our method, as they would not need to make use of a pseudoinverse calculation to derive the model's attention.

\subsection{Final Thoughts}

Our model has once again proven the power of machine learning methods when applied to astrophysical observations, with our approach allowing reliable uncertainty measurements to be determined through data-driven analysis. These uncertainties further enhance the applicability of our predictions to astrophysical problems and situations.

With \rf{precise} estimations of several significant chemical abundances, our method can be applied to identify and analyse a wide range of populations within the Milky Way. Especially notable is the potential to isolate stellar populations from accreted structures (such as the  `Gaia-Enceladus Sausage' \citep{Belokurov2018} and `Sequoia' \citep{Myeong2019} merger remnants) through their \cf{divergent abundance profiles} when compared to in-situ Galactic populations.

Future work will make great use of upcoming Gaia data releases, which will likely enhance the accuracy and observational depth available for BP/RP spectral observations. Methods such as ours have the potential to \rf{precisely} measure stellar atmospheric parameters for a huge sample of stars, especially those that may lie outside of the observational range of high-resolution spectrographs.

\section*{Acknowledgements}
We would like to thank our anonymous referee for their insightful guidance and corrections.

J.L.S. acknowledges support from the Royal Society (URF\textbackslash R1\textbackslash191555).

This paper made use of the Whole Sky Database (wsdb) created by Sergey Koposov and maintained at the Institute of Astronomy, Cambridge by Sergey Koposov, Vasily Belokurov and Wyn Evans with financial support from the Science \& Technology Facilities Council (STFC) and the European Research Council (ERC).

This research was supported in part at KITP by the Heising-Simons Foundation and the National Science Foundation under Grant No. NSF PHY-1748958.

This work has made use of data from the European Space Agency (ESA) mission
{\it Gaia} (\url{https://www.cosmos.esa.int/gaia}), processed by the {\it Gaia}
Data Processing and Analysis Consortium (DPAC,
\url{https://www.cosmos.esa.int/web/gaia/dpac/consortium}). Funding for the DPAC
has been provided by national institutions, in particular the institutions
participating in the {\it Gaia} Multilateral Agreement.

This publication makes use of data products from the Two Micron All Sky Survey, which is a joint project of the University of Massachusetts and the Infrared Processing and Analysis Center/California Institute of Technology, funded by the National Aeronautics and Space Administration and the National Science Foundation.

This publication makes use of data products from the Wide-field Infrared Survey Explorer, which is a joint project of the University of California, Los Angeles, and the Jet Propulsion Laboratory/California Institute of Technology, funded by the National Aeronautics and Space Administration. 

Funding for the Sloan Digital Sky Survey IV has been provided by the Alfred P. Sloan Foundation, the U.S. Department of Energy Office of Science, and the Participating Institutions. SDSS-IV acknowledges support and resources from the Center for High Performance Computing  at the University of Utah. The SDSS website is www.sdss.org.
SDSS-IV is managed by the Astrophysical Research Consortium for the Participating Institutions of the SDSS Collaboration including the Brazilian Participation Group, the Carnegie Institution for Science, Carnegie Mellon University, Center for Astrophysics | Harvard \& Smithsonian, the Chilean Participation Group, the French Participation Group, Instituto de Astrof\'isica de Canarias, The Johns Hopkins University, Kavli Institute for the Physics and Mathematics of the Universe (IPMU) / University of Tokyo, the Korean Participation Group, Lawrence Berkeley National Laboratory, Leibniz Institut f\"ur Astrophysik Potsdam (AIP),  Max-Planck-Institut f\"ur Astronomie (MPIA Heidelberg), Max-Planck-Institut f\"ur Astrophysik (MPA Garching), Max-Planck-Institut f\"ur Extraterrestrische Physik (MPE), National Astronomical Observatories of China, New Mexico State University, New York University, University of Notre Dame, Observat\'ario Nacional / MCTI, The Ohio State University, Pennsylvania State University, Shanghai Astronomical Observatory, United Kingdom Participation Group, Universidad Nacional Aut\'onoma de M\'exico, University of Arizona, University of Colorado Boulder, University of Oxford, University of Portsmouth, University of Utah, University of Virginia, University of Washington, University of Wisconsin, Vanderbilt University, and Yale University.

Guoshoujing Telescope (the Large Sky Area Multi-Object Fiber Spectroscopic Telescope LAMOST) is a National Major Scientific Project built by the Chinese Academy of Sciences. Funding for the project has been provided by the National Development and Reform Commission. LAMOST is operated and managed by the National Astronomical Observatories, Chinese Academy of Sciences.


\section*{Data Availability}

The datasets used in this article were derived from sources in the public domain: Gaia DR3, \url{https://gea.esac.esa.int/archive/}; UnWISE, \url{https://catalog.unwise.me/catalogs.html}; 2MASS, \url{https://irsa.ipac.caltech.edu/Missions/2mass.html}; LAMOST DR7, \url{http://dr7.lamost.org/catalogue}; SDSS DR17/APOGEE, \url{https://skyserver.sdss.org/dr17}.

The catalogue produced by this work is available at \url{https://zenodo.org/doi/10.5281/zenodo.10471095}.



\bibliographystyle{mnras}
\bibliography{MAIN} 




\appendix


\section{Pseudo-absolute magnitude}
\label{ap:pseudo_mag}
\cf{While the apparent brightness of a star may introduce unhelpful trends to our model's learning, we find there is utility in providing the NN with a sense of an object's \emph{absolute} brightness. Absolute magnitude is non-linear in the parallax, which adds significant complexity to propagating uncertainties between parallax and absolute magnitude.} Instead, we construct a `pseudo'-absolute magnitude \citep{Arenou1999} from Gaia parallaxes and photometric magnitudes. For a Gaia parallax, $\varpi$, and extinction-corrected photometric magnitude (in this case, the Gaia RP-band magnitude), the pseudo-absolute magnitude, $M_{\mathrm{RP},\mathrm{pseudo}}$, is defined as 
\begin{equation}
   M_{\mathrm{RP},\mathrm{pseudo}}=\varpi10^{\;0.2\;G_\mathrm{RP}}.
   \label{eq:pseudo}
\end{equation}
\cf{Note that here, unlike for the BP/RP coefficients, we do apply an extinction correction to the $RP$ photometry using the Rayleigh-Jeans Colour Excess (RJCE) method \citep{Majewski2011} (using the 2MASS $H$ and WISE $W2$ measurements for our sample) and the extinction coefficients provided by \citet{Wang2019}.}
This has an advantage over other measures of absolute brightness, as uncertainties relate linearly between Gaia parallax and pseudo-absolute magnitudes. Thus, our uncertainty in psuedo-absolute magnitude is exactly $\sigma_{\mathrm{RP,pseudo}}=(\sigma_\varpi/\varpi) M_{\mathrm{RP},\mathrm{pseudo}}$.

With the addition of the pseudo-absolute magnitude to the input features, we find the network's $\log g$ predictions are more accurate. This is due to absolute magnitude allowing easier differentiation between stellar types, such as separating giants and main-sequence objects. 

\begin{figure*}
	\centering
	\includegraphics[width=.49\textwidth, trim={0, 0, 40, 0}]{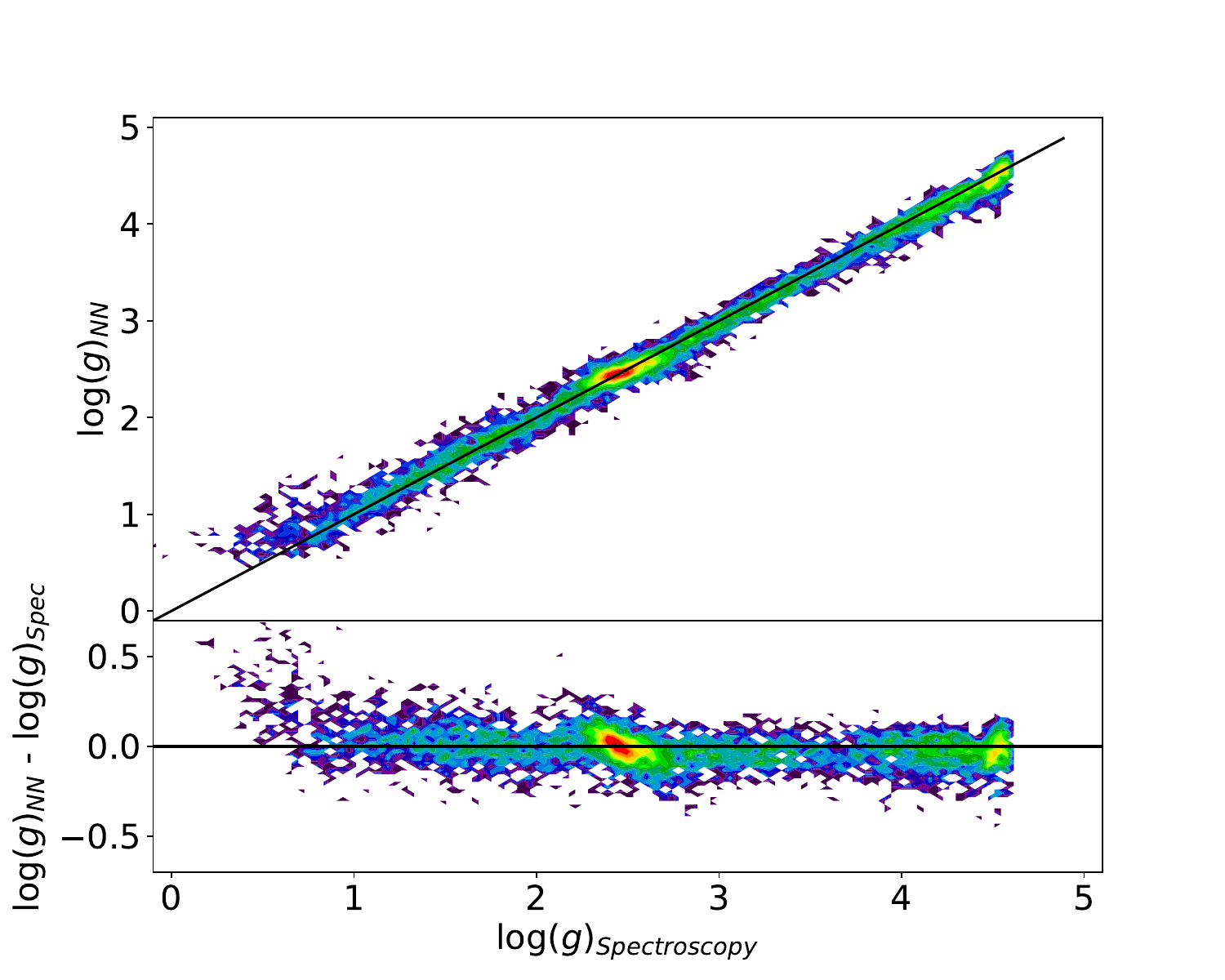}
	\includegraphics[width=.49\textwidth, trim={0, 0, 40, 0}, clip]{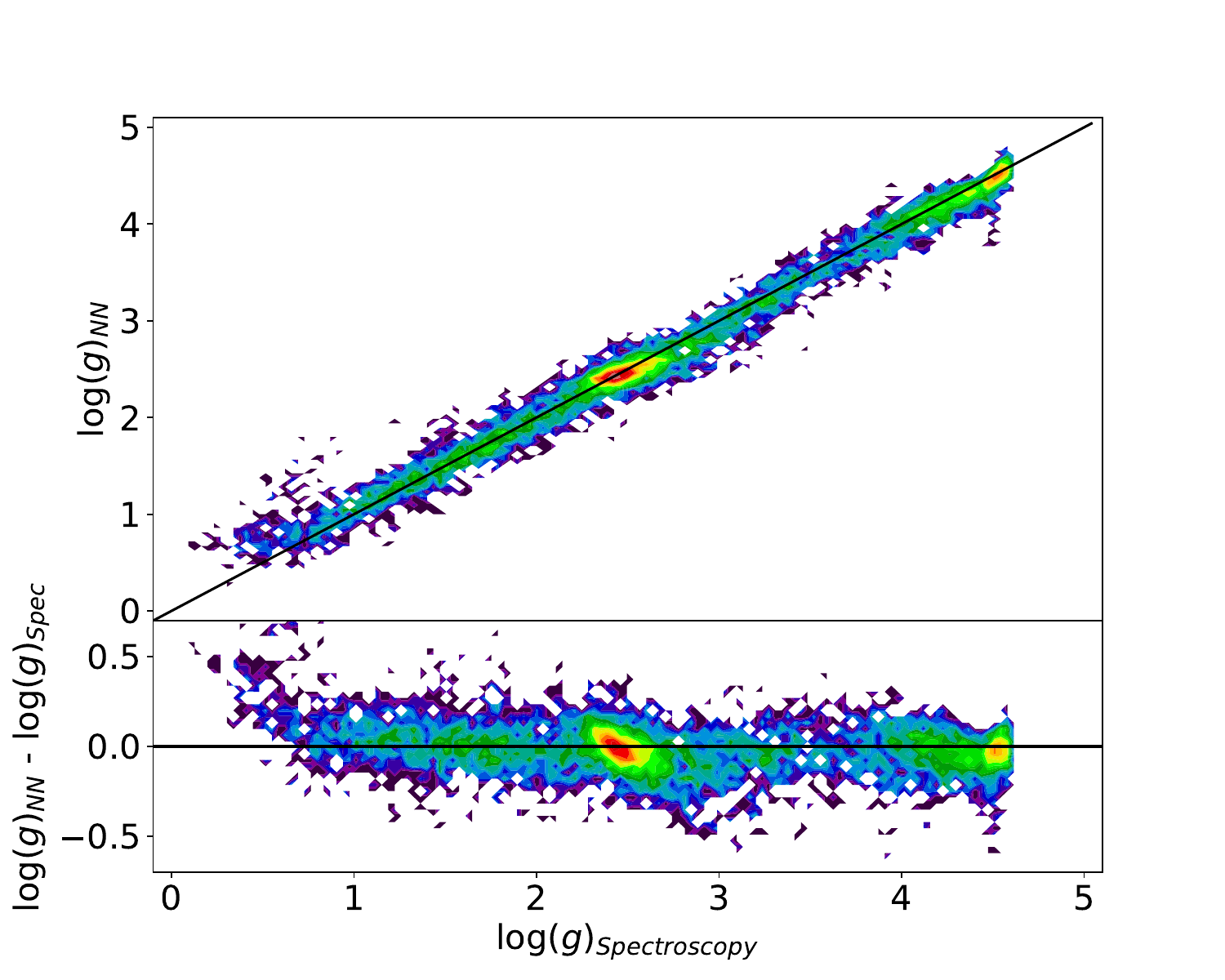}
    \caption{Comparison between the NN's $\log(g)$ predictions against values from APOGEE, for a model trained with (left) and without (right) pseudo-absolute magnitude as an input feature. Each figure shows a direct comparison in its upper panel, and the residual difference in its lower panel. Note that the contours on these plots are on a logarithmic scale. This comparison uses a subsample of $\sim$40,000 APOGEE objects for both figures. We see a reduction in the RMSE between the NN and APOGEE, from 0.10 without pseudo-absolute magnitudes, to 0.072 with them included.}
    \label{fig:LogG_Amk_comp_AmkInputs}
\end{figure*}

We plot this improvement in $\log(g)$ predictions in Fig.~\ref{fig:LogG_Amk_comp_AmkInputs}, and see a notable increase in the model's performance: \cf{a 29\% reduction in the residual RMSE (from 0.10 to 0.071), and a 12\% reduction in the uncertainty reported by the NN (from $\pm$0.145 to $\pm$0.127).} Notably, we see a reduction in the more significant divergences between the NN and APOGEE around $\log(g) \sim 3.0$ and $\sim 4.5$. This improvement suggests the NN's predictions are more informed with our pseudo-magnitude included, and are therefore less likely to be biased towards $\log(g)$ values corresponding to the two data overdensities ($\log(g) \sim 2.5$ for red giant stars, and $\sim4.5$ for main sequence). 

\bsp	
\label{lastpage}
\end{document}